\documentclass[prd,twocolumn,preprintnumbers,showpacs,floatfix
]{revtex4}
\usepackage[colorlinks=true,linkcolor=blue,citecolor=blue,urlcolor=blue]{hyperref}
\usepackage{gensymb,comment}
\usepackage{amsmath,amssymb,epsfig,bm,mathrsfs,color}
\usepackage{slashed}

\newcommand{\be}{\begin{equation}}
\newcommand{\ee}{\end{equation}}
\newcommand{\bea}{\begin{eqnarray}}
\newcommand{\eea}{\end{eqnarray}}
\newcommand{\half}{\frac1 2}

\newcommand{\ep}{\epsilon}

\newcommand{\vecq}{{\bm q}}
\newcommand{\vecp}{{\bm p}}

\newcommand{\veck}{{\bm k}}

\newcommand{\vectau}{{\bm \tau}}
\newcommand{\vecrho}{{\bm \rho}}

\newcommand{\ie}{{\it i.e.}}
\newcommand{\eg}{{\it e.g.}}

\newcommand{\MeV}{\text{MeV}}
\newcommand{\kHz}{\text{kHz}}
\newcommand{\aap}{Astron.\&~Astrophys.}
\newcommand{\mnras}{Mon.~Not.~RAS}
\newcommand{\apjl}{ApJ~Lett.}
\newcommand{\nphysa}{Nucl.~Phys.~A}

\definecolor{red}{rgb}{0.8,0,0}
\definecolor{orange}{rgb}{0.8,0.2,0.0}
\definecolor{blue}{rgb}{0.3,0.0,0.8}
\definecolor{green}{rgb}{0,0.5,0.0}
\definecolor{darkred}{rgb}{0.7,.1,.2}
\definecolor{bgred}{rgb}{1.,.95,.95}
\definecolor{bgblue}{rgb}{.95,.95,1.}
\definecolor{bluegreen}{rgb}{0.,.5,.3}
\definecolor{darkred}{rgb}{0.7,.1,.2}
\definecolor{darkgreen}{rgb}{0.1,.6,.0}
\definecolor{lightyellow}{rgb}{1.,1.,.8}
\definecolor{darkcyan}{rgb}{0.,.7,.9}
\definecolor{lightblue}{rgb}{0.6,0.8,1}
\definecolor{lightgreen}{rgb}{0.7,1.,.9}
\definecolor{money}{rgb}{0.4,0.8,0.}
\definecolor{purple}{rgb}{0.9,0.0,0.8}
\definecolor{orange}{rgb}{0.9,0.5,0.0}
\definecolor{newgr}{rgb}{0.2,0.8,0.2}
\definecolor{newbl}{rgb}{0.3,0.6,0.8}
\definecolor{newor}{rgb}{1.0,0.6,0.}

\usepackage[normalem]{ulem}  

\begin{document}

\title{ Bulk viscosity from Urca processes: $npe\mu$ matter in the neutrino-transparent regime}

\author{Mark Alford} \email{alford@physics.wustl.edu}
\affiliation{Department of Physics, Washington University, St.~Louis,
  Missouri 63130, USA}

\author{ Arus Harutyunyan} \email{arus@bao.sci.am}
\affiliation{Byurakan Astrophysical Observatory,
  Byurakan 0213, Armenia\\
  Department of Physics, Yerevan State University, Yerevan 0025, Armenia}

\author{ Armen Sedrakian}
\email{sedrakian@fias.uni-frankfurt.de}
\affiliation{Frankfurt Institute for Advanced Studies, D-60438
  Frankfurt am Main, Germany\\
 Institute of Theoretical Physics, University of Wroclaw,
50-204 Wroclaw, Poland}

\date{21 June 2023 } 

\begin{abstract}

  We study the bulk viscosity of moderately hot and dense,
  neutrino-transparent relativistic $npe\mu$ matter arising from
  weak-interaction direct Urca processes. This work parallels our
  recent study of the bulk viscosity of $npe\mu$ matter with a trapped
  neutrino component. The nuclear matter is modeled in a relativistic
  density functional approach with two different parametrizations --
  DDME2 (which does not allow for the low-temperature direct-Urca
  process at any density) and NL3 (which allows for low-temperature
  direct-Urca process above a low-density threshold).  We compute the
  equilibration rates of Urca processes of neutron decay and lepton
  capture, as well as the rate of the muon decay, and find that the
  muon decay process is subdominant to the Urca processes at
  temperatures $T\geq 3$~MeV in the case of DDME2 model and
  $T\geq 1$~MeV in the case of NL3 model. Thus, the
  Urca-process-driven bulk viscosity is computed with the assumption
  that pure leptonic reactions are frozen. As a result the electronic
  and muonic Urca channels contribute to the bulk viscosity
  independently and at 
  certain densities the bulk viscosity of
  $npe\mu$ matter shows  instead of the standard one-peak (resonant) form a ``flattened" shape.  In
  the final step, we estimate the damping timescales of density
  oscillations by the bulk viscosity. We find that, \eg, at a typical
  oscillation frequency $f=1$~kHz, the damping of oscillations is most
  efficient at temperatures $3\leq T\leq 5$~MeV and densities
  $n_B\leq 2n_0$ where they can affect the evolution of the
  post-merger object.

\end{abstract}

\maketitle

\section{Introduction}
\label{sec:intro}

The recent detections of gravitational waves and their electromagnetic
counterparts produced in binary neutron-star (BNS) mergers by the
LIGO-Virgo collaboration motivates studies of the properties of hot
and dense nuclear matter (for reviews see
\cite{Oertel2017,Lovato2022,Sedrakian2023}  and  input
models for simulations see \cite{Dexheimer2022}).  Numerical
simulations of BNS mergers performed in the framework of
nondissipative hydrodynamics~\cite{Perego:2019adq,
  Hanauske:2019qgs,Kastaun:2016elu,Bernuzzi:2015opx,
  Foucart:2015gaa,Kiuchi:2012mk,Sekiguchi:2011zd,Ruiz2016,East:2016,
  Most2019,Bauswein2019,Endrizzi2018,Ciolfi2019,Tsokaros2019} (for
reviews see \cite{Baiotti:2016qnr,Baiotti2019,Faber2012:lrr}) predict
large-amplitude density oscillations and intense gravitational wave
emission during the first tens of milliseconds of the post-merger
evolution. The density oscillations eventually will be damped by
dissipative processes in post-merger matter which will affect the
gravitational wave signal. Among various dissipative processes, bulk
viscous dissipation by weak interactions is likely to be the most
efficient mechanism in damping the density oscillations in post-merger
matter as it follows from initial estimates~\cite{Alford2018a} and
more recent implementations in the numerical
simulations~\cite{Celora2022,Most2022,Camelio2023a,Camelio2023b,Hammond2023}.
In the cold regime, relevant for mature compact stars, bulk viscosity
has been extensively studied following the seminal work of Ref.~\cite{Sawyer1989}.
Bulk viscosity of hot and dense matter in various regimes was computed
in several recent
works~\cite{Alford2019a,Alford2019b,Alford2021a,Alford2021c,Alford2022,Alford:2023gxq}
either in the neutrino transparent or trapped regimes. Results for the
bulk viscosity and damping timescales that interpolate between these
regimes and cover the entire temperature range were given recently in
Refs.~\cite{Alford2020,Alford2022}.

Here we extend our recent work~\cite{Alford2021c} on the influence of
the muonic component on the bulk viscosity of neutron-proton-electron
matter from the neutrino-trapped to the neutrino-transparent
regime. Matter is transparent to neutrinos at intermediate
temperatures $1\leq T\leq 10$~MeV and as already demonstrated in the
previous investigations~\cite{Alford2020,Alford2019a} the bulk viscous
damping is expected to be most efficient in this regime. 
The results that we present show the
likely importance of bulk viscous damping arising from
beta equilibration via weak interactions. Exactly how the physics of beta equilibration should be included in merger simulations is a separate question that we do not address here.

Some of the
results reported here were previewed in a review
article~\cite{Alford2022} which reported results for the entire range
of temperatures relevant for binary neutron star mergers by
interpolating between neutrino-transparent and neutrino-trapped
regimes using the DDME2 parametrization of the nuclear density
functional. Here we expand on this discussion by (a)~adding results
obtained with an alternative NL3 density functional which allows us to
assess the uncertainties associated with the choice of the density
functional; (b) focusing on the neutrino-transparent regime we provide
details of the derivations of the rates for processes involving muons
in Subsec.~\ref{sec:mu-decay} and \ref{sec:Muon_decay_rate} and bulk
viscosity in Subsec.~\ref{sec:bulk_muons}. Finally, the Appendices
contain details of the derivation of rates of processes together with
their low-temperature limits as well as the susceptibilities in the
isothermal and isentropic cases needed for the evaluation of the bulk
viscosity.

Below, we use the same formalism as in 
Ref.~\cite{Alford2021c} by keeping track of three types of processes:
(a) the nucleonic Urca process on electrons, (b) the nucleonic Urca
processes on muons, and (c) purely leptonic processes, all in the
neutrino-transparent regime. It will turn out that the muon decay
rate is much smaller than the Urca process rates on electrons and
muons in the entire temperature-density range. This simplifies the
treatment of the coupled network of reactions, as the purely leptonic
processes can be considered as decoupled on the time scales that are
characteristic for Urca processes. The importance of the extension to
the neutrino-transparent regime lies in the fact (confirmed by
explicit computations below) that in this regime the bulk viscous
damping timescale is short (in the range $1$-$10$ ms), therefore  
the bulk viscosity may have a significant impact
on the initial phase of post-merger dynamics which is characterized 
by a typical timescale
$\sim$10\,ms (see also the earlier work~\cite{Alford2020,Alford2019a}
where the muonic Urca processes were excluded).

This paper is organized as follows. In Sec.~\ref{sec:urca_rates} we discuss 
the rates of the weak processes, specifically, those of the direct Urca 
processes and the muon decay. In Sec.~\ref{sec:bulk} we briefly review 
the derivation of the bulk viscosity of $npe\mu$ matter. 
Section~\ref{sec:num_results} collects our results of the weak process rates, 
the bulk viscosity, and the damping timescales of density oscillations for 
two EoS models based on the density functional theory. Our results are summarized 
in Sec.~\ref{sec:conclusions}. Appendix~\ref{app:rates_low} provides the 
derivation of the weak process rates in the degenerate matter.  
Appendix~\ref{app:A_coeff} details the computation of the relevant 
susceptibilities in both cases of isothermal and adiabatic oscillations.
We use natural (Gaussian) units with $\hbar = c = k_B = 1$ and the metric 
$g_{\mu\nu} = \textrm{diag}(1, -1, -1, -1)$.

\begin{widetext}

\section{Weak processes in $npe\mu$ matter}
\label{sec:urca_rates}

Consider neutron-star matter composed of neutrons, protons, electrons,
and muons in the density range $0.5n_0\leq n_B\leq 5 n_0$ where $n_0$
is the nuclear saturation density (which is a parameter of the density
functionals considered) and the temperature range $1\leq T\leq 10$
MeV. In this temperature-density range, the matter is conjectured to
be transparent for neutrinos.

The simplest semi-baryonic beta equilibration processes are the 
direct Urca processes of neutron decay and lepton capture, respectively
\bea\label{eq:n_decay}
&& n\rightarrow p + l^-+\bar{\nu}_l,\\
\label{eq:l_capture}
&& p + l^-\rightarrow n+{\nu}_l,
\eea
where $l^-=\{e^-,\mu^-\}$ is electron or muon, $\nu_l$ is the corresponding 
neutrino. There are also modified Urca processes, which we discuss in Sec.~\ref{sec:bulk_muons}.

In addition, the purely leptonic muon-decay process 
\bea\label{eq:mu_decay}
&& \mu^- \rightarrow e^-+\bar{\nu}_e +{\nu}_\mu
\eea
takes place. 
The opposite process $e^- \rightarrow \mu^-+{\nu}_e +\bar{{\nu}}_\mu$ does not occur because it is forbidden by energy conservation: in the rest frame of the initial state electron there is
not enough energy to create the final state particles.
The processes~\eqref{eq:n_decay}--\eqref{eq:mu_decay}
proceed only in the direction from left to right because
in 
neutrino-transparent matter neutrinos/antineutrinos can appear only in final states.

\subsection{Urca processes}

The  rates of the processes~\eqref{eq:n_decay}
and \eqref{eq:l_capture} are given, respectively, by~(see~Ref.~\cite{Shapiro1983}, Chapt. 7)
\bea \label{eq:Gamma1_def}
\Gamma_{n\to pl\bar\nu} &=& \int\!\! \frac{d^3p}{(2\pi)^32p_0} \int\!\!
\frac{d^3p'}{(2\pi)^32p'_0} \int\!\! \frac{d^3k}{(2\pi)^32k_0}
\int\!\! \frac{d^3k'}{(2\pi)^32k'_0}\sum \vert 
{\cal M}_{\rm Urca}\vert^2 \nonumber\\
& \times & \bar{f}(k)\bar{f}(p) \bar{f}(k') f(p') 
(2\pi)^4\delta^{(4)}(k+p+k'-p'),\\
\label{eq:Gamma2_def}
\Gamma_{pl\to n\nu} &=& \int\!\! \frac{d^3p}{(2\pi)^32p_0} \int\!\!
\frac{d^3p'}{(2\pi)^32p'_0} \int\!\! \frac{d^3k}{(2\pi)^32k_0}
\int\!\! \frac{d^3k'}{(2\pi)^32k'_0}
\sum \vert {\cal M}_{\rm Urca}\vert^2 \nonumber\\
& \times & {f}(k) {f}(p) \bar{f}(k') \bar{f}(p')
(2\pi)^4\delta(k+p-k'-p').
\eea
where $f(p) = \{\exp[(E_p-\mu)/T+1\}^{-1}$ etc. are the Fermi
distribution functions of particles, with $E_p$ being the single-particle spectrum for momentum $p$, and
$\bar{f}(p)=1-f(p)$. The mapping between the particle labeling and
their momenta is as follows: $(l) \to k$,
$(\nu_l/\bar{\nu}_l) \to k'$, $(p) \to p$, and $(n) \to p'$. Note that
in neutrino-transparent matter $\bar f(k') = 1$ in
Eqs.~\eqref{eq:Gamma1_def} and \eqref{eq:Gamma2_def}.

The spin-averaged relativistic matrix element of
the Urca processes reads~\cite{Greiner2000gauge}
\be\label{eq:matrix_el_full}
\sum \vert {\cal M}_{\rm Urca}\vert^2 = 32 G_F^2\cos^2
\theta_c \left[(1+g_A )^2(k\cdot p) (k'\cdot p')
+(1-g_A)^2(k\cdot p') (k'\cdot p)
+(g_A^2-1)m^*_n m^*_p(k\cdot k')\right],
\ee 
where $G_F=1.166\cdot 10 ^{-5}$ GeV$^{-2}$ is the Fermi coupling
constant, $\theta_c$ is the Cabibbo angle with $\cos\theta_c=0.974$,
$g_A=1.26$ is the axial-vector coupling constant, and $m_n^*/m_p^*$ is
the effective neutron/proton mass. In our calculations, we will keep only the first
 term of this expression, which we expect to dominate because $g_A$ is close to 1.  The twelve-dimensional phase-space integrals in
Eqs.~\eqref{eq:Gamma1_def} and \eqref{eq:Gamma2_def} can then be reduced to
the following four-dimensional integrals which are then computed
numerically~\cite{Alford2021c}
\bea\label{eq:Gamma1_final} 
\Gamma_{n\to pl\bar\nu} (\mu_{\Delta_l})
&=& -\frac{{G}^2T^4}{(2\pi)^5} 
\int_{-\infty}^\infty\!\!\! dy\,
\!\int_0^\infty\!\! dx\, \left[(\mu_n^*+yT)^2 
 -m_n^{*2}-x^2T^2\right]\nonumber\\
&&\times \left[(\mu_l +\mu_p^* +\bar{y}_lT)^2
-m_l^2-m_p^{*2} -x^2T^2\right]\nonumber\\
&&\times \int_{m_l/T-\alpha_l}^{\alpha_p +\bar{y}_l}\! 
dz\, \bar{f}(z){f}(z-\bar{y}_l)\,\theta_x\! 
\int_{0}^\infty\! dz'\,f(z'+y)\,\theta_y,\\
\label{eq:Gamma2_final} 
\Gamma_{pl\to n\nu} (\mu_{\Delta_l}) 
&=& \frac{{G}^2T^4}{(2\pi)^5}\int_{-\infty}^\infty\!
dy\! \int_0^\infty\! dx\, \left[(\mu_n^*+yT)^2 
-m_n^{*2}-x^2T^2\right]\nonumber\\
&&\times \left[(\mu_l +\mu_p^* +\bar{y}_lT)^2
-m_l^2-m_p^{*2} -x^2T^2\right]\nonumber\\
&&\times \int_{m_l/T-\alpha_l}^{\alpha_p +\bar{y}_l}\! 
dz\, f(z)f(\bar{y}_l-z)\,\theta_x\! \int_{0}^{\alpha_n+y}\! 
dz'\, {f}(z'-y)\,\theta_z,
\eea
where $G=G_F \cos\theta_c(1+g_A^2)$, $m_l$ is the lepton mass,
$\alpha_l = \mu_l/T$, $\alpha_N= \mu_N^*/T$ for $N=\{n,p\}$ with $\mu_N^*$ being the effective nucleon mass, see Sec.~\ref{sec:rates}; 
$\bar{y}_l=y+\mu_{\Delta_l}/T$ with $\mu_{\Delta_l}=\mu_n
-\mu_p-\mu_l$, and $f(x) = (e^x+1)^{-1}$ is the Fermi distribution
function of dimensionless variable $x$.
The $\theta$-functions in Eqs.~\eqref{eq:Gamma1_final} and 
\eqref{eq:Gamma2_final} imply 
\bea\label{eq:thetax}
\theta_x &: &
(z_k-x)^2 \leq \left(z -\alpha_p 
-\bar{y}_l\right)^2 -m_p^{*2}/T^2\leq (z_k+x)^2,\\
\label{eq:thetay}
\theta_y &: &
(z'-x)^2 \leq \left(z' +\alpha_n+ y\right)^2 
-m_n^{*2}/T^2\leq (z'+x)^2,\\
\label{eq:thetaz}
\theta_z &: &
(z'-x)^2 \leq \left(z'-\alpha_n-y\right)^2 
-m_n^{*2}/T^2\leq (z'+x)^2.
\eea 
The integration variables $y$ and $x$ are normalized-by-temperature
transferred energy and momentum, respectively; the
variable $z$ is the normalized-by-temperature lepton energy, 
computed from its chemical potential, $z_k=\sqrt{(z+\alpha_l)^2
-m_l^2/T^2}$ is the normalized lepton momentum, and $z'$ is 
the normalized neutrino/antineutrino energy.

In beta-equilibrium the rates of the neutron decay and lepton capture 
should be equal: $\Gamma_{n\to pl\bar\nu}=\Gamma_{pl\to n\nu}$.  This is the case  in the 
low-temperature regime $T\ll \mu_i$ 
for  $\mu_{\Delta_l}=0$, \ie, $\mu_n=\mu_p+\mu_l$.
In that case the low-temperature limit of the Urca process rates~\eqref{eq:Gamma1_final}
and \eqref{eq:Gamma2_final}  are given by the
Fermi-surface approximation (see Appendix~\ref{app:rates_low})
\bea\label{eq:Gamma_lowT}
\Gamma_{n\to pl\bar\nu}=\Gamma_{pl\to n\nu}= \frac{\alpha}{2} 
{G}^2 T^5 \mu_n^* ({p}_{Fp}^2+p_{Fl}^2+2\mu_l \mu_{p}^*
-p_{Fn}^2)\theta(p_{Fl}+{p}_{Fp}-p_{Fn}),
\eea
where
$\alpha=3\left[\pi^2 \zeta(3) + 15 \zeta(5)\right]/16\pi^5 \simeq
0.0168$. 
However, at  higher temperatures, the Fermi-surface approximation is no longer valid;
  non-negligible neutrino momentum enters ~\eqref{eq:Gamma1_final} and
  \eqref{eq:Gamma2_final} with opposite signs. As a consequence matter is in beta-equilibrium at non-vanishing values of
$\mu_{\Delta_l}^{\rm eq}$~\cite{Alford2018b}.

For small departures from $\beta$-equilibrium $\mu_{\Delta_l}- 
\mu_{\Delta_l}^{\rm eq}\ll T$, and the net proton production rate 
can we approximated as $\Gamma_{n\to pl\bar\nu}-\Gamma_{pl\to n\nu}=\lambda_l
(\mu_{\Delta_l}- \mu_{\Delta_l}^{\rm eq})$ with the expansion coefficients 
\bea \label{eq:lambda_def}
\lambda_l = \left(\frac{\partial
\Gamma_{n\to pl\bar\nu}}
{\partial\mu_{\Delta_l}}-\frac{\partial\Gamma_{pl\to n\nu}}
{\partial\mu_{\Delta_l}}\right)\bigg\vert_{\mu_{\Delta_l}
=\mu_{\Delta_l}^{\rm eq}}.
\eea

The coefficients $\lambda_l$ in the low-$T$ limit of neutrino-transparent matter are given by
(see~Appendix~\ref{app:rates_low}) 
\bea\label{eq:lambda_lowT}
\lambda_{l} = \frac{17}{480\pi}{G}^2T^4 \mu_n^* 
({p}_{Fp}^2+p_{Fl}^2+2\mu_l \mu_{p}^*-p_{Fn}^2)
\theta(p_{Fl}+{p}_{Fp}-p_{Fn}).
\eea
In the limit of nonrelativistic nucleons
$p_{FN}\ll\mu_N^*\simeq m_N^*$ Eqs.~\eqref{eq:Gamma_lowT} and
\eqref{eq:lambda_lowT} reduce to our previous
results~\cite{Alford2019b} if the lepton mass is neglected, \ie,
$\mu_l=p_{Fl}$ (ultra-relativistic limit).

We will neglect the isospin chemical potentials
$\mu_{\Delta_l}^{\rm eq}$ below and employ the low-temperature
beta-equilibrium condition $\mu_n=\mu_p+\mu_l$.  Recent work on bulk
viscosity in muonless nuclear matter \cite{Alford:2023gxq} found that
inclusion of $\mu_{\Delta_l}^{\rm eq}$ does not affect the temperature
at which bulk viscosity achieves its maximum.

\subsection{Muon decay}
\label{sec:mu-decay}
The rate of the $\mu$-decay process~\eqref{eq:mu_decay}
is given by 
\bea \label{eq:Gamma_lep}
\Gamma_{\mu\to e\bar\nu\nu} &=&
\int\!\! \frac{d^3k_\mu}{(2\pi)^32k_{0\mu}} \int\!\! \frac{d^3k_e}
{(2\pi)^32k_{0e}} \int\!\! \frac{d^3k_{\bar{\nu}_e}}{(2\pi)^32
k_{0\bar{\nu}_e}}\int\!\! \frac{d^3k_{\nu_\mu}}{(2\pi)^32k_{0\nu_\mu}} 
\sum \vert {\cal M}_{\rm lep}\vert^2 \nonumber\\
&\times & f(k_\mu) \bar{f}(k_e)\bar{f}(k_{\bar{\nu}_e}) \bar{f}(k_{\nu_\mu}) 
(2\pi)^4\delta^{(4)}(k_e+k_{\bar{\nu}_e}+k_{\nu_\mu}-k_\mu),
\eea
with the spin-averaged scattering matrix element given by~\cite{Guo:2020tgx}
\be\label{eq:matrix_lep}
\sum \vert {\cal M}_{\rm lep}\vert^2 = 128 
G_F^2 \left(k_e \cdot k_{\nu_\mu}\right) 
\left(k_\mu \cdot k_{\bar{\nu}_e}\right).
\ee 
The final rate is given by the expression
\bea\label{eq:Gamma_lep_final} 
\Gamma_{\mu\to e\bar\nu\nu} (\mu_\Delta^L)
&=& -\frac{4{G}^2_FT^4}{(2\pi)^5} 
\int_{-\infty}^\infty\!\!\! dy\,\!\int_0^\infty\!\! dx
\int_{m_e/T-\alpha_e}^{\tilde{y}}\!\!\! dz\, 
\bar{f}(z)\,\tilde{\theta}_x\!\int_{0}^\infty\! 
dz'\,f(z'+y)\,\tilde{\theta}_y \nonumber\\
&&\times \left[(\mu_e+\tilde{y}T)^2-m_e^2-x^2T^2\right]
\left[(\mu_\mu+yT)^2 -m_\mu^{2}-x^2T^2\right],
\eea
where $\mu_{\Delta}^L\equiv \mu_\mu-\mu_e=\mu_{\Delta_e}-\mu_{\Delta_\mu}$,
$\tilde{y}=y+\mu_{\Delta}^L/T$, and the $\theta$-functions imply 
\bea\label{eq:thetaxL}
\tilde{\theta}_x &: & (z_k-x)^2 \leq 
\left(z-\tilde{y}\right)^2 \leq (z_k+x)^2,\\
\label{eq:thetayL}
\tilde{\theta}_y &: & (z'-x)^2 \leq \left(z' 
+\alpha_\mu+ y\right)^2 -m_\mu^{2}/T^2\leq (z'+x)^2,
\eea 
with $z_k=\sqrt{(z+\alpha_e)^2-m_e^2/T^2}$. In the low-temperature limit, we find
\bea\label{eq:Gamma_lep_lowT}
\Gamma_{\mu\to e\bar\nu\nu} = 
\frac{\alpha}{2}{G}^2T^5\mu_\mu 
(p_{Fe}^2-p_{F\mu}^2)\theta(p_{Fe}-p_{F\mu}).
\eea
Note that there are also ``modified Urca-type'' leptonic reactions
involving electromagnetic interaction with spectator
leptons~\cite{Alford:2010jf}. However, the total rate of these
processes is found to be at least three orders of magnitude smaller
than the rate~\eqref{eq:Gamma_lep_final}.

\section{Bulk viscosity of $npe\mu$ matter}
\label{sec:bulk}

In this section, we analyze the bulk viscosity coefficient of 
neutrino-transparent $npe\mu$ matter arising from the Urca 
processes~\eqref{eq:n_decay} and \eqref{eq:l_capture}. We
consider small-amplitude density oscillations with a frequency 
$\omega$ following the approach first proposed in
  Ref.~\cite{Sawyer1989}. Separating the oscillating parts from the static 
equilibrium values of particle densities we can write $n_j(t)=
n_{j0}+\delta n_j(t)$, where $\delta n_j(t)\sim e^{i\omega t}$, 
where $j=\{n,p,e^-,\mu^-\}$ labels the particles.

Oscillations drive the system out of chemical equilibrium 
leading to nonzero chemical imbalances $\mu_{\Delta_l}=
\delta\mu_n-\delta\mu_p-\delta\mu_l$, which can be written as
\bea\label{eq:delta_mu_l} 
\mu_{\Delta_l} &=& A_n \delta n_n 
 -A_p \delta n_p -A_l \delta n_l,
\eea 
where the particle susceptibilites are defined as $A_n=A_{nn}-A_{pn}$, 
$A_p=A_{pp}-A_{np}$, and $A_l=A_{ll}$ with 
\bea\label{eq:A_f}
A_{ij} = \frac{\partial \mu_i}{\partial n_j}, \quad
A_j = \frac{\partial \mu_j}{\partial n_j},
\eea 
and the derivatives are computed in the static equilibrium state.
The off-diagonal elements $A_{np}$ and $A_{pn}$ are 
nonzero because of the cross-species strong interaction between 
neutrons and protons. The computation of particle susceptibilities 
$A_i$ is performed in Appendix~\ref{app:A_coeff}.

If the weak processes were switched off, then the number of 
all particle species would conserve separately, which implies
\bea\label{eq:cont_j}
\frac{\partial}{\partial t} \delta {n}^0_j(t)+ \theta n_{j0} =0
\quad\Rightarrow\quad \delta {n}^0_j(t) = -\frac{\theta}{i\omega}\, n_{j0},
\eea 
where $\theta=\partial_i v^i$ is the fluid velocity divergence.
Once the weak reactions~\eqref{eq:n_decay}, \eqref{eq:l_capture} 
and \eqref{eq:mu_decay} are switched on, there is a net production 
of particles that should be included in the balance equations. 
To linear order in chemical imbalances, these equations read
\bea\label{eq:cont_n_slow}
\frac{\partial}{\partial t}\delta n_n(t) &=& 
-\theta  n_{n0} -\lambda_e\mu_{\Delta_e}(t)
-\lambda_{\mu} \mu_{\Delta_\mu}(t),\\
\label{eq:cont_p_slow}
\frac{\partial}{\partial t}\delta n_p(t) &=& 
-\theta n_{p 0} +\lambda_e\mu_{\Delta_e}(t)
+\lambda_{\mu} \mu_{\Delta_\mu}(t),\\
\label{eq:cont_e_slow}
\frac{\partial}{\partial t}\delta n_e(t) &=& 
-\theta n_{e 0} +\lambda_e\mu_{\Delta_e}(t)
+\lambda_L\mu_{\Delta}^L(t),\\
\label{eq:cont_mu_slow}
\frac{\partial}{\partial t}\delta n_\mu(t) &=& 
-\theta n_{\mu 0} +\lambda_{\mu}\mu_{\Delta_\mu}(t)
-\lambda_L\mu_{\Delta}^L(t),
\eea 
where $\lambda_l$ are defined in \eqref{eq:lambda_def} and
$\lambda_L$ is defined analogously to $\lambda_l$, \ie, 
\bea \label{eq:lambda_lep_def}
\lambda_L = \frac{\partial
\Gamma_{\mu\to e\nu\bar\nu}}
{\partial\mu_{\Delta_L}}
\bigg\vert_{\mu_{\Delta}^L=0} \ .
\eea

To proceed further we need to specify how the muon decay
reaction~\eqref{eq:mu_decay} affects the bulk viscosity from the 
Urca processes~\eqref{eq:n_decay} and \eqref{eq:l_capture}. As we 
show below, we deal typically with one of these two limiting cases: 
\begin{equation}
\begin{array}{rl}
    \text{(a) slow lepton equilibration:} & \lambda_{L}\ll \lambda_{e}, 
\lambda_\mu \\
\text{(b) slow muon equilibration:} & \lambda_L, \lambda_\mu\ll 
\lambda_e
\end{array}
\label{eq:regimes}
\end{equation}
In the case (a) the muon decay rate is
much slower than the Urca process rates, \ie, $\lambda_{L}\ll \lambda_{e}, 
\lambda_\mu$. In the case (b) the processes 
involving muons (\ie, muon decay and muonic Urca reactions) are much 
slower than electron Urca process rates $\lambda_L, \lambda_\mu\ll 
\lambda_e$. In this limiting case muons can be simply neglected and 
the bulk viscosity arises only from electronic Urca reactions. 
Below we derive the bulk viscosity in terms of equilibration 
rates and particle susceptibilities for case (a).

\subsection{Bulk viscosity in slow lepton-equilibration limit}

In this limit muon decay is too slow to contribute, so we drop the
terms proportional to $\lambda_L$ when substituting
Eq.~\eqref{eq:delta_mu_l} in Eqs.~\eqref{eq:cont_n_slow} and
\eqref{eq:cont_e_slow}.  We obtain
\bea
i\omega\delta n_n &=& 
-n_{n0}\theta -(\lambda_e+\lambda_{\mu}) A_n\delta n_n
+(\lambda_e+\lambda_{\mu}) A_p\delta n_p
+\lambda_e A_e\delta n_e+\lambda_{\mu} A_\mu\delta n_\mu,\\
i\omega\delta n_e &=& -n_{e0}\theta +\lambda_e A_n\delta n_n
- \lambda_e A_p\delta n_p-\lambda_e A_e\delta n_e.
\eea 
We close the system exploiting the relations 
$\delta n_p+\delta n_n=\delta n_B$, $\delta n_e+\delta n_\mu=
\delta n_p$, which lead us to ($\lambda\equiv\lambda_e+\lambda_{\mu}$)
\bea\label{eq:delta_ne}
\delta n_e &=& \frac{-n_{e0}\theta +\lambda_e (A_n+A_p)
\delta n_n- \lambda_e A_p \delta n_B}
{i\omega+\lambda_e (A_e+A_{\nu_e})},\\
\label{eq:delta_nn}
i\omega\delta n_n &=& -n_{n0}\theta -
(\lambda A_n+\lambda A_p+\lambda_{\mu} A_\mu)\delta n_n\nonumber\\
&& +(\lambda_e A_e-\lambda_{\mu} A_\mu)\delta n_e 
 +(\lambda A_p+\lambda_{\mu} A_\mu)\delta n_B .
\eea
Solving the coupled Eqs.~\eqref{eq:delta_ne} and \eqref{eq:delta_nn} to find 
\bea\label{eq:delta_nn1}
D\delta n_n &=&
 -\frac{\theta}{i\omega}\bigg\{i\omega \Big[n_{n0}
 (i\omega+\lambda_e A_e) +n_{e0}(\lambda_e A_e- \lambda_{\mu} 
 A_\mu )\Big]\nonumber\\
 &+&n_{B0}\Big[i\omega(\lambda A_p+ \lambda_{\mu} A_\mu)
 +\lambda_e \lambda_{\mu} (A_p A_e+ A_pA_\mu
 +A_e A_\mu)\Big]\bigg\},\\
\label{eq:delta_ne1}
D\delta n_e &=&
-\frac{\theta}{i\omega}\bigg\{i\omega n_{e0} 
\Big[i\omega +\lambda_{\mu} A_2+\lambda_e (A_n+A_p)\Big] 
+i\omega n_{n0} \lambda_e (A_n+A_p) \nonumber\\
&-&  \lambda_e  n_{B0} \Big[A_p (i\omega+\lambda_{\mu} A_2)
-\lambda_{\mu}(A_n+A_p)(A_p+A_\mu)\Big]\bigg\},
\eea 
where we used the baryon conservation
$\delta n_B =-n_{B0}(\theta/i\omega)$ and defined
\bea\label{eq:D}
D=(i\omega+\lambda_e A_1)(i\omega+\lambda_{\mu} A_2)
-\lambda_e \lambda_{\mu}(A_n+A_p)^2,
\eea 
and 
\bea
\label{eq:def_A1}
A_1 &=& A_n+A_p+A_{e},\\
\label{eq:def_A2}
A_2 &=& A_n+A_p+A_{\mu}.
\eea

To find the bulk viscosity we still need to separate 
the instantaneous equilibrium parts of particle densities from 
Eqs.~\eqref{eq:delta_nn1} and \eqref{eq:delta_ne1}. As discussed 
in Ref.~\cite{Alford2020}, the equilibrium shifts $\delta n_j^{\rm eq}$ 
are the solutions of the balance equations~\eqref{eq:cont_n_slow} 
and \eqref{eq:cont_e_slow} in the case if the Urca processes are 
infinitely fast such that the $\beta$-equilibrium is restored instantly. 
This implies that $\delta n_j^{\rm eq}$ can be obtained by letting
$\lambda_{e,\mu}\to \infty$ in Eqs.~\eqref{eq:delta_nn1} and 
\eqref{eq:delta_ne1}. However, as we argued in Ref.~\cite{Alford2020}, 
one can use the opposite limit $\lambda_{e,\mu}\to 0$ with 
quasi-equilibrium solutions given by Eq.~\eqref{eq:cont_j}
$\delta n_j^{0}=-{\theta} n_{j0}/{i\omega}$ 
instead of $\delta n_j^{\rm eq}$ as both choices lead to the same 
result for the bulk viscosity. Subtracting the local quasi-equilibrium 
parts $\delta n_j^{0}$ from Eqs.~\eqref{eq:delta_nn1} and 
\eqref{eq:delta_ne1} we find the relevant nonequilibrium parts 
$\delta n'_j = \delta n_j-\delta n^{0}_j$
\bea\label{eq:delta_nn2}
\delta n'_n 
 &=&\frac{\theta}{i\omega}
\frac{i\omega (\lambda_e {C_1} +\lambda_{\mu}{C_2})
+\lambda_e \lambda_{\mu} \big[{C_2}  (A_e+A_{\nu_e}) 
+{C_1} (A_\mu+A_{\nu_\mu})\big]}
{(i\omega+\lambda_e A_1)(i\omega+\lambda_{\mu} A_2)
-\lambda_e \lambda_{\mu}(A_n+A_p)^2},\\
\label{eq:delta_ne2}
\delta n'_e 
&=&-\frac{\theta}{i\omega}\frac{i\omega\lambda_e C_1
+\lambda_e \lambda_{\mu} \big[A_2 C_1 - (A_n+A_p)C_2\big]}
{(i\omega+\lambda_e A_1)(i\omega+\lambda_{\mu} A_2)
-\lambda_e \lambda_{\mu}(A_n+A_p)^2}.
\eea 
Then the nonequilibrium part of the pressure, 
referred to as bulk viscous pressure, will be given by
\bea\label{eq:Pi}
\Pi =\sum_j
c_j\delta n'_j,
\eea 
with
\bea\label{eq:c_j_def}
c_j &\equiv &
\frac{\partial p}{\partial n_j}
=\sum_i n_{i0} \frac{\partial \mu_i}{\partial n_j}=\sum_i n_{i0}A_{ij}.
\eea
Here we used the definitions~\eqref{eq:A_f} and the Gibbs-Duhem relation $dp=n_BsdT+\sum_i n_i d \mu_i
\approx \sum_i n_i d \mu_i$ (the term with $dT$ is small in the 
parameter range considered here), where $s$ is the entropy per  baryon. The bulk 
viscous pressure then reads
\bea\label{eq:Pi_slow1}
\Pi &=& (c_n-c_p-c_\mu)\delta n'_n +(c_e-c_\mu)\delta n'_e
= C_2 \delta n'_n + (C_2-C_1)\delta n'_e\nonumber\\
&=& \frac{\theta}{i\omega}
\frac{i\omega (\lambda_e C_1^2+\lambda_{\mu} C_2^2)
+\lambda_e \lambda_{\mu} \big[A_1 C_2^2 +A_2 C_1^2-2(A_n+A_p)C_1C_2\big]}
{(i\omega+\lambda_e A_1)(i\omega+\lambda_{\mu} A_2)
-\lambda_e \lambda_{\mu}(A_n+A_p)^2},
\eea 
where we defined
\bea
\label{eq:def_C1}
c_n-c_p-c_e =n_{n0}A_n-n_{p0}A_{p}-
n_{e0}A_e\equiv C_1,\\ 
\label{eq:def_C2}
c_n-c_p-c_\mu =n_{n0}A_n-n_{p0}A_{p}-
n_{\mu0}A_\mu\equiv C_2.
\eea
Extracting the real part of Eq.~\eqref{eq:Pi_slow1} and recalling the
definition of the bulk viscosity ${\rm Re}\Pi=-\zeta \theta$ we find
\bea\label{eq:zeta_slow}
\zeta(\omega) = \frac{n_1+n_2\omega^2 }{(d_1-\omega^2)^2 +d_2 \omega^2 },
\eea
where we defined 
\bea\label{eq:num_denom}
n_1 &=& \lambda_e \lambda_{\mu}\Big[\lambda_e 
\left[(A_n+A_p) C_1- A_1 C_2\right]^2\nonumber\\
&&+\lambda_{\mu} \left[(A_n+A_p) C_2- A_2 C_1\right]^2 \Big],\\
n_2 &=& \lambda_e C_1^2+\lambda_{\mu} C_2^2,\\
d_1 &=& \lambda_e \lambda_{\mu}\left[A_1A_2-(A_n+A_p)^2\right],\\
d_2 &=& (\lambda_e A_1+\lambda_{\mu} A_2)^2.
\eea

The slow muon-equilibration limit can be obtained by dropping 
the terms $\propto\lambda_\mu$ in Eq.~\eqref{eq:zeta_slow}
\bea\label{eq:zeta_slow1}
\zeta_e = \frac{C_1^2}{A_1}\frac{\gamma_e}
{\omega^2+\gamma_e^2},
\eea
with $\gamma_e=\lambda_e A_1$, which coincides with the 
result of our previous work~\cite{Alford2019b}.

In the limit of high frequencies $\omega\gg \lambda A$ we find
from Eq.~\eqref{eq:zeta_slow}
\bea\label{eq:zeta_slow2}
\zeta = \frac{\lambda_e C_1^2+\lambda_{\mu} C_2^2}
{\omega^2}=\zeta_e+\zeta_\mu,
\eea
where $\zeta_e$ and $\zeta_\mu$ are the contributions
by electrons and muons, respectively.

\section{Numerical results}
\label{sec:num_results}


The numerical evaluation of equilibration rates~\eqref{eq:Gamma1_final}, 
\eqref{eq:Gamma2_final} and \eqref{eq:Gamma_lep_final} is performed 
within the framework of covariant density functional approach to the 
nuclear matter. The Lagrangian density reads
\bea\label{eq:lagrangian} 
{\cal L} & = &
\sum_N\bar\psi_N\bigg[\gamma^\mu \left(i\partial_\mu-g_{\omega}
\omega_\mu - \half g_{\rho }\vectau\cdot\vecrho_\mu\right)
- m^*_N\bigg]\psi_N + \sum_{l}\bar\psi_l
(i\gamma^\mu\partial_\mu - m_l)\psi_l,\\
\nonumber & + & \half \partial^\mu\sigma\partial_\mu\sigma-
\half m_\sigma^2\sigma^2 -U(\sigma)- \frac{1}{4}\omega^{\mu\nu}
\omega_{\mu\nu} + \half m_\omega^2\omega^\mu\omega_\mu -
\frac{1}{4}\vecrho^{\mu\nu}\vecrho_{\mu\nu} + \half
m_\rho^2\vecrho^\mu\cdot\vecrho_\mu, 
\eea 
where $N$ sums over nucleons, $\psi_N$ are the nucleonic Dirac 
fields, $m_N^*=m_N - g_{\sigma}\sigma$ are the nucleon effective
masses, with $m_N$ being the nucleon mass in the vacuum. Next,
$\sigma,\omega_\mu$, and $\vecrho_\mu$ are the scalar-isoscalar, 
vector-isoscalar and vector-isovector meson fields, respectively; 
$\omega_{\mu\nu}=\partial_\mu\omega_\nu-\partial_\nu\omega_\mu$ 
and $\vecrho_{\mu\nu}=\partial_\mu \vecrho_{\nu}-\partial_\nu 
\vecrho_{\mu}$ are the field strength tensors of vector mesons;
$m_{i}$ are the meson masses and $g_{i}$ are the baryon-meson 
couplings with $i=\sigma,\omega,\rho$, and $U(\sigma)$ is the 
self-interaction of scalar meson field. Next, $\psi_l$ are the 
leptonic free Dirac fields with masses $m_\lambda$ where $l=\{e,\mu\}$. 
We adopt two different parametrizations of Lagrangian~\eqref{eq:lagrangian}, 
specifically, the model DDME2~\cite{Lalazissis2005} with density-dependent 
nucleon-meson couplings and with $U(\sigma)=0$, and the model 
NL3~\cite{Lalazissis1997}, which has density-independent 
nucleon-meson couplings but contains self-interaction terms of 
$\sigma$-meson fields given by $U(\sigma)=g_2\sigma^3/3+g_3\sigma^4/4$.

The spectrum of nucleonic excitations derived from Eq.~\eqref{eq:lagrangian} 
in the mean-field approximation is given by~\cite{Glendenning_book} 
\bea\label{eq:spectrum}
E_k = \sqrt{k^2+m^{* 2}_N} + g_{\omega}\omega_0 + 
I_{3N} g_{\rho}\rho_{03} +\Sigma_r,
\eea
where $I_{3N}$ is the third component of the nucleon isospin, and
$\Sigma_r$ is the so-called rearrangement self-energy~\cite{Typel1999}
which is introduced to maintain the thermodynamic
consistency  in the case where the nucleon-meson 
couplings are density-dependent. 

Introducing the nucleon effective chemical potentials as
$\mu^*_N = \mu_N-g_{\omega }\omega_0 - I_{3N} g_{\rho
}\rho_{03}-\Sigma_r$ one can write the argument of nucleon
Fermi-functions as $E_k-\mu_N =\sqrt{k^2+m^{* 2}_N}-\mu_N^*$ which
formally coincides with the spectrum of free nucleons with effective
masses and effective chemical potentials.

The composition of $\beta$-equilibrated matter at the given baryon
density $n_B$ and temperature should be determined by imposing the
$\beta$-equilibrium conditions, the charge neutrality condition
$n_p=n_e+n_\mu$ and the baryon number conservation $n_B=n_n+n_p$.  As
discussed above, we adopt for the unperturbed background
$\beta$-equilibrium conditions $\mu_{\Delta_l}=\mu_n-\mu_p-\mu_l=0$,
with $l=\{e,\mu\}$ which are valid in the low-temperature limit.

\begin{figure}[t] 
\begin{center}
\includegraphics[width=0.45\columnwidth, keepaspectratio]{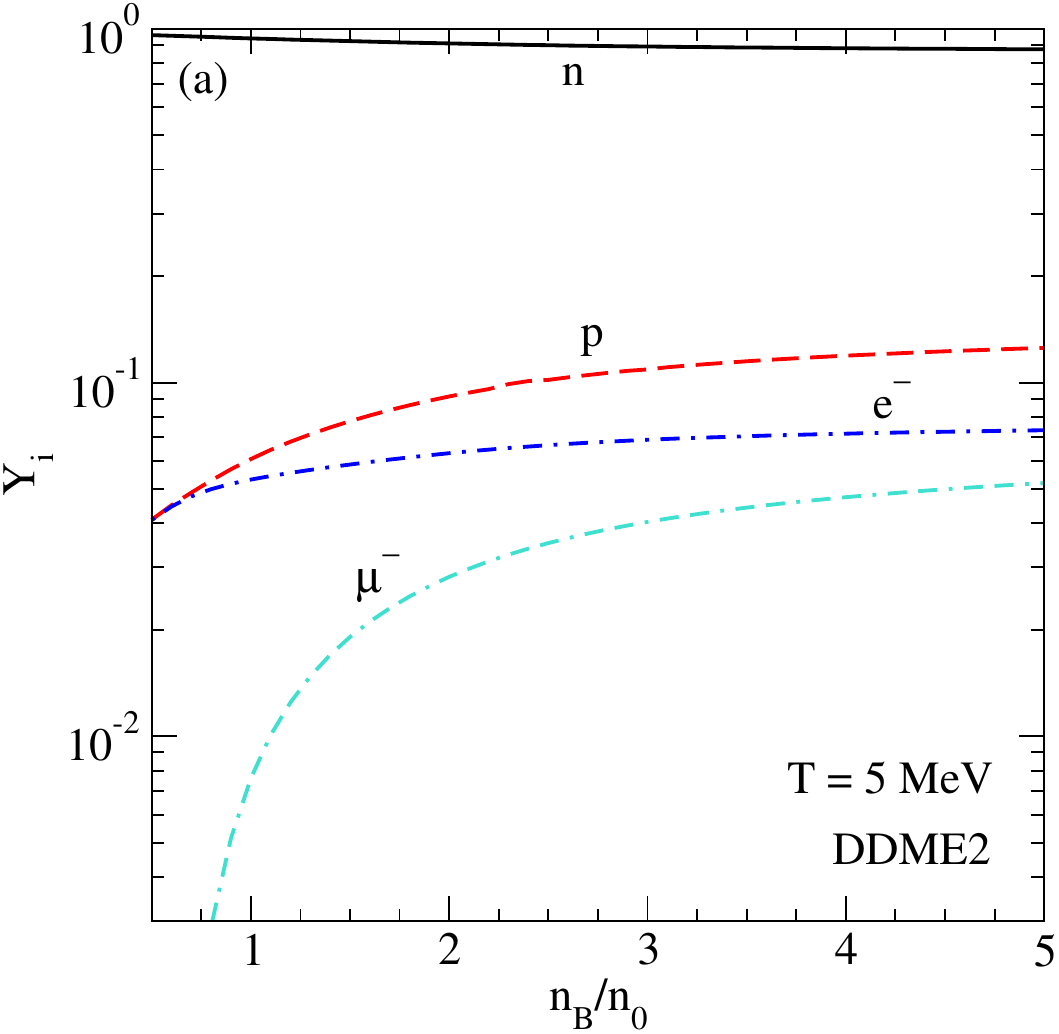}
\hspace{0.5cm}
\includegraphics[width=0.45\columnwidth, keepaspectratio]{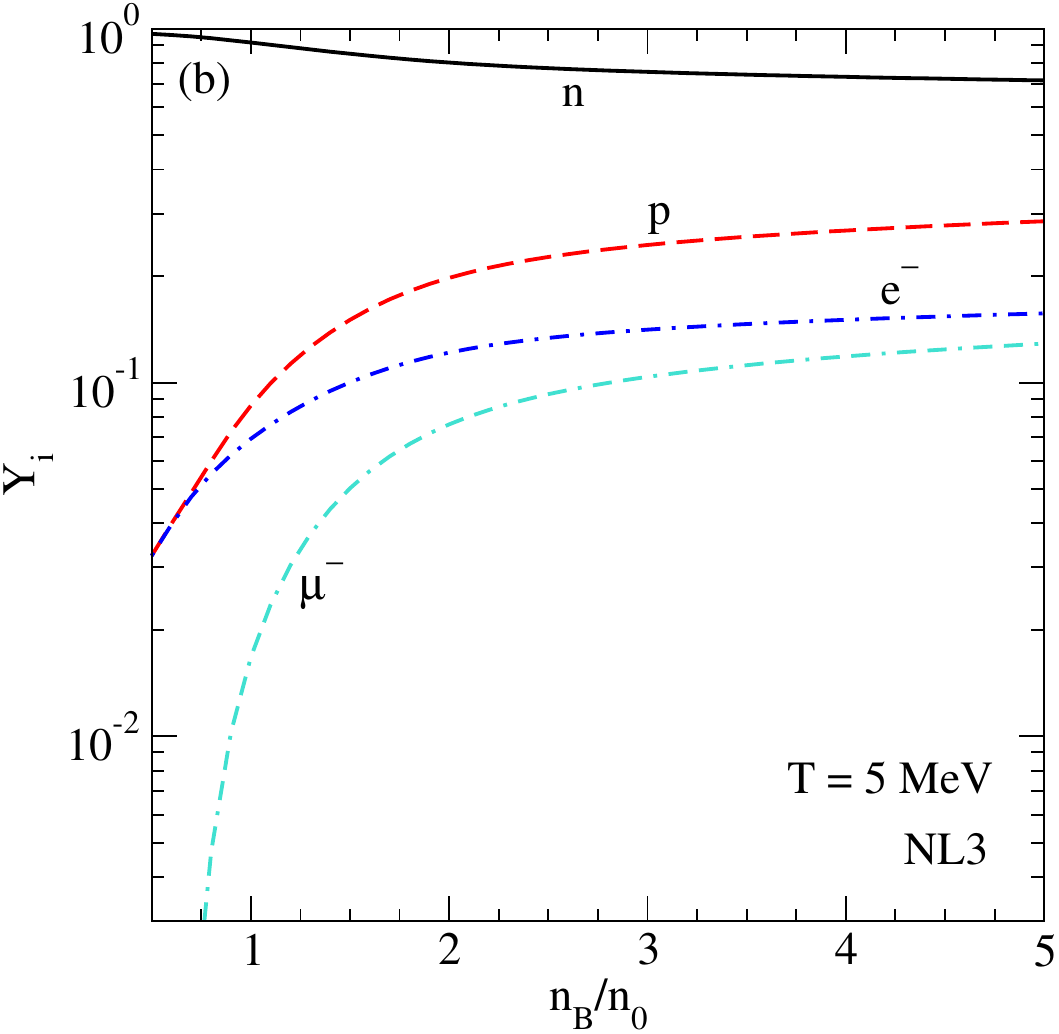}
\caption{ Particle fractions in finite-temperature $\beta$-equilibrated matter  as 
  functions of the baryon density $n_B$ (in units of nuclear 
  saturation density $n_0$) for models DDME2 (a) and NL3 (b) at  fixed temperature $T=5$ MeV. }
\label{fig:fractions} 
\end{center}
\end{figure}
\begin{figure}[!] 
\begin{center}
\includegraphics[width=0.45\columnwidth, keepaspectratio]{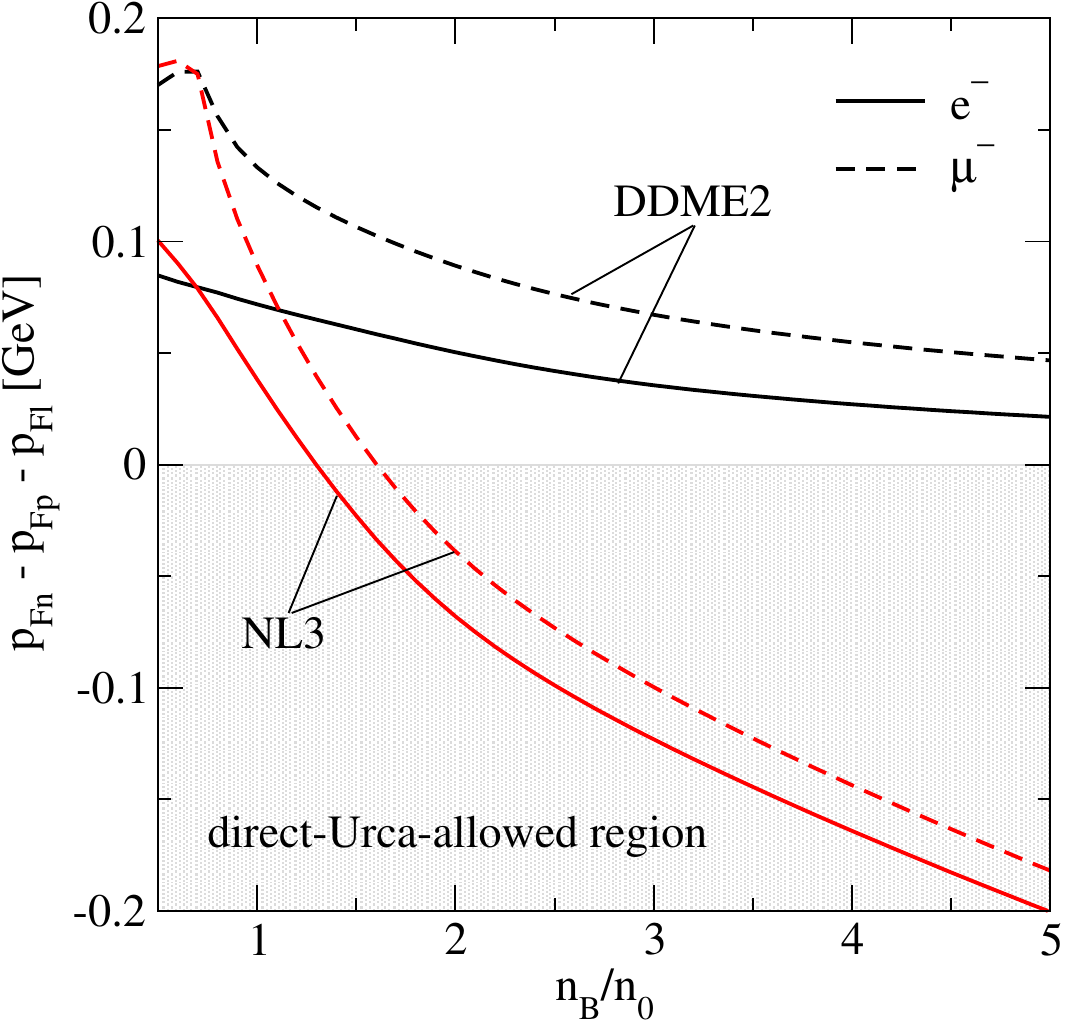}
\caption{ The sum $p_{Fn}-p_{Fp}-p_{Fl}$, $l=\{e,\mu\}$, in
  zero-temperature $npe\mu$ matter for DDME2 and NL3 models. At $T=0$
  the direct Urca process is only allowed in the regions where
  $p_{Fn}-p_{Fp}-p_{Fl}\leq 0$. Thus for DDME2 the direct Urca
  processes on electrons and muons are Boltzmann-suppressed at all the
  densities plotted, whereas for NL3 the threshold density for
  electronic and muonic Urca processes are $\approx 1.3 n_0$ and
  $\approx 1.6 n_0$, respectively.  Above these the direct Urca
  processes are unsuppressed at low temperatures.}
\label{fig:pfermi} 
\end{center}
\end{figure}

Particle fractions in $\beta$-equilibrated $npe\mu$ matter for the two 
parametrizations are shown in Fig.~\ref{fig:fractions}. The main 
difference between these two models is the larger proton and lepton 
fractions in the NL3 model. As a result, NL3 has direct electronic Urca threshold 
at $n_B\simeq 1.3n_0$ and muonic Urca threshold at $n_B\simeq 1.6n_0$,
see Fig.~\ref{fig:pfermi}, with $n_0$ being the nuclear saturation 
density which has the values $n_0=0.152$ fm$^{-3}$ for model DDME2 
and $n_0=0.153$ fm$^{-3}$ for model NL3. The model DDME2 instead 
does not reach the direct Urca thresholds up to baryon density $n_B=5n_0$.
In contrast to the case of neutrino-trapped matter~\cite{Alford2021c}, 
in the neutrino-transparent matter muons appear only above a certain
baryon density $n_B\gtrsim n_0$, where the condition $\mu_e\geq 
m_\mu\simeq 106$ MeV is satisfied.

\subsection{Beta-equilibration rates}
\label{sec:rates}

\subsubsection{Urca process rates}

\begin{figure}[t] 
\begin{center}
\includegraphics[width=0.45\columnwidth,keepaspectratio]{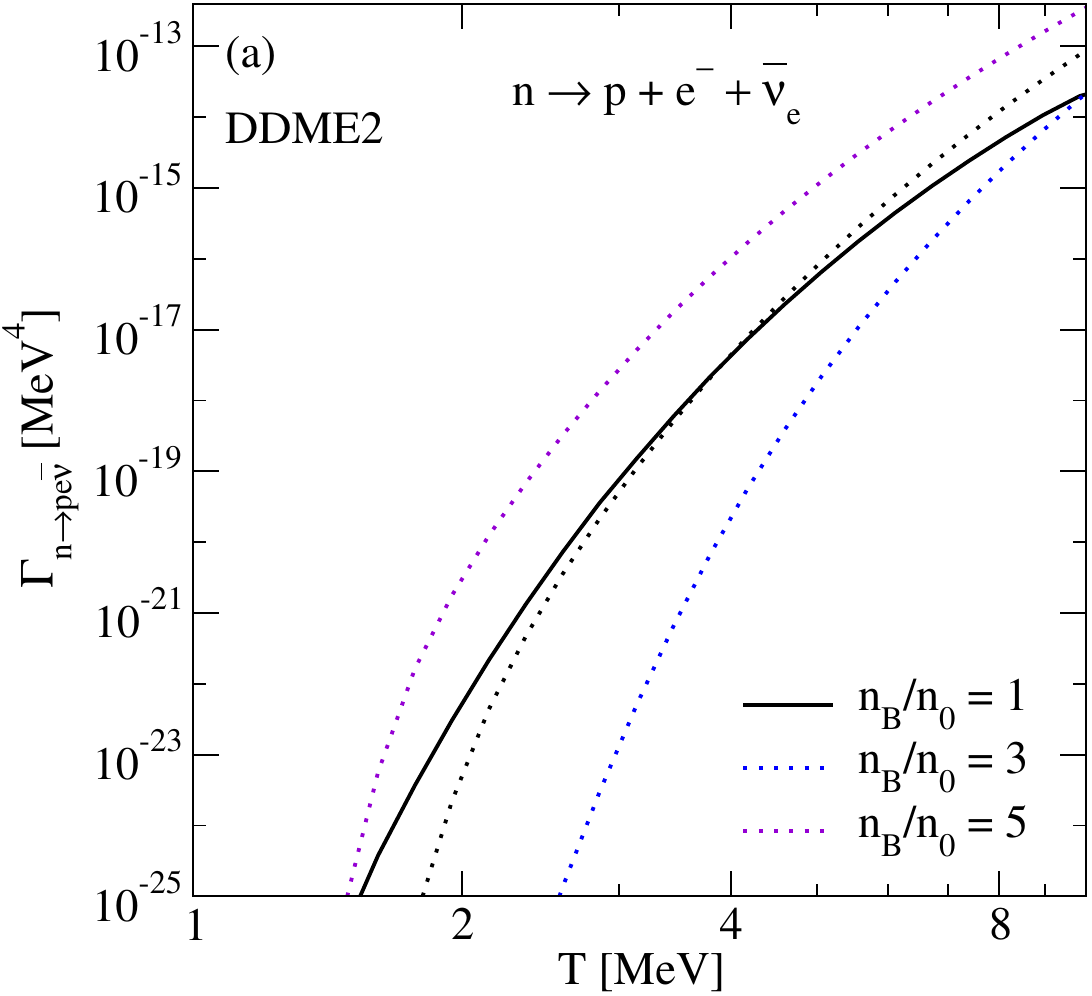}
\hspace{0.5cm}
\includegraphics[width=0.45\columnwidth,keepaspectratio]{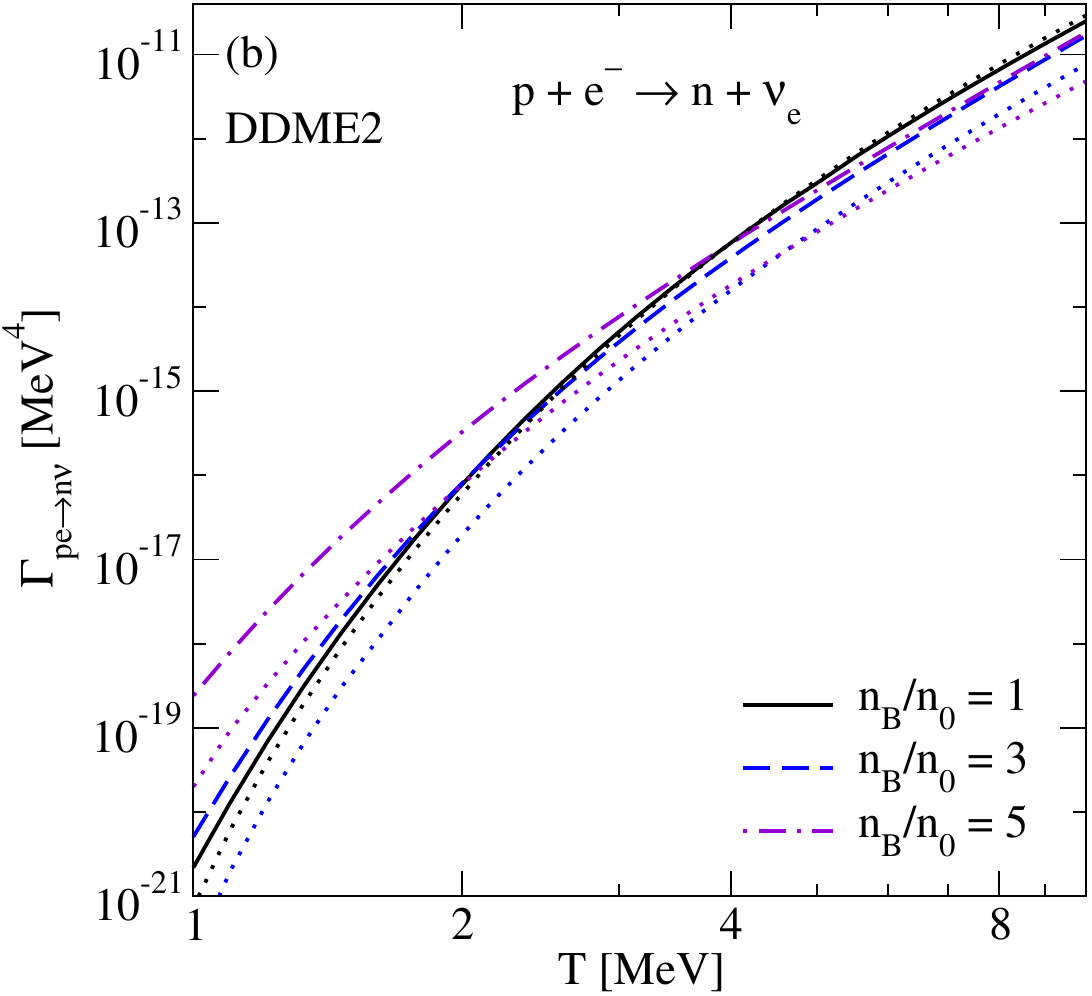}
\caption{ The 
rates for (a) neutron decay to an electron and (b) electron 
capture direct Urca processes as functions of the temperature for various densities 
for the DDME2 model. The  dotted lines show the Urca process rates computed in 
Ref.~\cite{Alford2019b} within the approximation of nonrelativistic nucleons.}
\label{fig:Gamma12_e_DDME2} 
\end{center}
\end{figure}

The direct Urca  neutron-to-electron 
decay and electron-capture rates are shown in 
Figs.~\ref{fig:Gamma12_e_DDME2} and \ref{fig:Gamma12_e_NL3} as 
functions of the temperature for models DDME2 and NL3, respectively. 
The modified Urca contribution is discussed at the end of Sec.~\ref{sec:bulk_results}.

In the DDME2 model, the densities we study are all below the direct
Urca threshold, so direct Urca rates are Boltzmann-suppressed at low
temperatures. We see this in the rapid dropping off of both the
neutron decay and electron capture rates as $T$ decreases.  In fact,
for densities $3n_0$ and $5n_0$ the suppression of the neutron decay
rate is so strong that those curves are not visible on the plot.

Comparing panels (a) and (b) of Fig.~\ref{fig:Gamma12_e_DDME2} we see
that the electron capture rate, although Boltzmann suppressed, is much
faster than the neutron decay rate, and much less dependent on
density.  At saturation density, it is about three orders of magnitude
faster than neutron decay and remains about the same as the density
increases.  Similar behavior of neutron decay rate was also found and
discussed in Ref.~\cite{Alford2021b}.

Figure~\ref{fig:Gamma12_e_DDME2} shows in addition the neutron decay
and electron capture rates computed in Ref.~\cite{Alford2019b} in the
approximation of nonrelativistic nucleons. We see that the electron
capture rates for nonrelativistic nucleons are smaller than the
relativistic ones, the difference being as large as an order of
magnitude at $n_B=5n_0$. The non-relativistic treatment of the neutron
decay process, instead, strongly overestimates the rates above the
saturation density, as the relativistic rates are strongly damped in
this regime, as already mentioned above.
\begin{figure}[t] 
\begin{center}
\includegraphics[width=0.45\columnwidth,keepaspectratio]{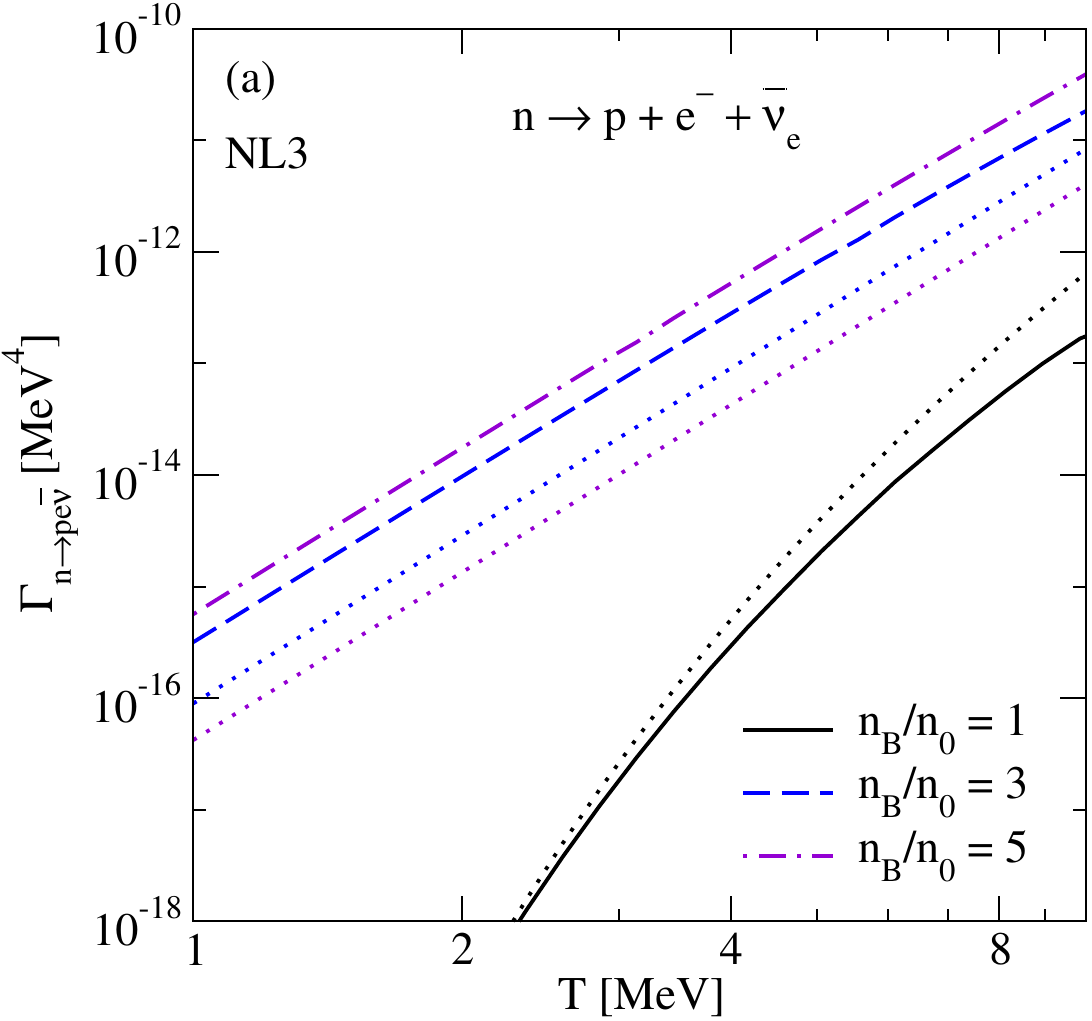}
\hspace{0.5cm}
\includegraphics[width=0.45\columnwidth,keepaspectratio]{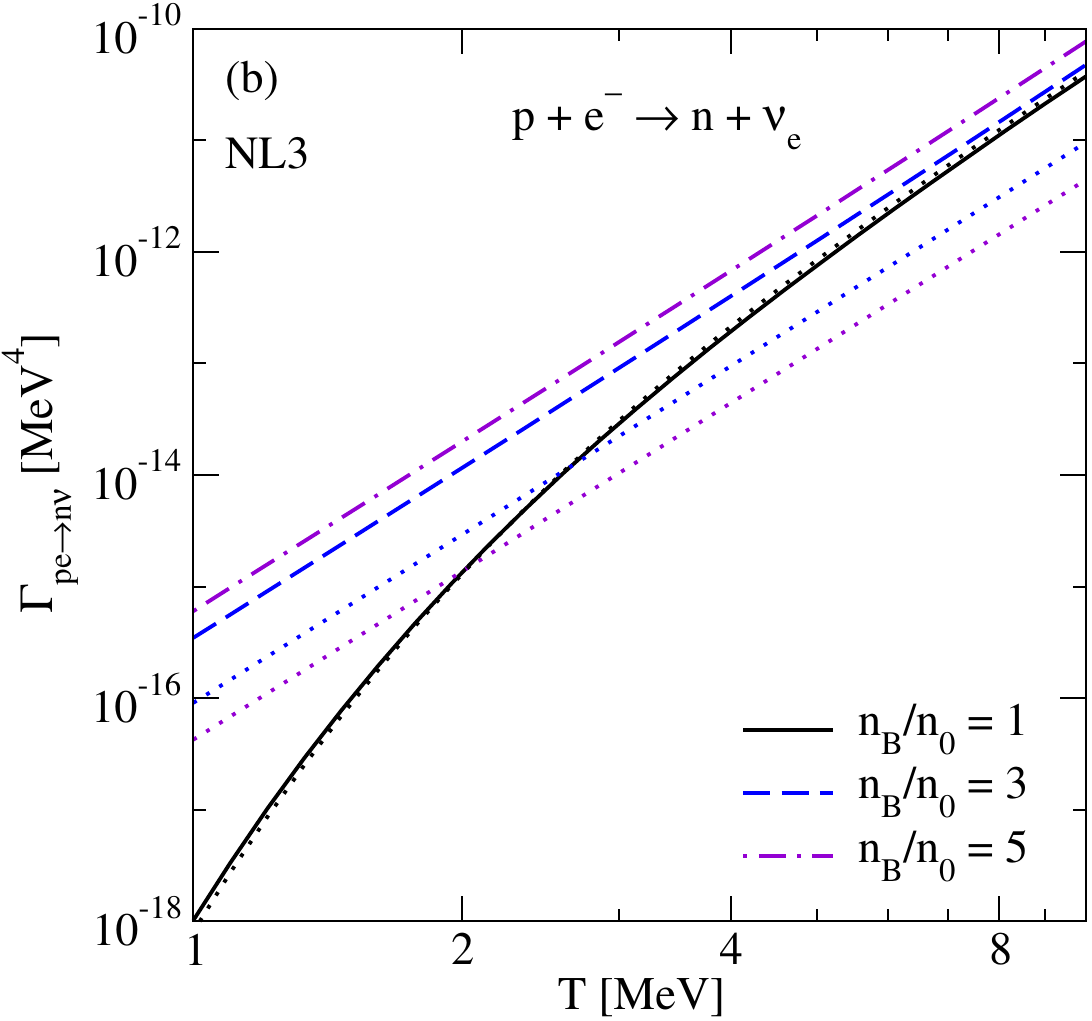}
\caption{ The rates for (a) neutron-to-electron decay  and (b) electron 
capture direct Urca processes as functions of the temperature for various densities 
for the  NL3 model. The  dotted lines show the Urca process rates computed in 
Ref.~\cite{Alford2019b} within the approximation of nonrelativistic nucleons.}
\label{fig:Gamma12_e_NL3} 
\end{center}
\end{figure}
\begin{figure}[!] 
\begin{center}
\includegraphics[width=0.45\columnwidth,keepaspectratio]{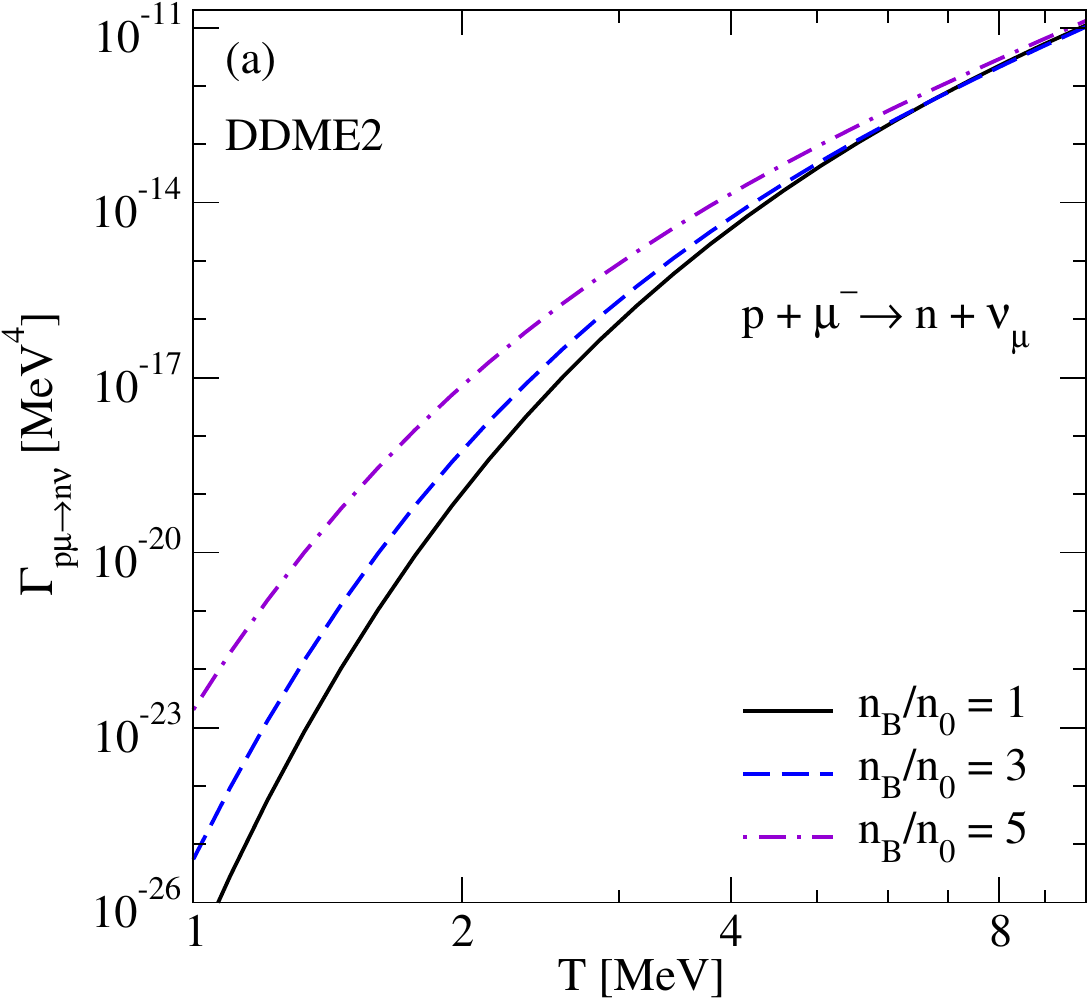}
\hspace{0.5cm}
\includegraphics[width=0.45\columnwidth,keepaspectratio]{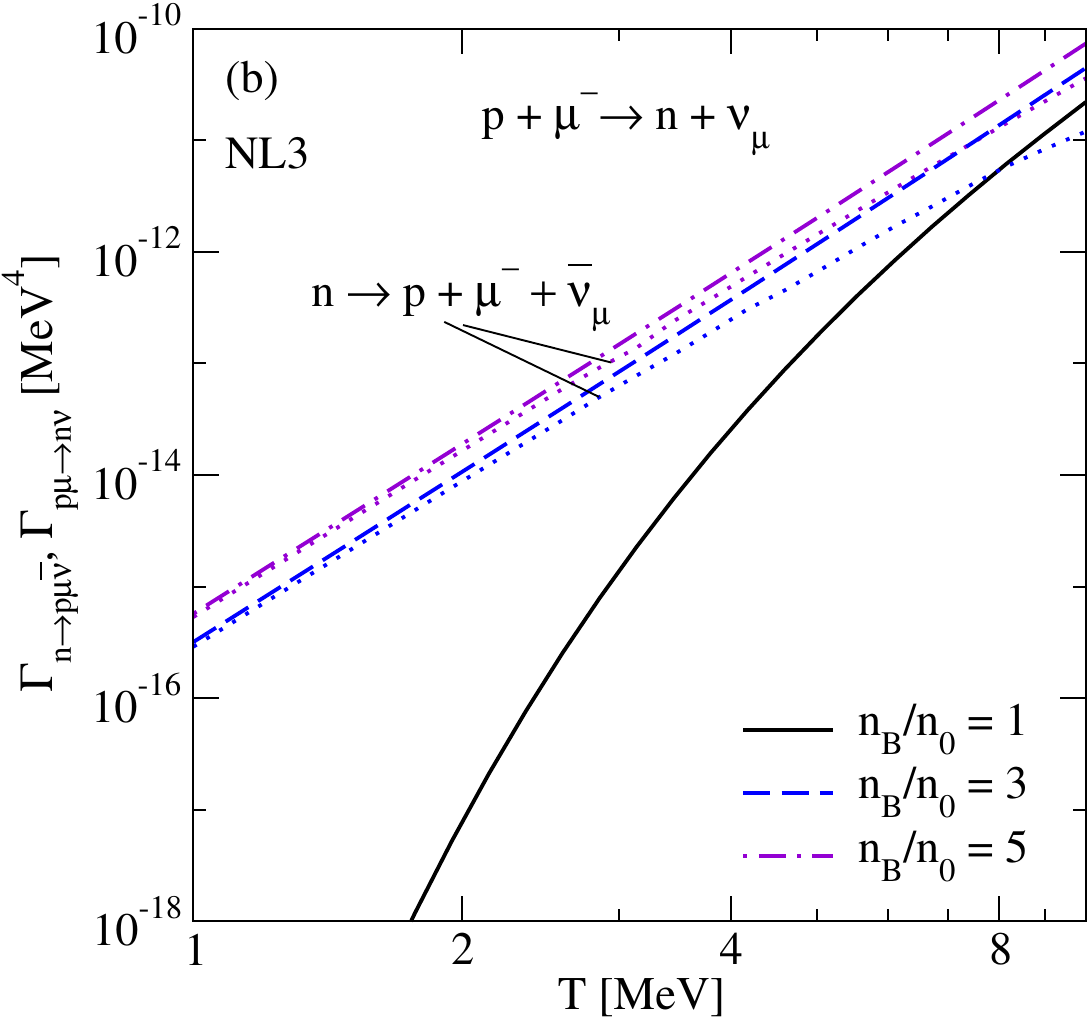}
\caption{ The muonic direct Urca processes rates for the DDME2 model (a) and the NL3  model (b). 
The neutron-to-muon decay is allowed only for the NL3 model above the direct Urca 
threshold  (dotted lines).}
\label{fig:Gamma12_mu} 
\end{center}
\end{figure}

Turning to the NL3 model, we see that at density $n_B=n_0$, which is
below the direct Urca threshold, both neutron decay and electron
capture rates show the expected Boltzmann suppression at low $T$, but
the rates for NL3 are significantly faster than for DDME2.  At higher
densities, the direct Urca channel is open for NL3 where the neutron
decay and the electron capture are almost equal and closely follow
their low-temperature scaling
$\Gamma_{n\to pe\bar{\nu}}=\Gamma_{pe\to n\nu}\propto T^5$ given by
Eq.~\eqref{eq:Gamma_lowT}. The discrepancy between the relativistic
and nonrelativistic calculations is within an order of magnitude also
in this case. Note that the Urca process rates increase with the
density in the case of NL3 model, but are non-monotonic in the case of
DDME2.

The rates of muonic direct Urca processes are shown in
Fig.~\ref{fig:Gamma12_mu}.  Panel (a) shows the results for muon
capture for the model DDME2.  The general behavior of the muon capture
rates is similar to electron capture rates; however, quantitatively
the muon capture rate is much smaller at low temperatures and becomes
comparable to the electron capture above $T\geq 5$ MeV. The
neutron-to-muon decay is strongly suppressed in the whole density
range for the DDME2 model.

The muon capture rates for the NL3 model are shown in
Fig.~\ref{fig:Gamma12_mu}\,(b). As in the case of DDME2, the muon
capture rate is much slower than the electron capture rate at low
temperatures $T\leq 5$ MeV below the direct Urca threshold, \ie, at
$n_B=n_0$, whereas the electron and muon capture rates are almost
equal above the threshold at all temperatures. We see also that the
neutron-to-muon decay rate is nonvanishing only above the threshold
where it is close to the muon capture rate. The difference between
these rates increases with the temperature.

\subsubsection{Muon decay rate}
\label{sec:Muon_decay_rate}
\begin{figure}[t]  
\begin{center}
\includegraphics[width=0.45\columnwidth,keepaspectratio]{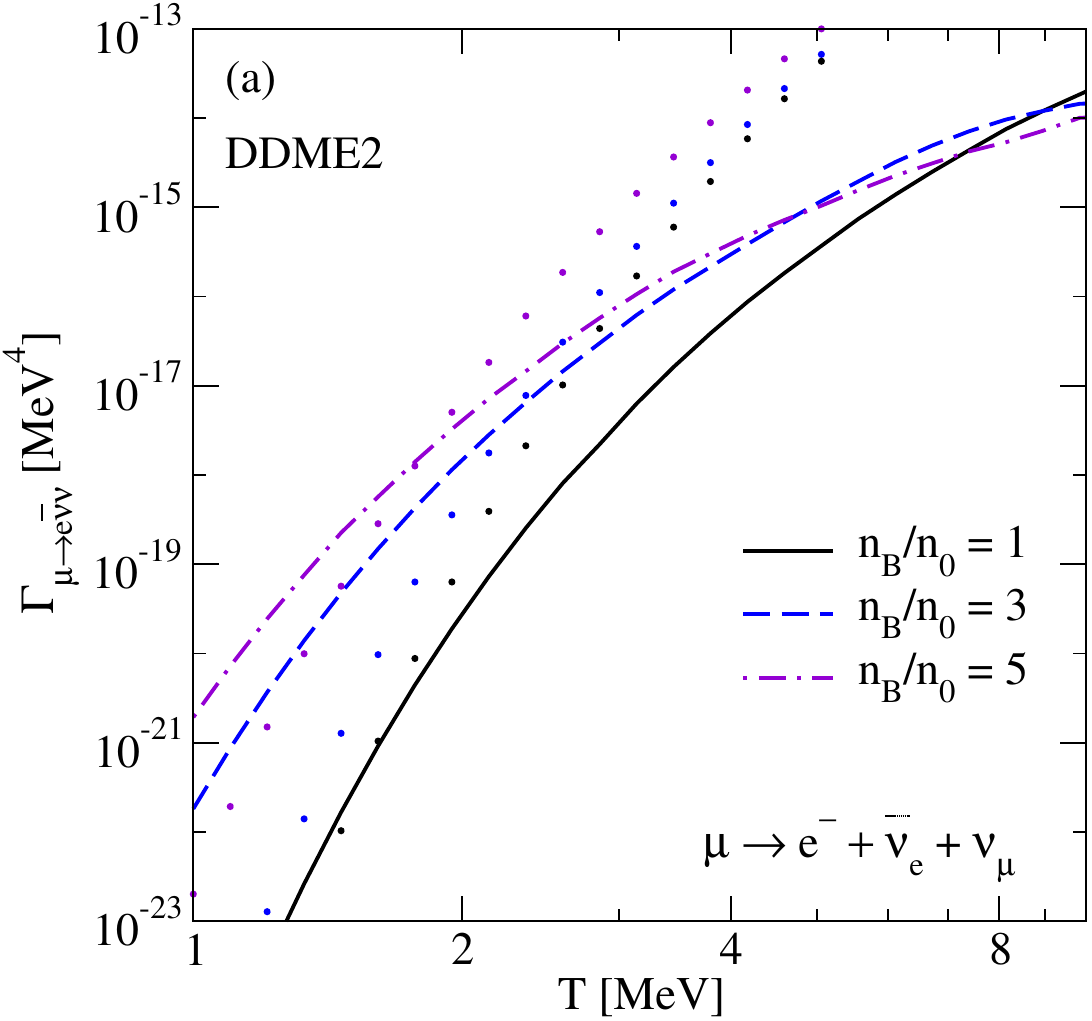}
\hspace{0.5cm}
\includegraphics[width=0.45\columnwidth,keepaspectratio]{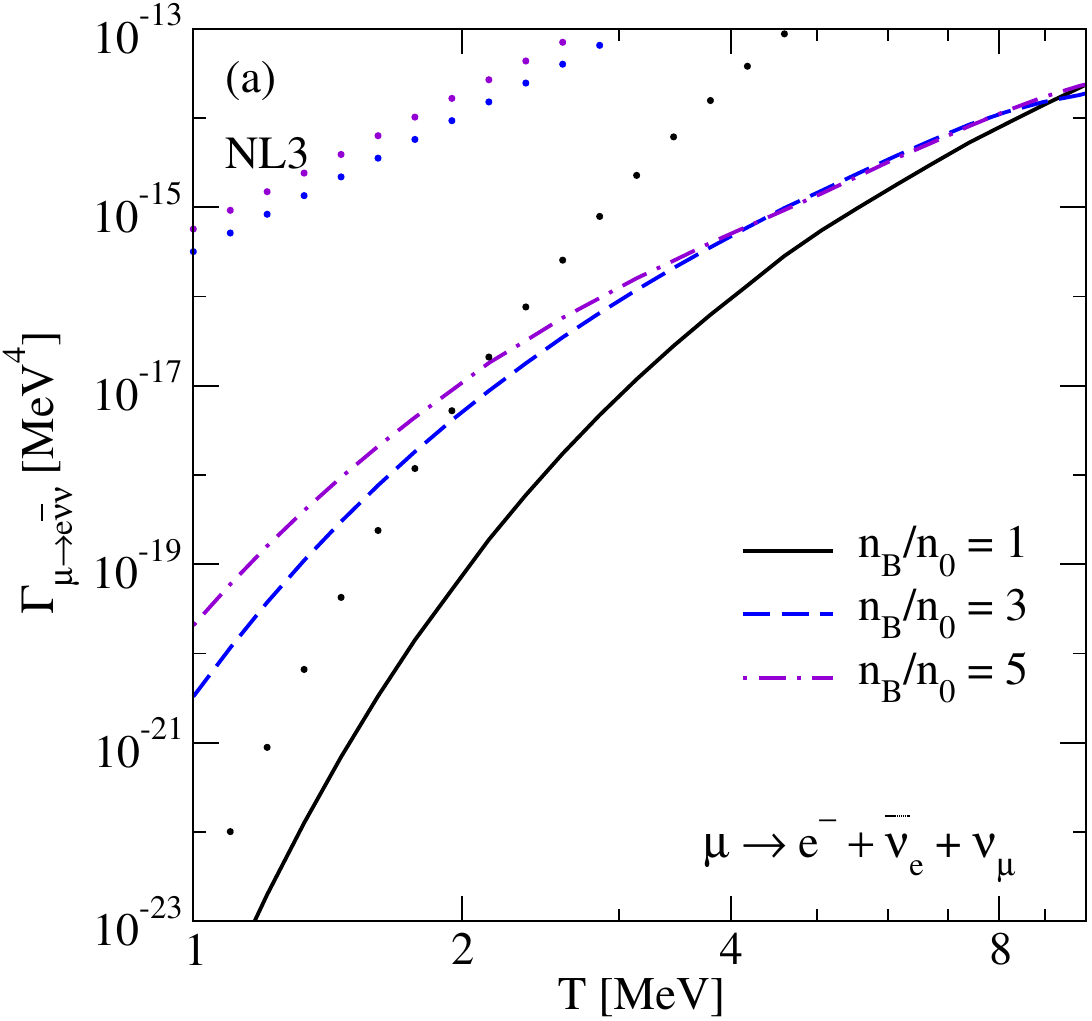}
\caption{ The muon decay rates as functions of the temperature for different 
densities for (a) the DDME2 model; (b) the NL3 model. The muon capture 
($p\ +\mu^- \to n\ +\nu_\mu)$ rates 
are shown by the dotted lines for comparison.
}
\label{fig:Gamma_lep}  
\end{center}
\end{figure}
Figure~\ref{fig:Gamma_lep} shows the muon decay rates given by
Eq.~\eqref{eq:Gamma_lep_final}.  Muon decay is Boltzmann suppressed at
low temperatures because, like neutron decay in the DDME2 model, at
all densities, it is Pauli blocked for particles on their Fermi
surfaces. A muon on its Fermi surface has just enough energy but
insufficient momentum to create a final state electron on its Fermi
surface, so it lacks the extra energy to create neutrinos to help with
momentum conservation.

To decide whether we are in the slow lepton equilibration limit or the
slow muon equilibration limit \eqref{eq:regimes} we compare the muonic
Urca rate to the electronic Urca and muon decay rates. The electron
capture rates are always found to exceed the muon decay rates at least
by an order of magnitude.  In Fig.~\ref{fig:Gamma_lep} where the Urca
muon capture rates are shown by dotted lines, we see that the Urca
muon capture rate is comparable to the muon decay rate only in the low
temperature domain $T\lesssim 2$ MeV in the case of DDME2 model,
indicating that the matter is in the slow-muon-equilibration regime,
where the muonic component can be simply neglected when computing the
bulk viscosity, as discussed in Sec.~\ref{sec:bulk}. At higher
temperatures $T\geq 3$ MeV the system is in the limit of slow lepton
equilibration
$\Gamma_{pe\to n\nu}\geq\Gamma_{p\mu\to n\nu}\gg \Gamma_{\mu \to
  e\bar\nu\nu}$.  (Note that if one includes the modified Urca
processes, then the muon decay rate will be always smaller than the
sum of the direct and modified Urca rates.)

In the case of NL3 model, the lepton capture rates are always larger
than the muon decay rates; they differ at least by an order of
magnitude below the direct Urca threshold and at least three orders of
magnitude above the threshold. Thus, the bulk viscosity of $npe\mu$
matter for the NL3 model should be computed under the
slow-lepton-equilibration assumption in the whole temperature-density
range of interest.

\subsection{Bulk viscosities}
\label{sec:bulk_results}

\subsubsection{Bulk viscosity of relativistic $npe$ matter}
\label{sec:bulk_electrons}

\begin{figure}[t]  
\begin{center}
\includegraphics[width=0.45\columnwidth,keepaspectratio]{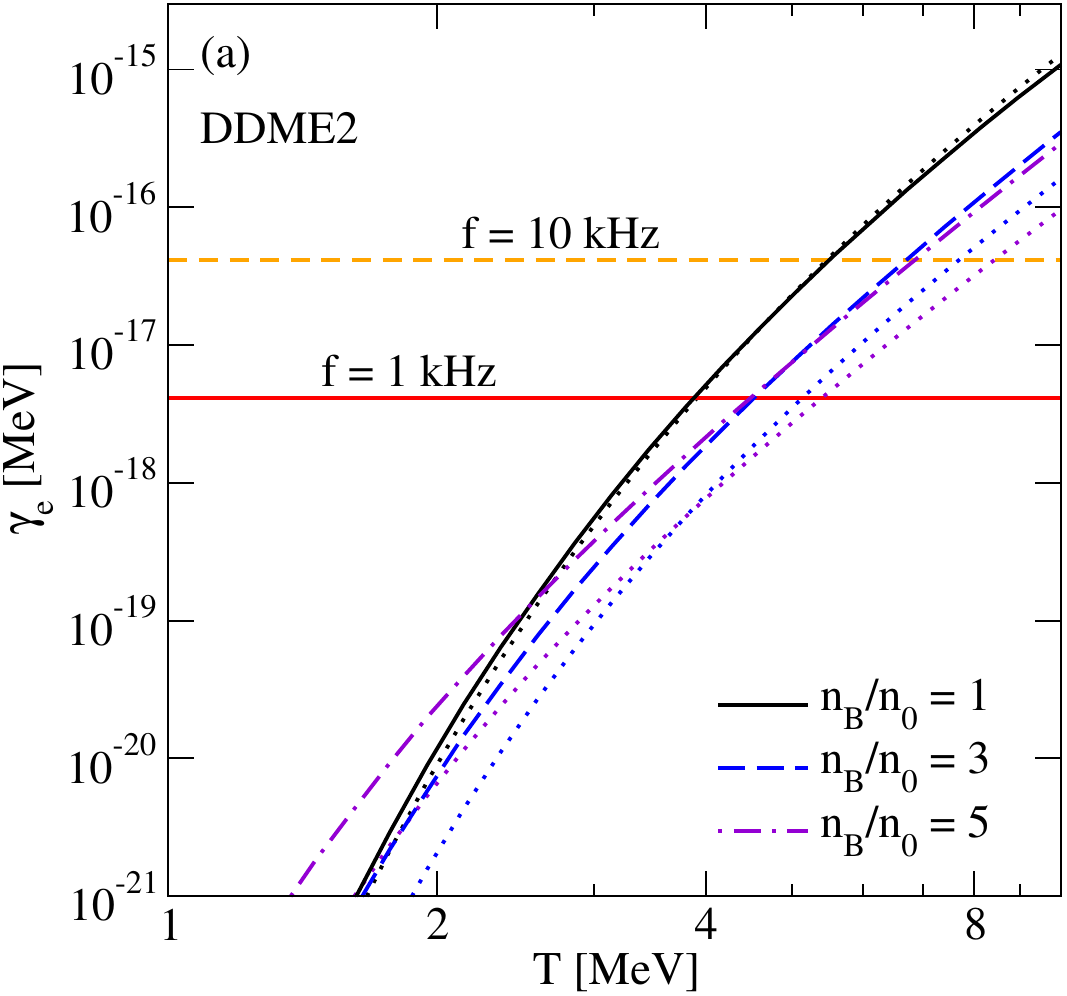}
\hspace{0.5cm}
\includegraphics[width=0.45\columnwidth,keepaspectratio]{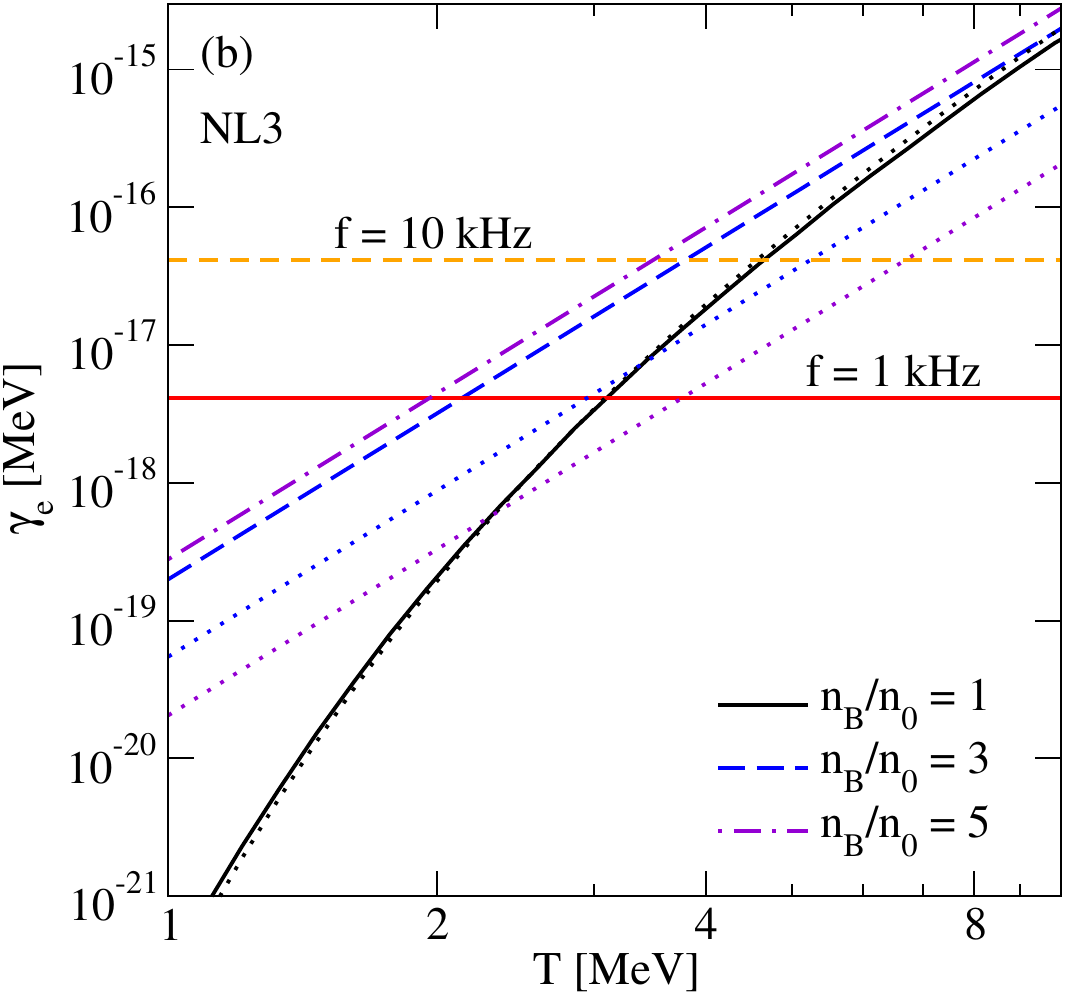}
\caption{ The $\beta$-relaxation rate $\gamma_e$ as a function of the
  temperature for fixed values of the density for (a) the DDME2 model;
  (b) the NL3 model. The dotted lines show the relaxation rates
  computed in Ref.~\cite{Alford2019b} within the approximation of
  nonrelativistic nucleons. The horizontal lines show where
  $\gamma_e=2\pi f$ for selected values of oscillation frequency
  $f= 1$ kHz (solid lines) and $f = 10$ kHz (dashed lines).  }
\label{fig:gamma_e_temp} 
\end{center}
\end{figure}
\begin{figure}[!] 
\begin{center}
\includegraphics[width=0.45\columnwidth,keepaspectratio]{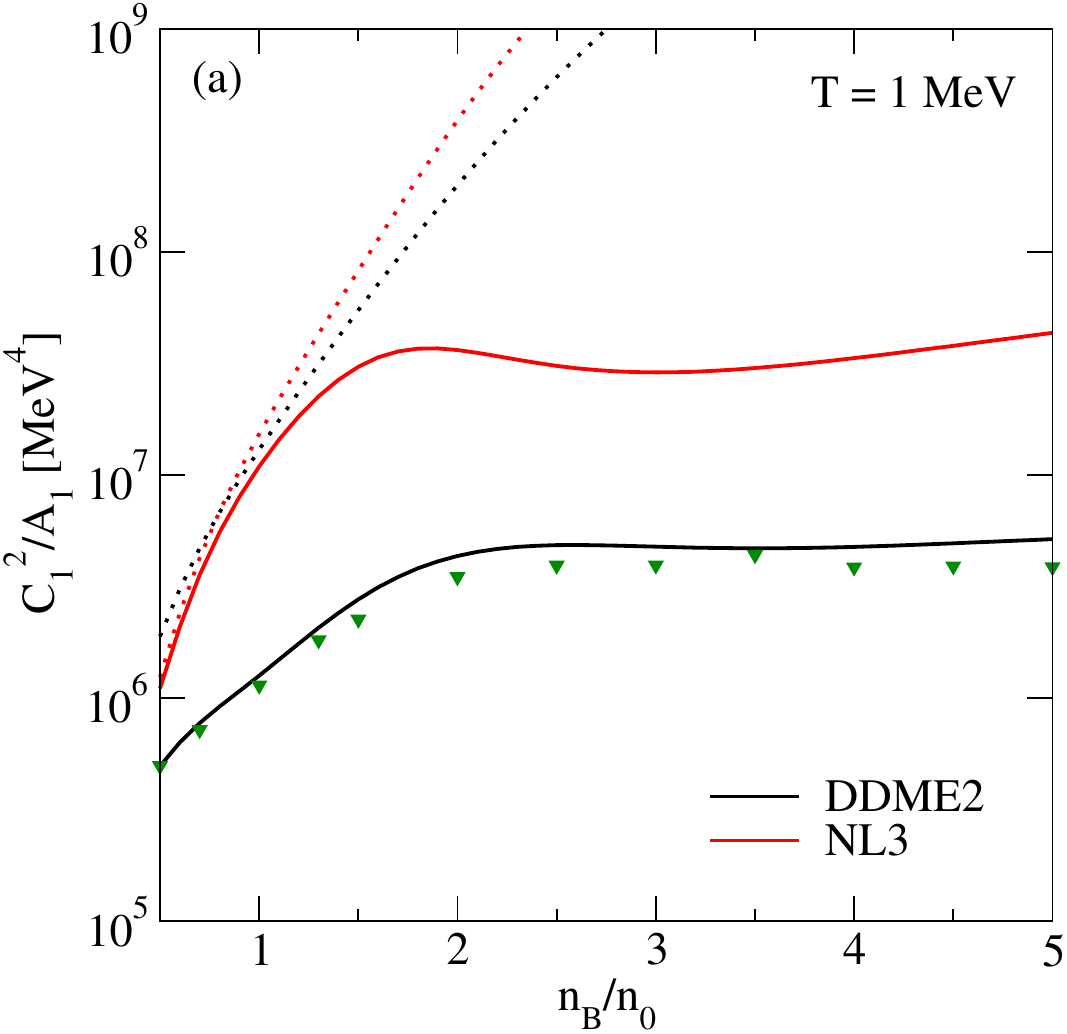}
\hspace{0.5cm}
\includegraphics[width=0.45\columnwidth,keepaspectratio]{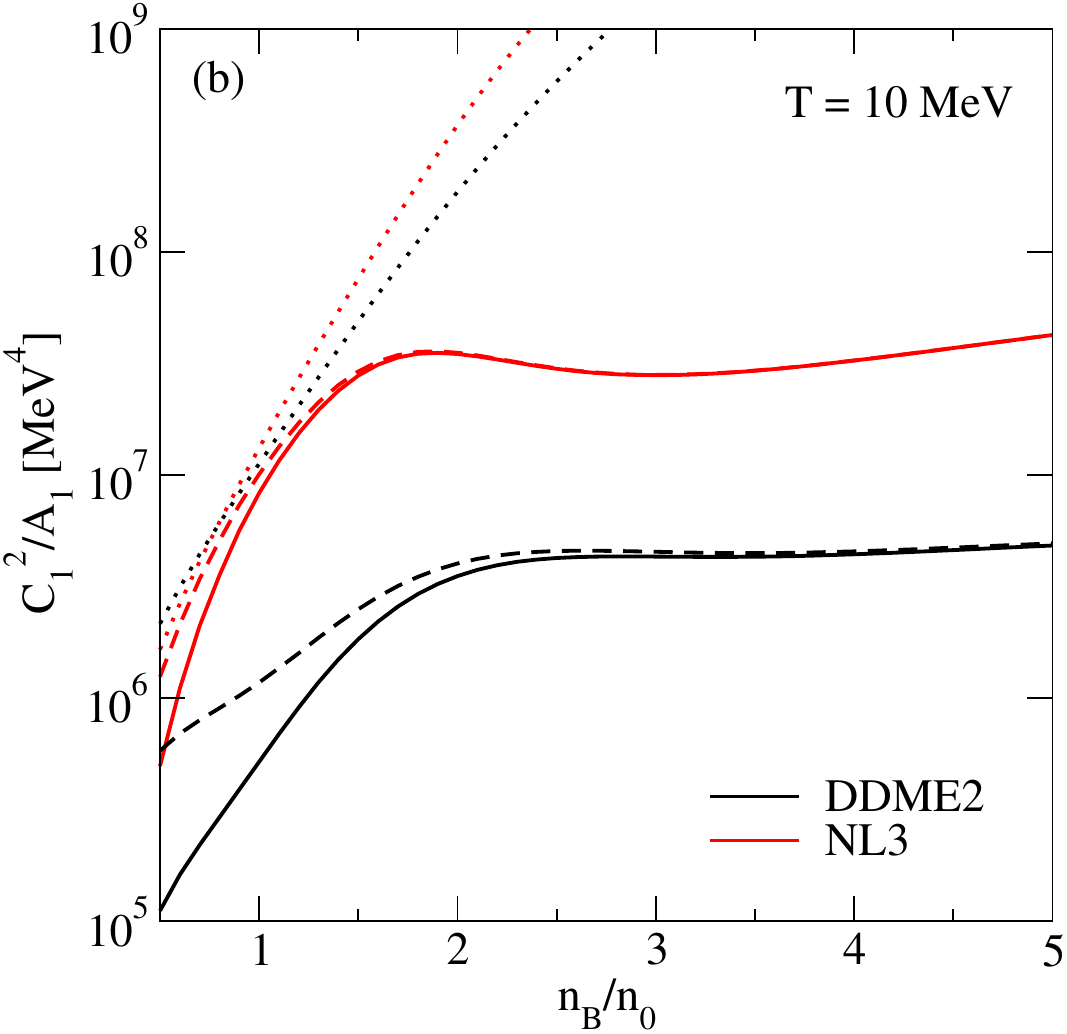}
\caption{ The susceptibility $C^2_1/A_1$ of $npe$ matter as 
a function of the baryon density for the DDME2 and NL3 models 
and 
fixed temperature  (a) $T=1$~MeV and (b) $T=10$~MeV.
The solid lines show the isothermal susceptibilities of relativistic matter, and the dotted lines show the 
isothermal susceptibilities computed in Ref.~\cite{Alford2019b} within the 
approximation of nonrelativistic nucleons. The green triangles in 
panel (a) show the result of Ref.~\cite{Alford2019a} for adiabatic susceptibility for the model 
DD2 at $T=1$~MeV. The dashed lines in panel (b) show the adiabatic 
susceptibilities at $T=10\,\MeV$ as computed in Appendix~\ref{app:A_ad}. At $T=1\,\MeV$ (panel (a)) the difference 
between the isothermal and adiabatic susceptibilities 
is very small and is invisible in the plot.
}
\label{fig:C2A_dens} 
\end{center}
\end{figure}

In this subsection, we will assume that muons are absent and discuss
the bulk viscosity of relativistic $npe$ matter given by
Eq.~\eqref{eq:zeta_slow1}.  This improves on our previous treatments
in Refs.~\cite{Alford2019a,Alford2019b} where we used non-relativistic
dispersion relations for nucleons in computing the rates of processes
(but not in computing the background nuclear equilibrium) and on
Ref.~\cite{Alford2022} by showing the results for the NL3 density
functional. For parallel developments which also used relativistic
dispersion relations for nucleons with alternative background nuclear
models see Ref.~\cite{Alford:2023gxq}.

The coefficients 
$\lambda_l$, defined by Eqs.~\eqref{eq:lambda_def}, were computed
by taking numerical derivatives of off-equilibrium Urca process rates. 
At densities $n_B\geq n_0$ we find approximately $\lambda_l\simeq 
c\Gamma_{pl\to n\nu}/T$, where the number $c$ varies in the range 
$0.3\leq c\leq 2$ (in the low-temperature limit $c\simeq 1.34$, see 
Eqs.~\eqref{eq:Gamma_lowT_trans} and \eqref{eq:lambda12_lowT_trans}).

The susceptibility $A_1$ given by Eq.~\eqref{eq:def_A1} is insensitive
both to the temperature and the density, therefore the relaxation rate
$\gamma_e=\lambda_e A_1$ scales as
$\gamma_e\propto \Gamma_{pe\to n\nu}/T$, see
Fig.~\ref{fig:gamma_e_temp}. The relaxation rate $\gamma_e$ crosses
the line of the constant angular frequency
$\omega=2\pi\times 1\,\kHz = 4.14\cdot 10 ^{-18}$\,MeV at temperatures
$3\div 4$ MeV if the density is below the direct Urca threshold and
around $T=2$ MeV for densities above the threshold, where
$\gamma_e\propto T^4$. Consequently, the bulk viscosity attains its
maximum at the temperature defined by the crossing. Compared to the
nonrelativistic treatment, the full relativistic calculation predicts
the point of the maximum of the bulk viscosity at lower temperatures,
because it predicts faster equilibration rates.

Figure~\ref{fig:C2A_dens} shows the combination of (isothermal)
susceptibilities $C_1^2/A_1$ relevant to the bulk viscosity (which in
$npe$ matter takes the form of Eq.~\eqref{eq:zeta_slow1}) at two fixed
temperatures $T=1$~MeV and $T=10$~MeV (panels (a) and (b),
respectively). In full relativistic calculation $C_1^2/A_1$ is almost
density-independent above $n_B=2n_0$ in contrast to its
nonrelativistic counterpart which monotonically increases and strongly
overestimates the bulk viscosity already at density $n_B=2n_0$. The
temperature dependence of isothermal susceptibility $C_1^2/A_1$ is
very weak in the range $1\leq T\leq 10$ MeV almost at all
densities. The only exception is the density range below the nuclear
saturation density. The green triangles in panel (a) show the results
of Ref.~\cite{Alford2019a} for the DD2 model at $T=1$~MeV which were
obtained by direct numerical differentiation of chemical imbalance
$\mu_{\Delta}$. (Note that Refs.~\cite{Alford2019a,Alford2019b} define
the susceptibilities $A_1$ and $C_1$ via alternative expressions
$A_1=-n_B^{-1} (\partial\mu_\Delta/\partial Y_p)_{n_B}$,
$C_1=n_B(\partial\mu_\Delta/ \partial n_B)_{Y_p}$). It is seen that
the results of our analytic expressions for relativistic
susceptibilities agree quite well with the results of
Ref.~\cite{Alford2019a}.

For the sake of completeness, we compute also the adiabatic
susceptibilities in addition to the isothermal ones.  The dashed lines
in panel (b) show the adiabatic susceptibilities as computed in
Appendix~\ref{app:A_ad}. (Note that, at low temperatures
$T\simeq 1$~MeV the difference between the isothermal and adiabatic
susceptibilities is very small and so is not visible on the left panel
of the plot). We see that at high temperatures the adiabaticity
enhances the susceptibility $C_1^2/A_1$ by a factor of a few at low
densities $n_B\leq 2n_0$. Comparing the two panels of
Fig.~\ref{fig:C2A_dens}, we see also that the adiabatic
susceptibilities are practically temperature-independent in the whole
range of densities $0.5n_0\leq n_B\leq 5n_0$.

\begin{figure}[t] 
\begin{center}
\includegraphics[width=0.45\columnwidth, keepaspectratio]{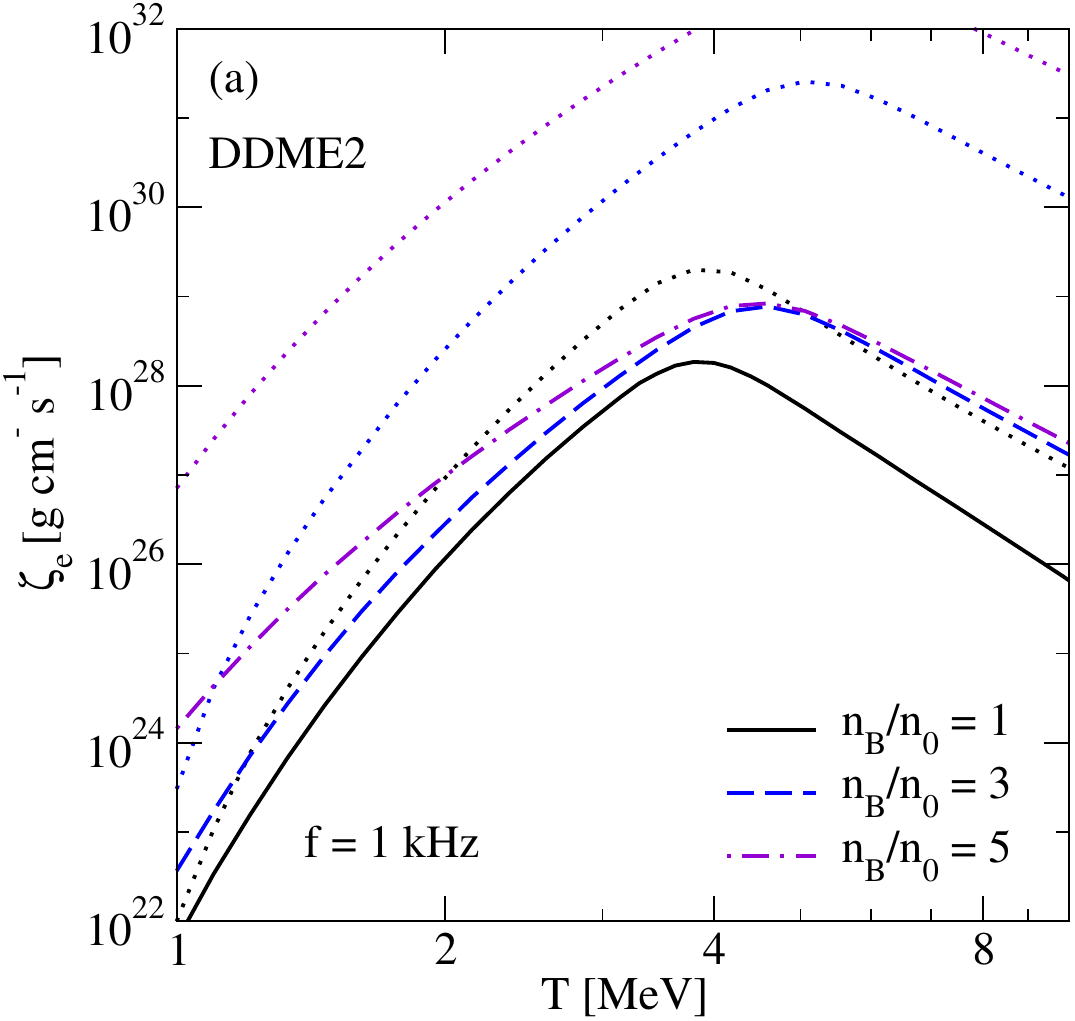}
\hspace{0.5cm} 
\includegraphics[width=0.45\columnwidth, keepaspectratio]{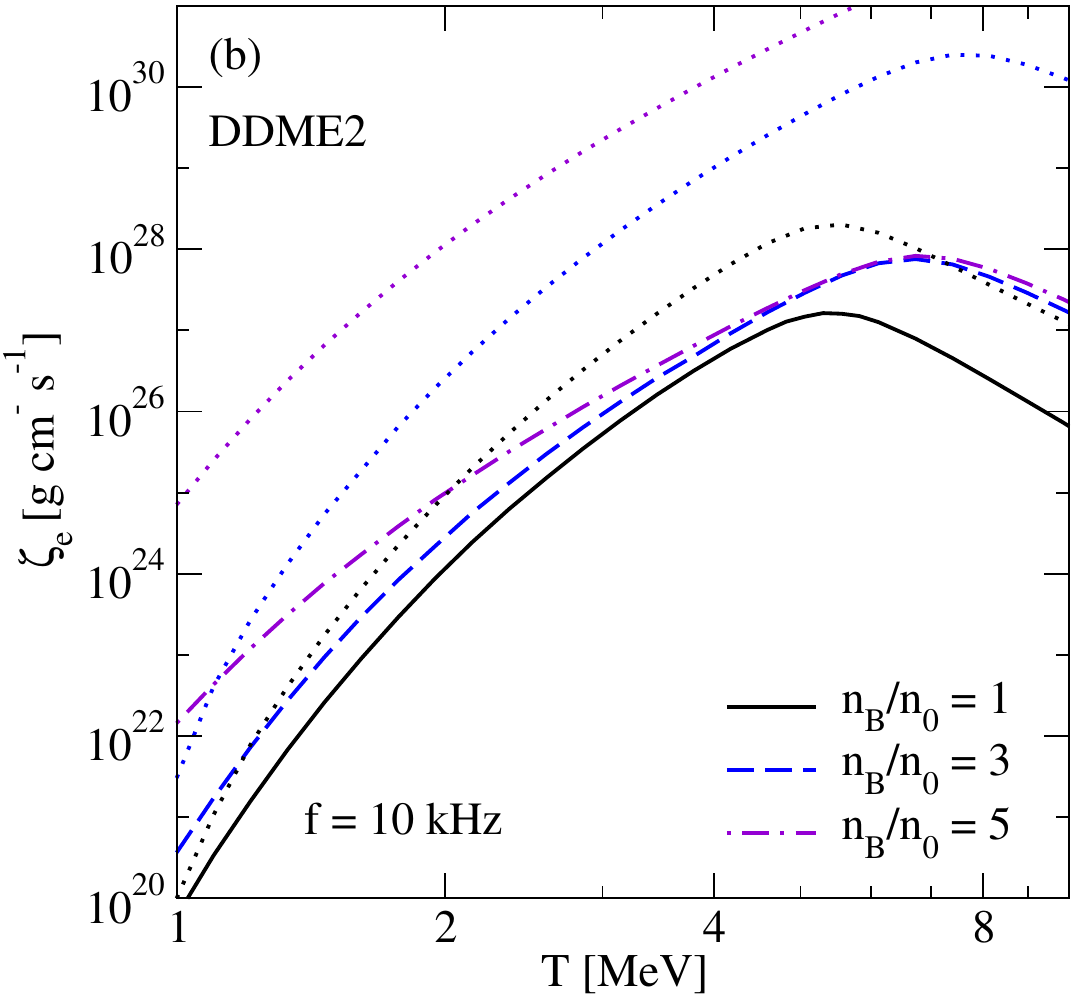}
\caption{Each panel shows the bulk viscosity of relativistic $npe$
  matter as a function of temperature for three values of baryon
  density for DDME2 model.  The left panel is for oscillations of
  frequency $f=1$ kHz; the right panel is for $f=10$ kHz.  The dotted
  lines show the results of Ref.~\cite{Alford2019b} obtained within
  the approximation of nonrelativistic nucleons. }
\label{fig:zeta_e_temp_DD} 
\end{center}
\end{figure}
\begin{figure}[!] 
\begin{center}
  \includegraphics[width=0.45\columnwidth,
  keepaspectratio]{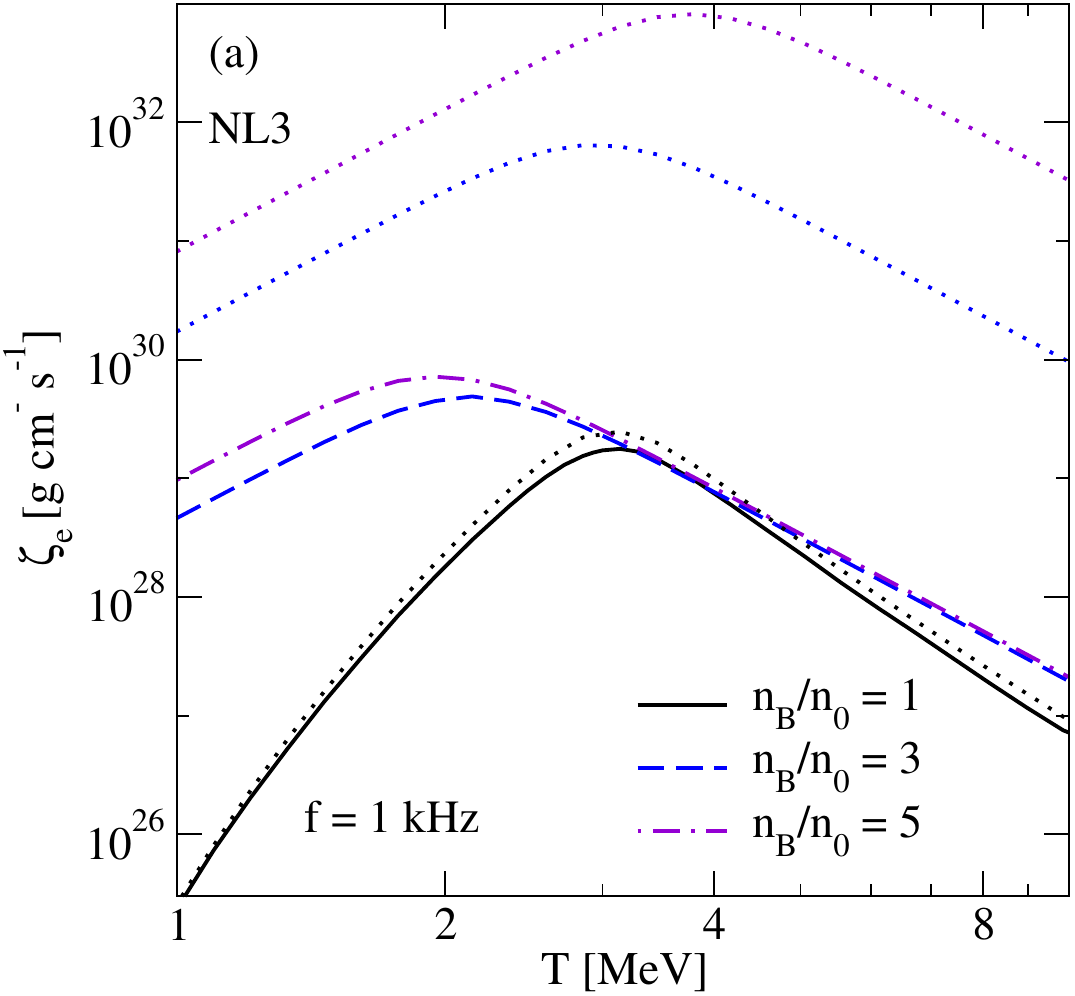} \hspace{0.5cm}
  \includegraphics[width=0.45\columnwidth,
  keepaspectratio]{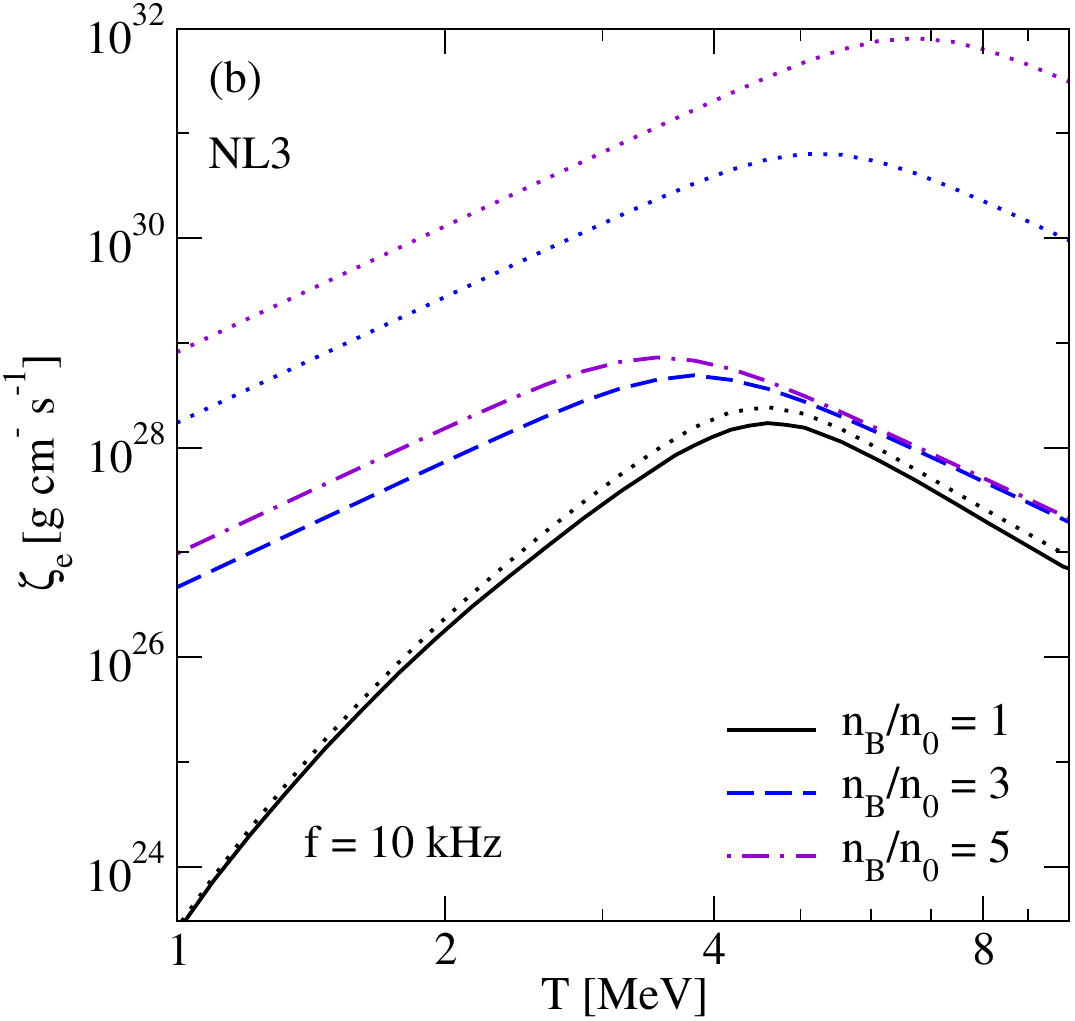}
  \caption{ The bulk viscosity of relativistic $npe$ matter as a
    function of temperature for three values of baryon density for the
    NL3 model at (a) $f=1$ kHz; (b) $f=10$ kHz. The dotted lines show
    the results of Ref.~\cite{Alford2019b} obtained within the
    approximation of nonrelativistic nucleons. }
\label{fig:zeta_e_temp_NL3} 
\end{center}
\end{figure}

Figures~\ref{fig:zeta_e_temp_DD} and \ref{fig:zeta_e_temp_NL3} show
the temperature dependence of the bulk viscosity of $npe$ matter for
DDME2 and NL3 models, respectively, computed according to
Eq.~\eqref{eq:zeta_slow1}. The results for the bulk viscosity in the
isothermal case are shown for two frequencies $f=1$ kHz and $f=10$ kHz
which bracket the typical range of frequencies of density oscillations
in BNS mergers. As discussed above, for any given frequency $\zeta_e$
has a maximum at the temperature where $\omega=\gamma_e(T_{\rm max})$,
and $T_{\rm max}$ increases with the frequency. The maximum value of
the bulk viscosity for the given density decreases with the frequency
as $\zeta_{e\rm max} = C_1^2/(A_1\omega)$.  At temperatures below the
resonant maximum, chemical equilibration is slower than density
oscillations, \ie, $\gamma_e\ll\omega$, and the bulk viscosity drops
rapidly as frequency rises $\zeta_e\propto \omega^{-2}$. At
temperatures above the resonant maximum, chemical equilibration is
faster than the oscillations and we have
$\zeta_e = C_1^2/(A_1\gamma_e)$, which is independent of the
frequency.

\begin{figure}[t] 
\begin{center}
\includegraphics[width=0.45\columnwidth, keepaspectratio]{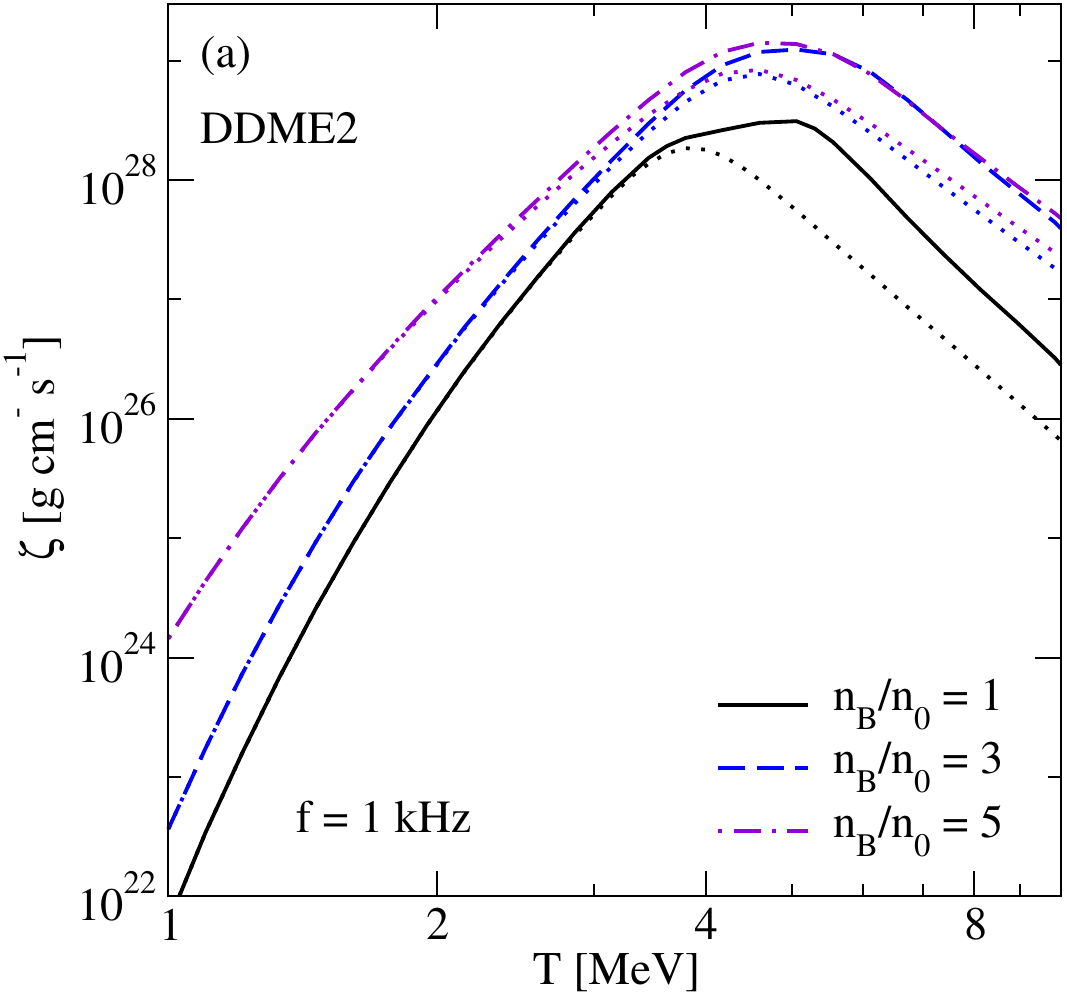}
\hspace{0.5cm} 
\includegraphics[width=0.45\columnwidth, keepaspectratio]{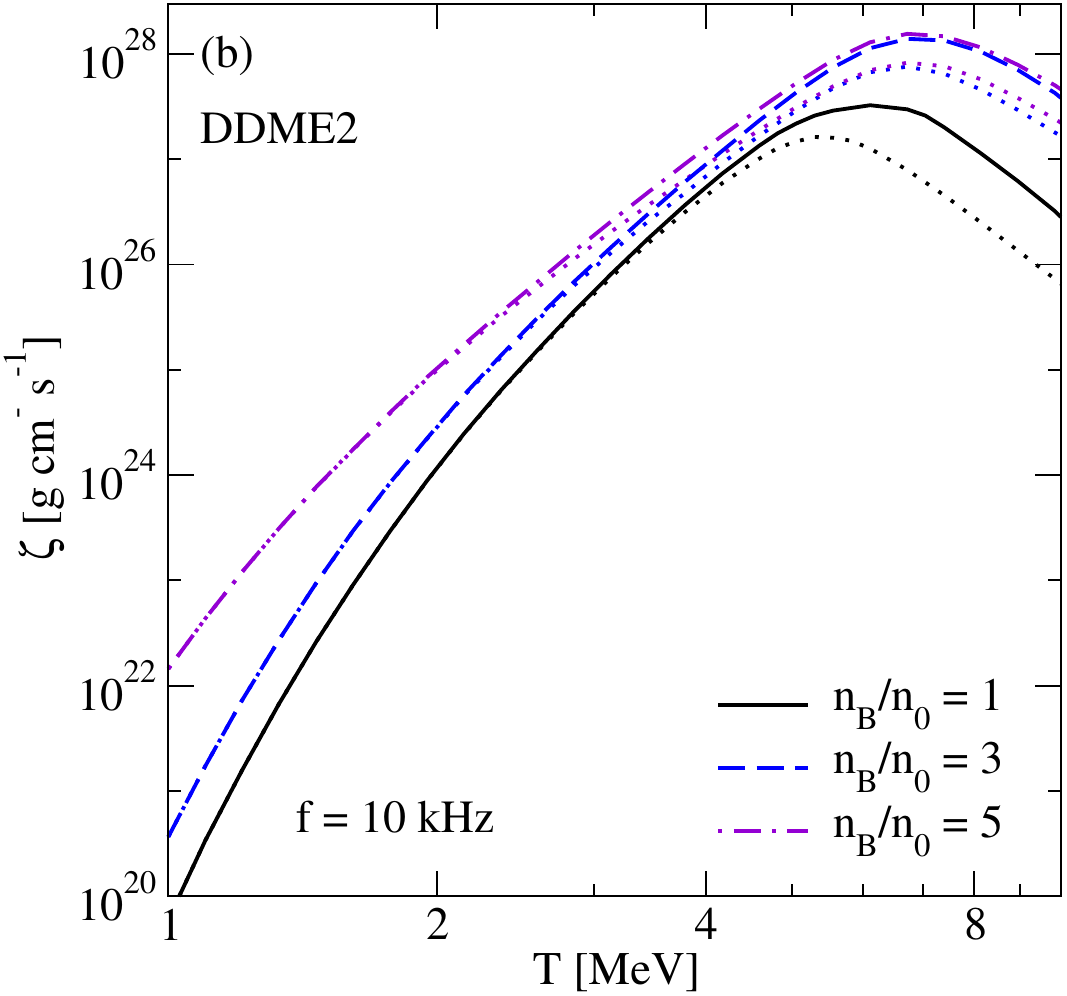}
\caption{ The bulk viscosity of relativistic $npe\mu$ matter as a
  function of temperature for three values of baryon density for the
  DDME2 model at (a) $f=1$ kHz; (b) $f=10$ kHz.  The dotted lines show
  the bulk viscosities of relativistic $npe$ matter. }
\label{fig:zeta_slow_temp_DDME2} 
\end{center}
\end{figure}
\begin{figure}[!] 
\begin{center}
  \includegraphics[width=0.45\columnwidth,
  keepaspectratio]{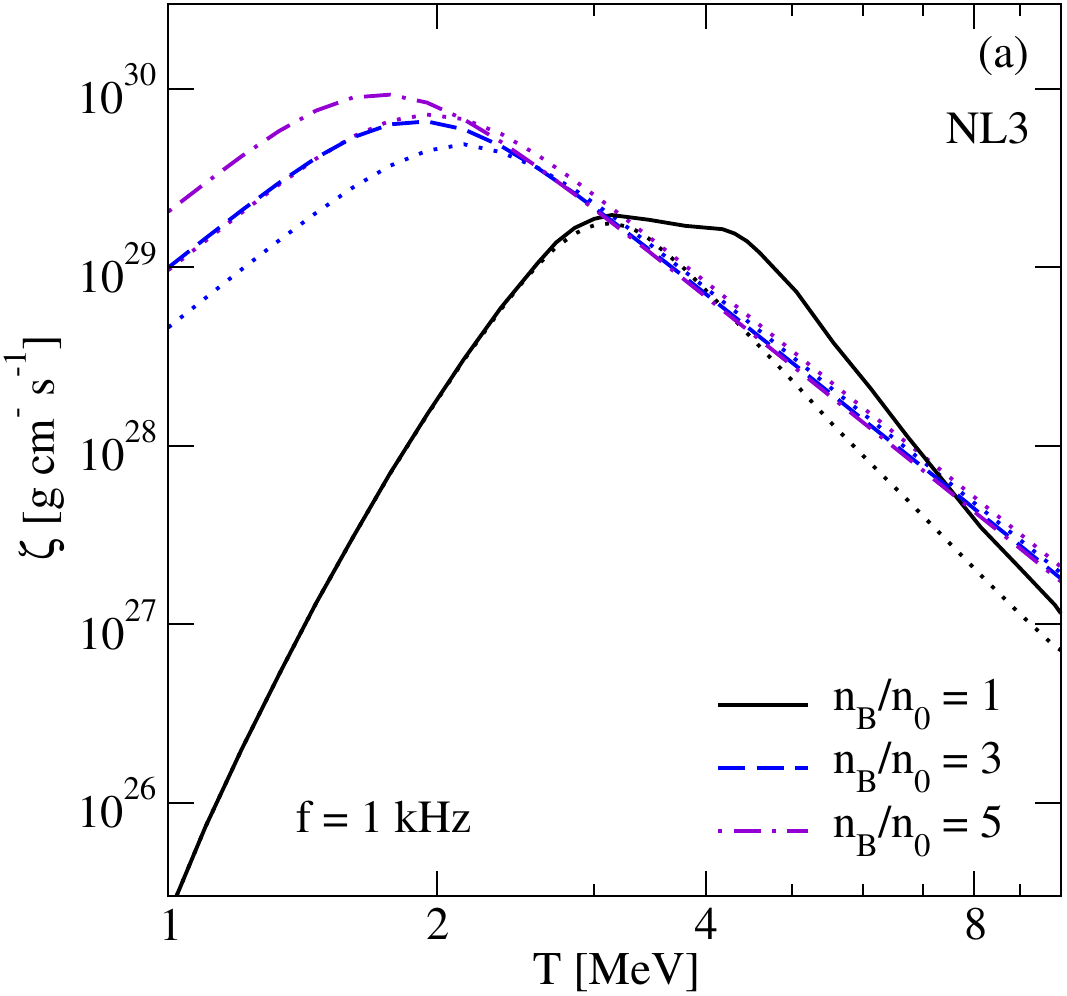} \hspace{0.5cm}
  \includegraphics[width=0.45\columnwidth,
  keepaspectratio]{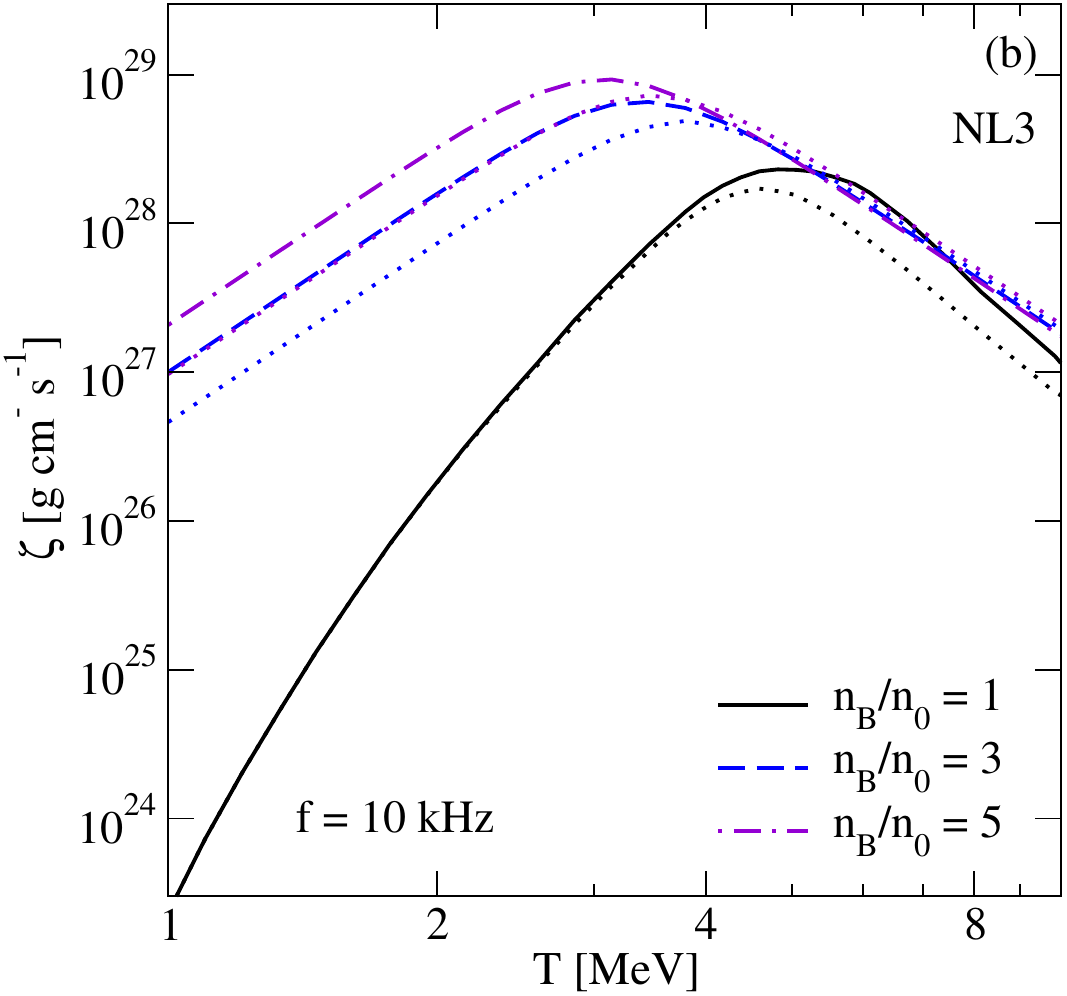}
  \caption{ The bulk viscosity of relativistic $npe\mu$ matter as a
    function of temperature for three values of baryon density for the
    NL3 model at (a) $f=1$ kHz; (b) $f=10$ kHz. The dotted lines show
    the bulk viscosities of relativistic $npe$ matter. }
\label{fig:zeta_slow_temp_NL3} 
\end{center}
\end{figure}

In the case of the DDME2 model, in which direct Urca processes are
kinematically forbidden at low temperatures (\ie, the relevant
densities are always below the threshold density) the maximum of the
bulk viscosity moves to a higher temperature as density increases from
$n_0$ to $3n_0$.  This is consistent with Fig.~\ref{fig:gamma_e_temp}
where we see that for DDME2 $\gamma_e$ drops as density rises from
$n_0$ to $3n_0$ at fixed $T$. In general, one expects Urca rates to
increase with density (as seen for NL3) but this can be offset by
other factors such as changes in the dispersion relations that affect
the density of states at the Fermi surface. We already know from
Fig.~\ref{fig:Gamma12_e_DDME2} that for DDME2 the Urca rates drop
slightly with increasing density at $T\gtrsim 3$~MeV.

The NL3 model, in which particles near the Fermi surfaces can undergo
direct Urca, shows the opposite behavior: the maximum is shifted to
lower temperatures once the direct Urca threshold is achieved. This is
expected since the rates rise with density because of increasing phase
space at the Fermi surfaces, so $\gamma=1$\,kHz is achieved at lower
temperatures. Comparing these results with the ones obtained within
the nonrelativistic approximation for nucleons we observe two
characteristic features: (i) the maximum is shifted to lower
temperatures in the relativistic calculation, the shift being larger
above the direct Urca threshold; (ii) the approximation of
nonrelativistic nucleons overestimates the bulk viscosity by orders of
magnitude for DDME2 models and by an order of magnitude for NL3 model.

Note that, according to the susceptibilities shown in
Fig.~\ref{fig:C2A_dens}, the bulk viscosities computed in the
adiabatic and isothermal cases will differ appreciably only below the
saturation density.

\subsubsection{Bulk viscosity of relativistic $npe\mu$ matter}
\label{sec:bulk_muons}

\begin{figure}[t] 
\begin{center}
\includegraphics[width=0.65\columnwidth, keepaspectratio]{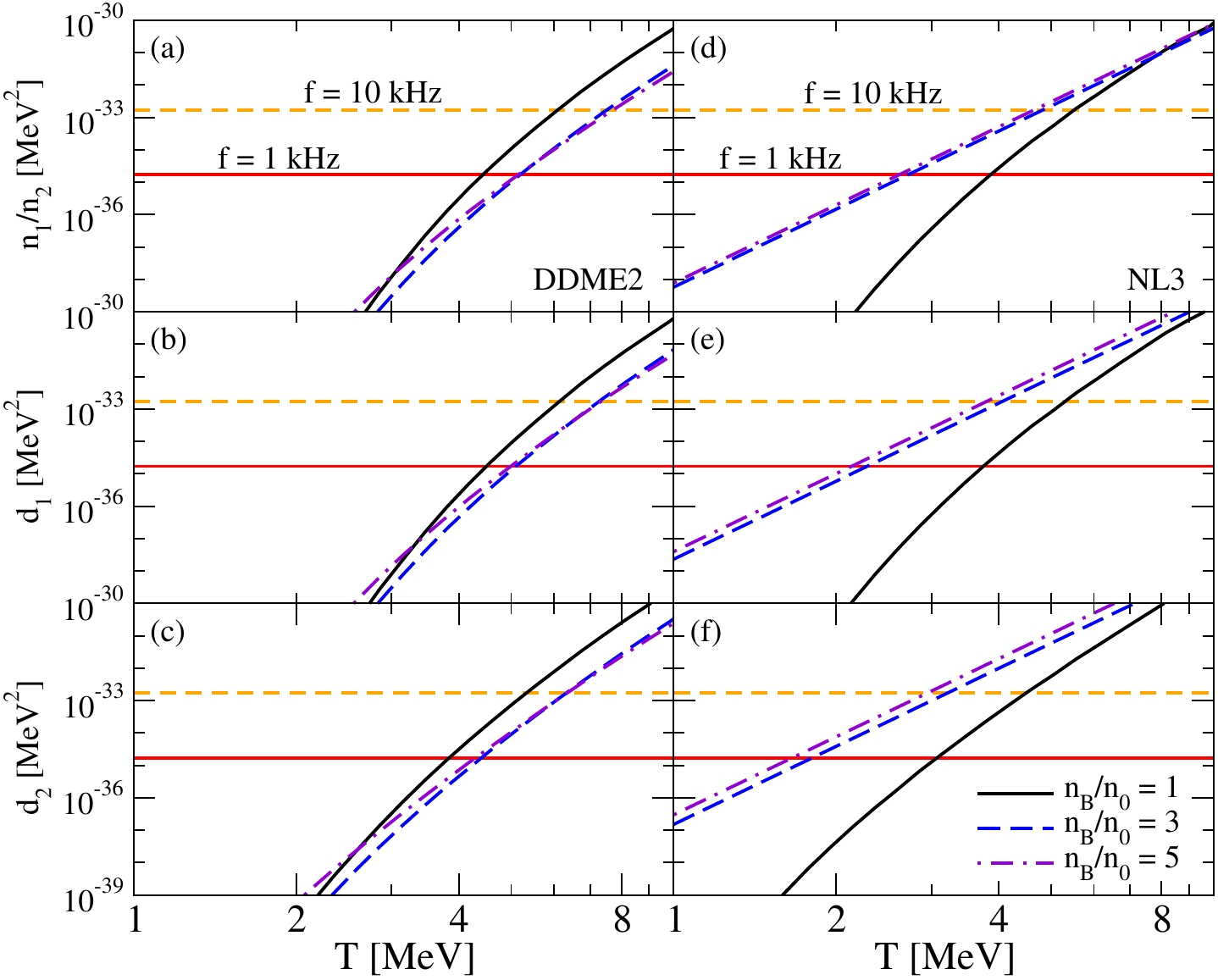}
\caption{ The quantities $n_1/n_2$, $d_1$ and $d_2$ entering in the
  expression of the bulk viscosity~\eqref{eq:zeta_slow} as functions
  of the temperature for fixed values of density for the DDME2 model
  (left panels) and the NL3 model (right panels).  The horizontal
  lines correspond to the squares of the oscillation frequencies fixed
  at $f = 1$ kHz (solid lines) and $f = 10$ kHz (dashed lines).}
\label{fig:num_denom} 
\end{center}
\end{figure}

The bulk viscosity of relativistic $npe\mu$ matter computed in the
slow-lepton equilibration limit~\eqref{eq:zeta_slow} is shown in
Figs.~\ref{fig:zeta_slow_temp_DDME2} and \ref{fig:zeta_slow_temp_NL3},
for models DDME2 and NL3, respectively.  The bulk viscosity of $npe$
matter $\zeta_e$ is shown for comparison by dotted lines. The
qualitative behavior of $\zeta$ is similar to that of $\zeta_e$.

In the following, we first focus on the DDME2 density functional model
(which does not reach the low-temperature Urca threshold at any
density) and discuss first the low-temperature regime which is
followed by a discussion of the high-temperature regime.  At low
temperatures, where $\lambda_iA_j\ll\omega$, we have
$n_1/n_2, d_1, d_2\sim \lambda_i A_j\ll \omega^2$, and the bulk
viscosity is given by
$\zeta\simeq n_2/\omega^2=(\lambda_e C_1^2 +\lambda_\mu
C_2^2)/\omega^2=\zeta_e+\zeta_\mu$~\cite{Haensel2000}.  In this regime
$\zeta_\mu$ is much smaller than $\zeta_e$, therefore the bulk
viscosity of $npe\mu$ matter practically coincides with that of $npe$
matter. As shown above, in the case of DDME2 model the muons should be
neglected in the evaluation of the bulk viscosity in the
low-temperature sector, where all muonic processes are suppressed
compared to the electronic Urca processes. However, because
$\zeta_\mu\ll \zeta_e$ in this regime, the muonic contribution
automatically drops, therefore the bulk viscosity of $npe\mu$ matter
in the whole regime can be computed from Eq.~\eqref{eq:zeta_slow}.

At high temperatures, where equilibration is fast compared to the
oscillation frequency, $\lambda_iA_j\gg\omega$, we approach the
low-frequency limit where the bulk viscosity becomes
frequency-independent and is equal to
$\zeta = n_1/d_1^2\sim 1/\lambda_i$ which decreases with the
temperature. In this regime, the bulk viscosity of $npe\mu$ matter
exceeds the bulk viscosity of $npe$ matter by factors between 2.5
and 8 for the model DDME2.  At intermediate temperatures, where
$\lambda_i A_j\approx T$, the bulk viscosity obtains a
maximum. However, as the quantities $n_1/n_2, d_1$ and $d_2$ reach
their maxima at slightly different temperatures, see
Fig.~\ref{fig:num_denom}, there is a broadened maximum or a
``flattened" structure in the temperature dependence of $\zeta$, which
is clearly pronounced at density $n_B=n_0$, see left panels
  of Figs.~\ref{fig:zeta_slow_temp_DDME2} and
  \ref{fig:zeta_slow_temp_NL3}. The maximum of the bulk viscosity of
$npe\mu$ matter is located at a slightly higher temperature as
compared to the bulk viscosity of $npe$ matter.

In the case of NL3 model, the effect of the inclusion of muons on the
bulk viscosity below the direct Urca threshold is the same as in the
case of DDME2 model: muons enhance $\zeta$ by up to a factor of 3 at
temperatures above the maximum, whereas they almost do not affect the
bulk viscosity below the maximum. At densities $n_B=3n_0$ and
$n_B=5n_0$, which are above the threshold, the electronic and muonic
Urca rates are almost equal, see Figs.~\ref{fig:Gamma12_e_NL3} and
\ref{fig:Gamma12_mu}, resulting in almost equal contributions of
electrons and muons to the bulk viscosity. Thus, to the left side of
the maximum, where the bulk viscosity is proportional to the reaction
rates, see Eq.~\eqref{eq:zeta_slow2}, we have
$\zeta\simeq \zeta_e+\zeta_\mu\simeq 2\zeta_e$. At higher
temperatures, the total bulk viscosity is slightly smaller than that
of $npe$ matter. Note also that above the direct Urca threshold, the
inclusion of muons moves the location of the resonant maximum to
smaller temperatures, whereas below the threshold the location of the
maximum remains nearly unchanged.

\begin{figure}[t]  
\begin{center}
\includegraphics[width=0.45\columnwidth,keepaspectratio]{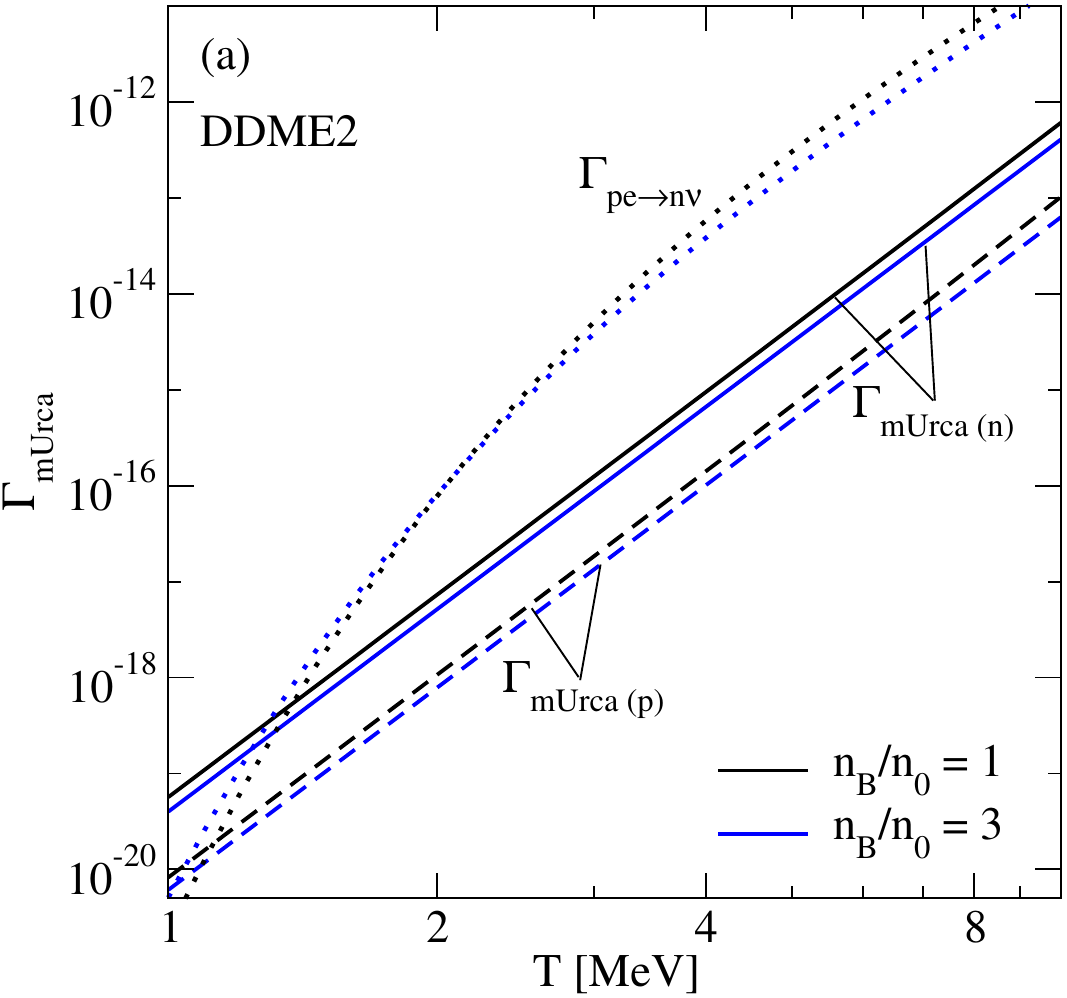}
\hspace{0.5cm} 
\includegraphics[width=0.45\columnwidth,keepaspectratio]{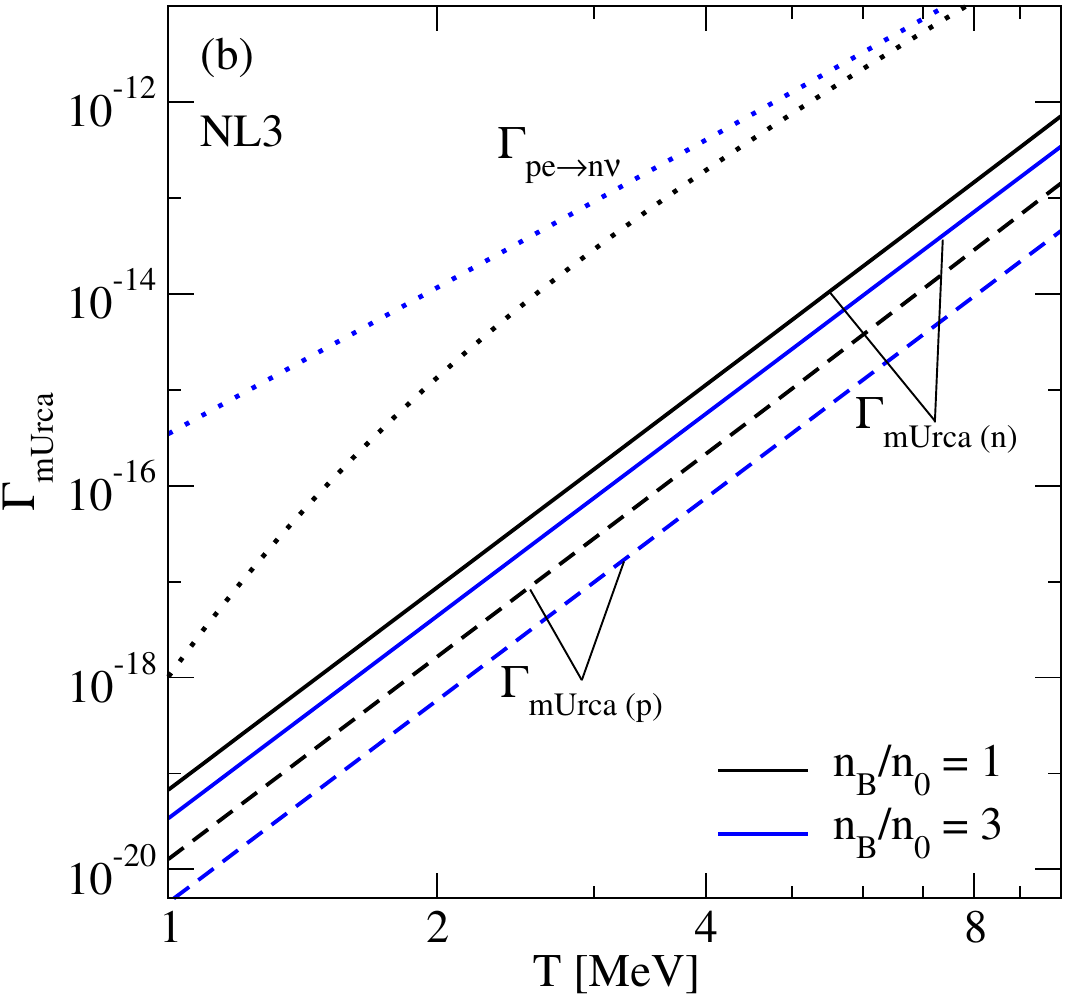}
\caption{ The rates of modified Urca process involving electrons at
  two fixed densities and for (a) DDME2 and (b) NL3 model. The direct
  Urca electron capture rates are shown by the dotted lines for
  comparison.}
\label{fig:Gamma_mod}  
\end{center}
\end{figure}
\begin{figure}[]  
\begin{center}
\includegraphics[width=0.45\columnwidth,keepaspectratio]{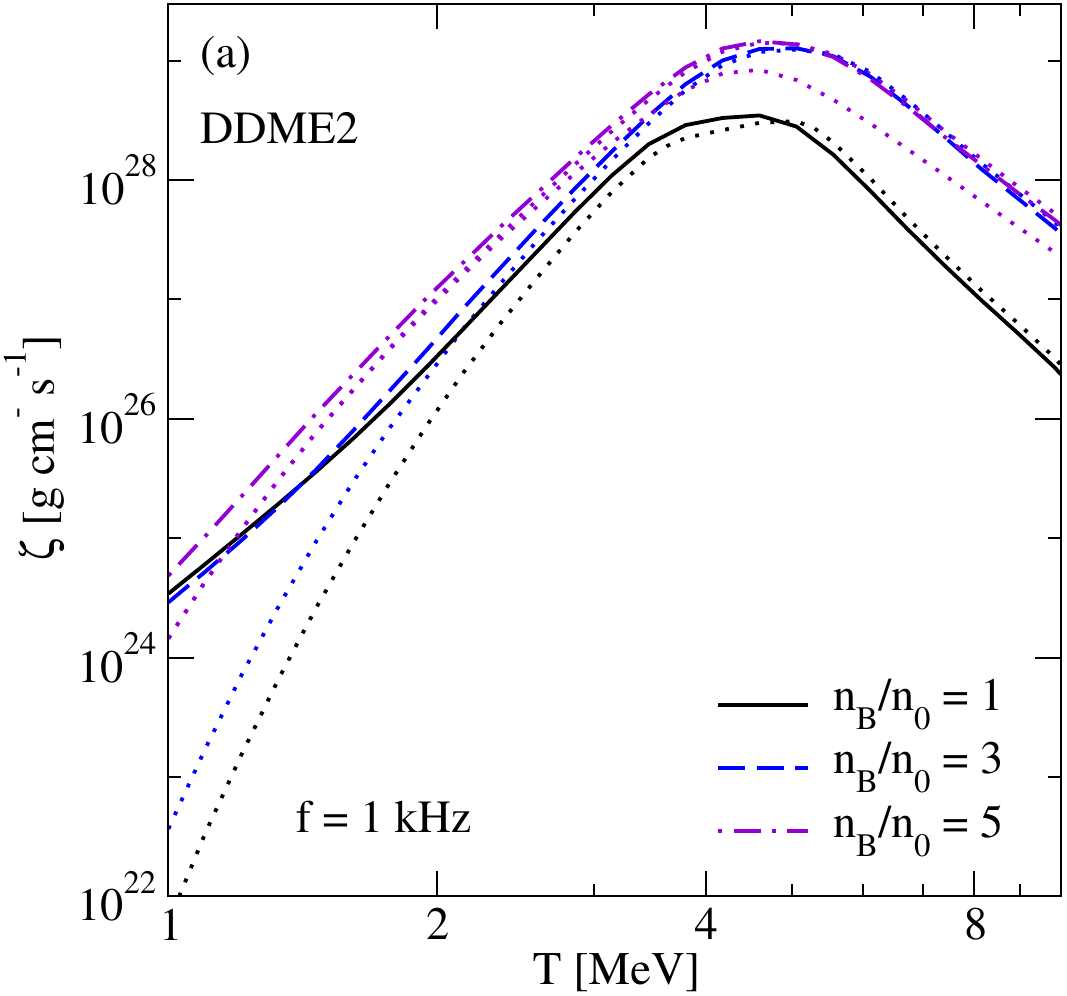}
\hspace{0.5cm} 
\includegraphics[width=0.45\columnwidth,keepaspectratio]{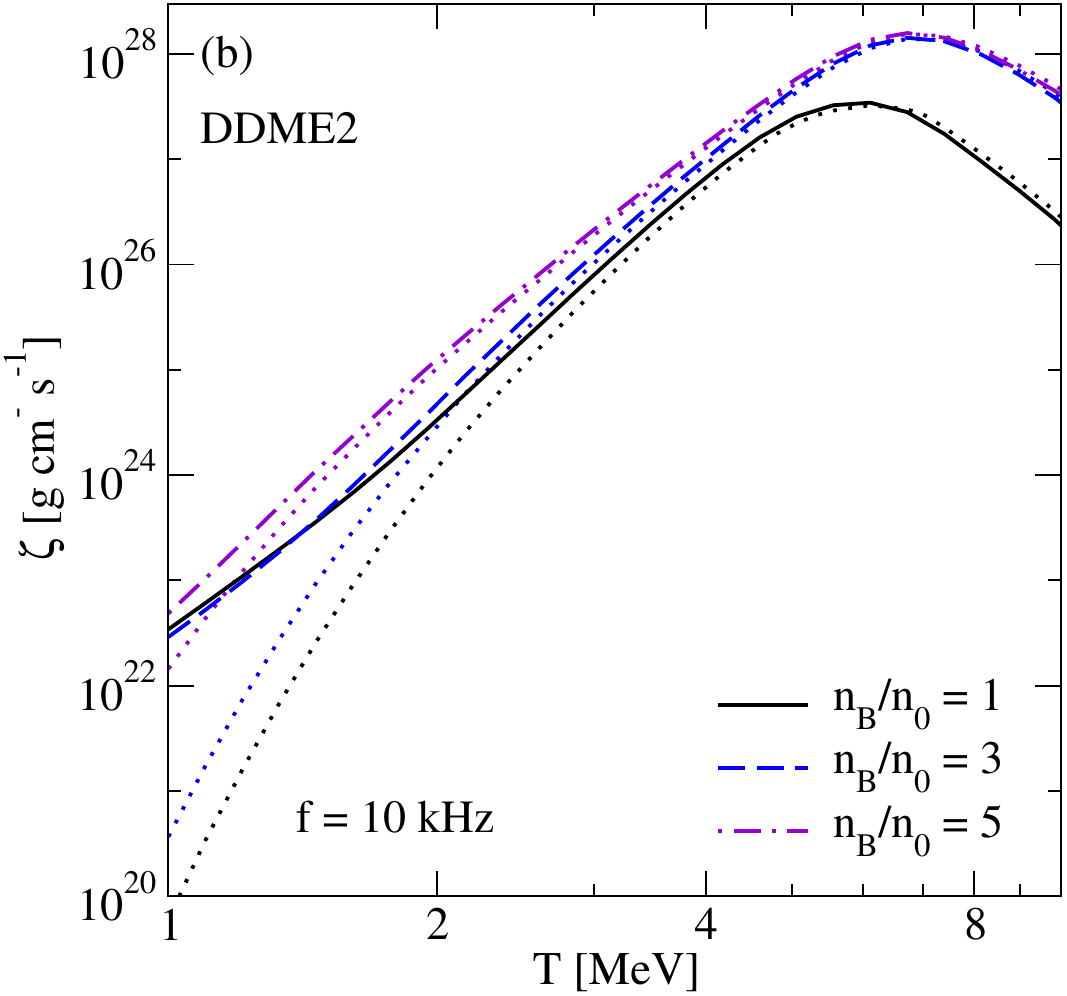}
\caption{ The bulk viscosity of relativistic $npe\mu$ matter with the
  inclusion of modified Urca processes for the DDME2 model at (a)
  $f=1$ kHz; (b) $f=10$ kHz.  The dotted lines reproduce the bulk
  viscosities shown in Fig.~\ref{fig:zeta_slow_temp_DDME2} which were
  obtained neglected modified Urca processes.}
\label{fig:zeta_mod}  
\end{center}
\end{figure}

For the sake of completeness, we also investigate how the modified
Urca processes $N+n\rightarrow N+p + l^-+\bar{\nu}_l$ and
$N+p + l^-\rightarrow N+ n+{\nu}_l$, $N\in n,p$, affect the bulk
viscosity (Fig.~\ref{fig:zeta_mod}).  For that purpose, we use the
low-temperature modified Urca rates from Ref.~\cite{Alford2019a}. Note
that there is no threshold for these processes.

The modified Urca rates for electronic processes for two models are
shown in Fig.~\ref{fig:Gamma_mod}. Note that the low-temperature
modified Urca rates are equal for neutron decay and electron capture
processes when $\mu_n=\mu_p+\mu_e$.  The rates of the muonic-modified
Urca processes are very close to these and are not shown. The dotted
lines show the rates of the direct Urca electron capture rates for
comparison.  Because the direct Urca neutron decay is strongly damped
at densities below the direct Urca threshold, its rate is much smaller
than that of the summed modified Urca rate at those densities. Below
the direct Urca threshold and at moderate temperatures $T\geq 3$~MeV,
the direct Urca lepton capture rates exceed the modified Urca rates by
at least an order of magnitude.  The modified Urca process rates
become comparable to the direct Urca electron capture rates at
$T\simeq 1.5$~MeV and the direct Urca muon capture rates at
$T\simeq 3$~MeV for the model DDME2.  Above the direct Urca threshold
which is realized only in the case of NL3 model at densities
$n_B\geq 1.5n_0$ both direct Urca rates are higher than the modified
Urca rates by at least an order of magnitude. In the case of the NL3
model, the direct Urca electron capture rate is always at least two
orders of magnitude larger than that of the modified Urca, whereas the
direct muon capture rate becomes smaller than the modified process
rate at $T\leq 2$~MeV below the threshold, \eg, at $n_B=n_0$. When the
modified Urca processes are included, the summed Urca process rates
are always much higher than the muon decay rates. Thus, the bulk
viscosity of $npe\mu$ matter can be computed according to the
slow-lepton-equilibration limit in the whole temperature-density range
of interest.  Figure~\ref{fig:zeta_mod} shows the bulk viscosity of
the $npe\mu$ matter with the inclusion of modified Urca processes for
the model DDME2. The bulk viscosity computed only with the direct Urca
is shown for comparison with the dotted lines.  We see that the
inclusion of modified Urca processes becomes important at densities
below the direct Urca threshold in the low-temperature regime
$T\leq 3$~MeV.  For NL3 model, the modified Urca processes do not have
any significant impact on the bulk viscosity. Also, note that the
modified Urca processes do not change the location of the maximum bulk
viscosity.

\subsection{Damping of density oscillations}
\label{sec:damping}

\begin{figure}[t] 
\centering
\includegraphics[width=0.45\columnwidth, keepaspectratio]{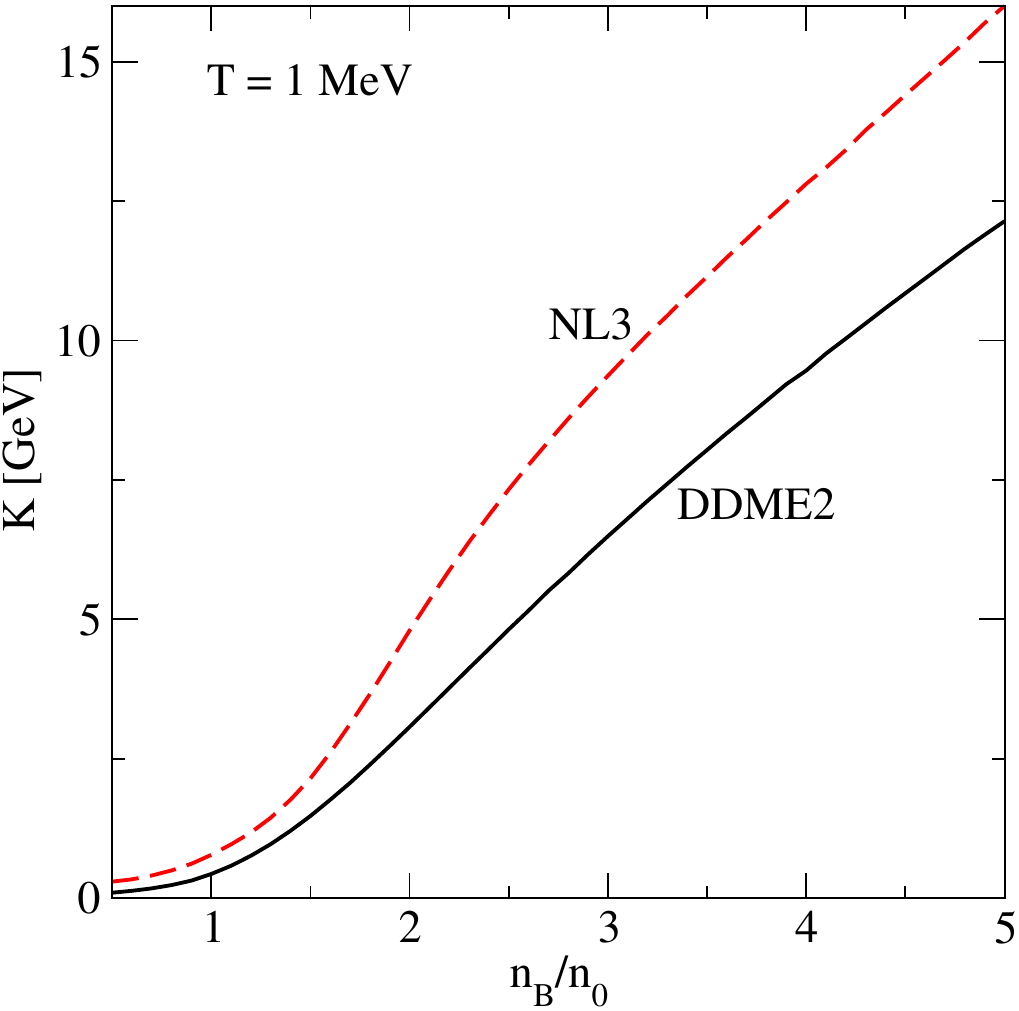}
\caption{ The incompressibility of nuclear matter for the DDME2 
and NL3 models. The temperature is fixed at $T=1$~MeV.}
\label{fig:compress} 
\end{figure}

Now we estimate the bulk viscous damping timescale in relativistic
$npe\mu$ matter. The damping timescale is the decay time for a density
oscillation and is given by~\cite{Alford2018a,Alford2019a,Alford2020}
\bea\label{eq:damping_time}
\tau_{\zeta} =\frac{1}{9}\frac{Kn_B}{\omega^2\zeta},
\eea 
where the incompressibility of nuclear matter is
\bea\label{eq:compress}
K=9n_B\frac{\partial^2\epsilon}{\partial n_B^2},
\eea 
and $\epsilon$ is the energy density.  The incompressibility is
plotted in Fig.~\ref{fig:compress}. It is not sensitive to the
temperature in the range $1\leq T\leq 10$~MeV, therefore the damping
timescale shows temperature dependence inverse to that of the bulk
viscosity and attains its minimum value at the temperature where the
bulk viscosity has a maximum, see Figs.~\ref{fig:tau_DDME2} and
\ref{fig:tau_NL3}. The damping timescale is frequency-independent in
the low-temperature regime, but is inversely proportional to
$\omega^2$ in the high-temperature regime above the minimum.  For the
minimal value, we have $\tau_\zeta\propto 1/\omega$.

\begin{figure}[t] 
\centering
\includegraphics[width=0.45\columnwidth, keepaspectratio]{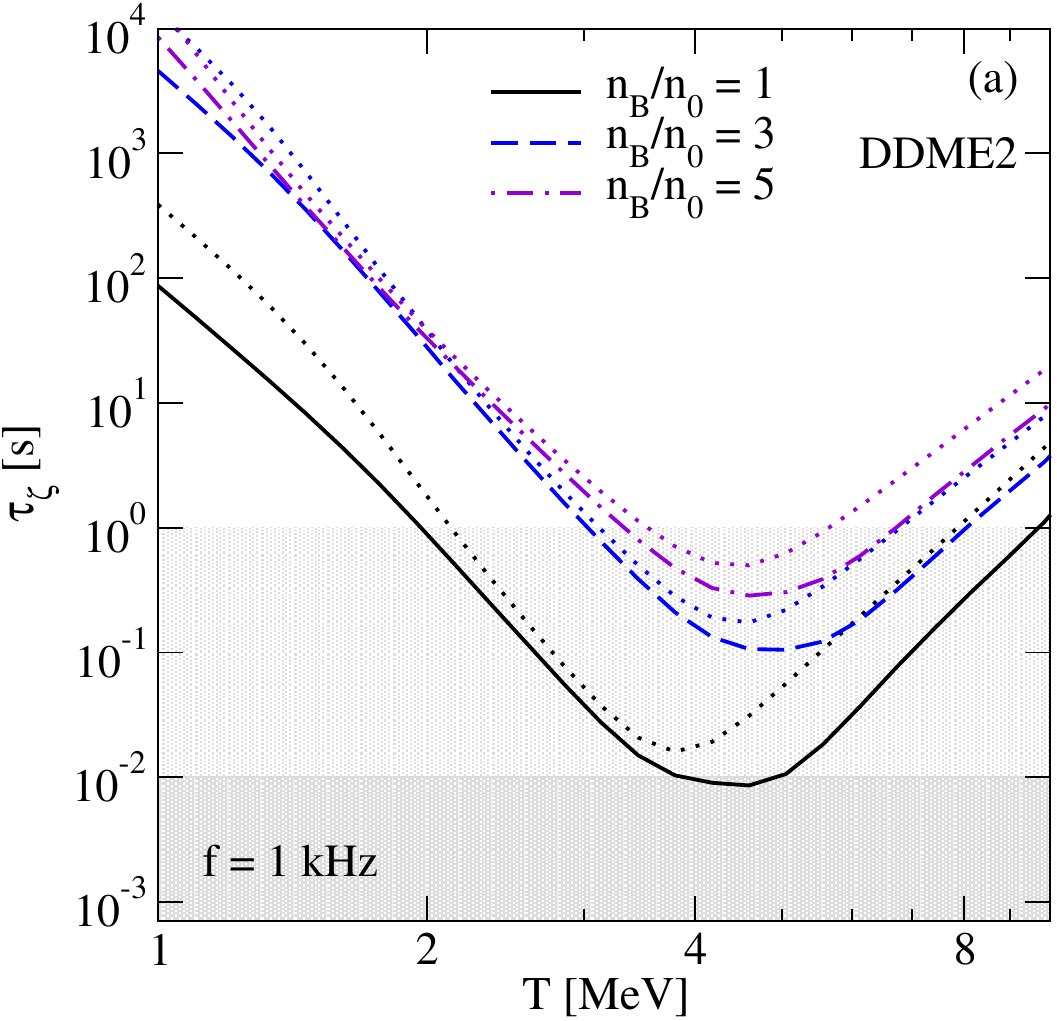}
\hspace{0.5cm}
\includegraphics[width=0.45\columnwidth, keepaspectratio]{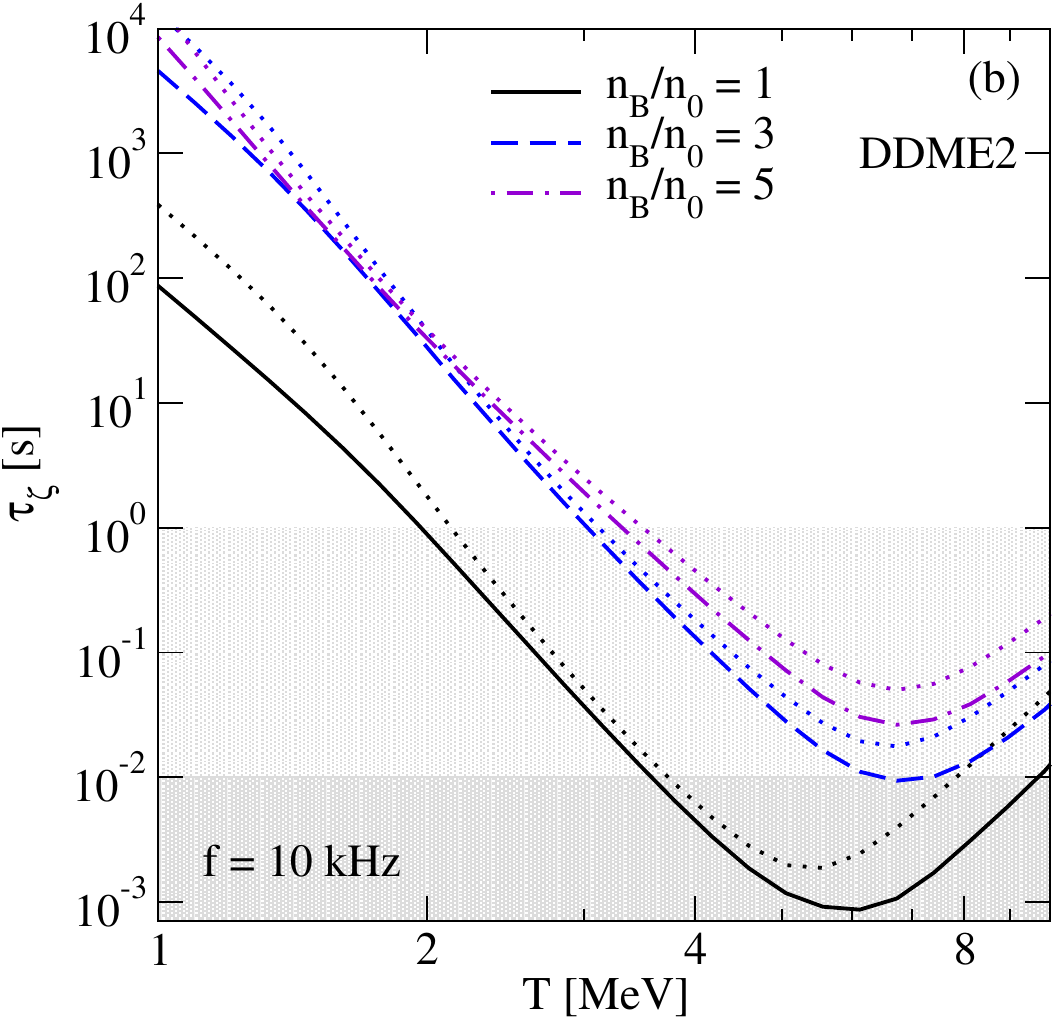}
\caption{ The damping timescale of oscillations as a function of
  temperature for various densities for the DDME2 model for (a)
  $f=1$~kHz and (b) $f=10$~kHz. The dotted lines show the damping
  timescales in $npe$ matter. }
\label{fig:tau_DDME2} 
\end{figure}
\begin{figure}[!] 
\centering
\includegraphics[width=0.45\columnwidth, keepaspectratio]{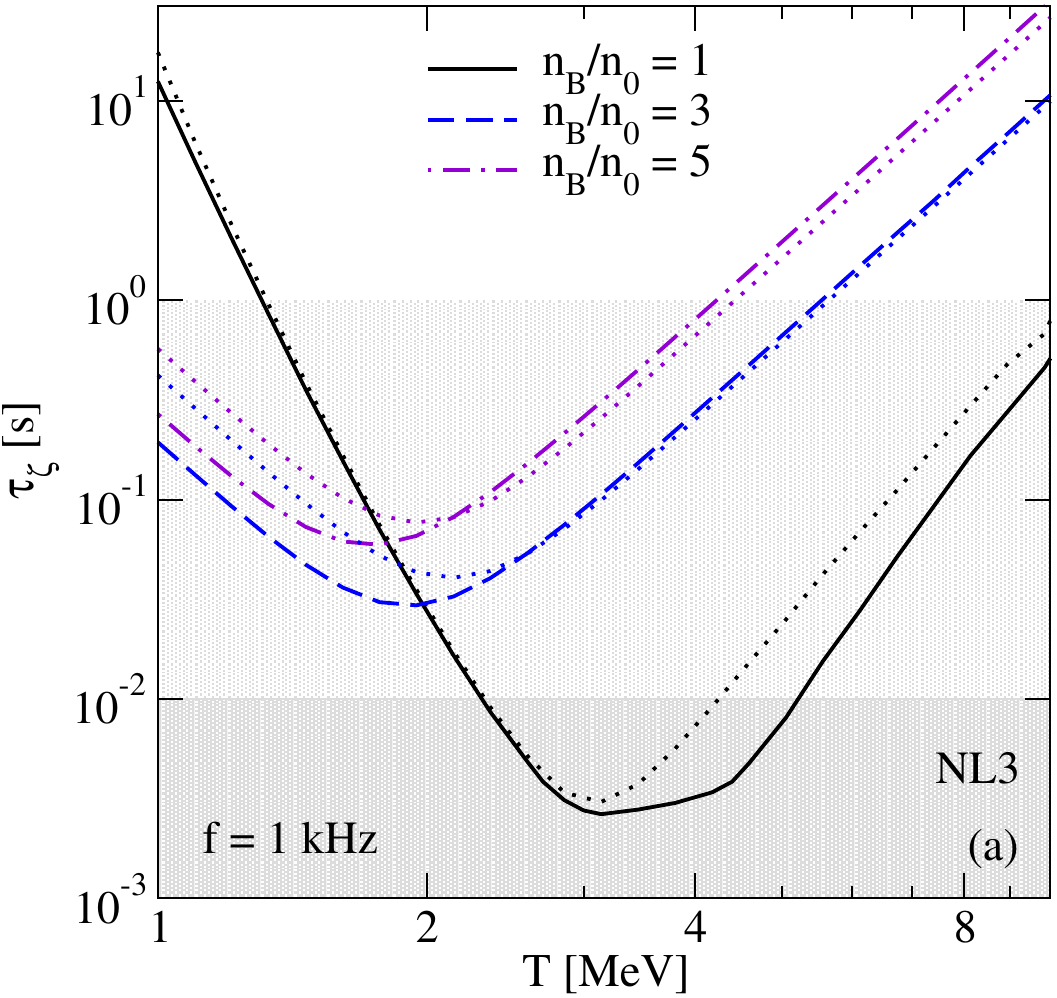}
\hspace{0.5cm}
\includegraphics[width=0.45\columnwidth, keepaspectratio]{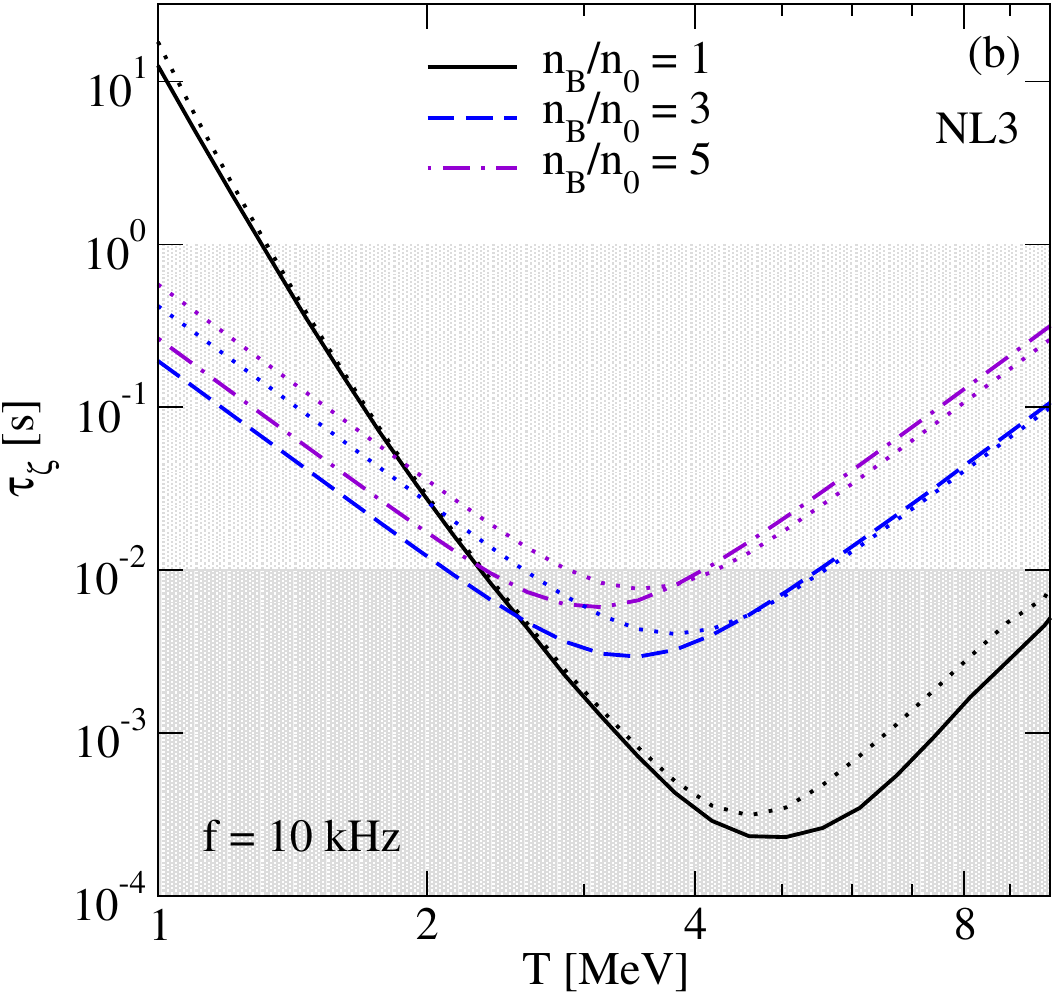}
\caption{ The damping timescale of oscillations as a function of
  temperature for various densities for the NL3 model for (a)
  $f=1$~kHz and (b) $f=10$~kHz. The dotted lines show the damping
  timescales for $npe$ matter. }
\label{fig:tau_NL3} 
\end{figure}

As seen from Eq.~\eqref{eq:damping_time}, the dependence of
$\tau_{\zeta}$ on the density arises from three factors: nuclear
incompressibility, the baryon density, and the inverse bulk viscosity.
Here we will use the bulk viscosities (plotted in
Fig.~\ref{fig:zeta_mod} for the model DDME2) which include both direct
and modified Urca processes. We see that the density dependence of the
maximum bulk viscosity roughly follows $\zeta\propto n_B$, therefore
the density dependence of minimal $\tau_\zeta$ just follows that of
nuclear incompressibility. Thus, the damping timescales are smaller,
and, therefore, the bulk viscous dissipation is more efficient at
lower densities. This result is in contrast to our previous
nonrelativistic treatment~\cite{Alford2020}, where the damping
timescale showed a decreasing behavior with the density as a result of
the overestimation of the bulk viscosity at high densities.

The shaded regions in Figs.~\ref{fig:tau_DDME2} and \ref{fig:tau_NL3}
show where the damping timescale becomes smaller than the short-term
($\simeq$10\,ms, dark shaded areas) and long-term ($\simeq$1 s,
lightly shaded areas) evolution timescales of a BNS merger remnant
object. For a typical oscillation frequency $f=1$~kHz, the model DDME2
predicts that the bulk viscous damping would be marginally relevant in
the short term, and noticeable for long-living remnants with
$\tau_\zeta\geq 10$~ms at any density and in the temperature range
$2\leq T\leq 10$~MeV.  The damping timescale reaches its minimum at
$n_B\leq n_0$ and $T\simeq 5$~MeV, where the damping time $\tau_\zeta$
reaches the short-term ($10\,$ms) evolution timescale. For higher
frequencies, there is already a window of densities and temperatures
where the damping timescales are shorter than the short-term evolution
timescale of BNS mergers. For $f=10$~kHz the short-term damping is
noticeable at densities $n_B\geq 3n_0$ and for temperatures between
$4\leq T\leq 10$~MeV.

In the case of NL3 model, there is always a range of densities and
temperatures where the bulk viscous damping time is comparable to the
short-term evolution timescale. For oscillations of frequency
$f=1$~kHz, the relevant parameter range is $n_B\lesssim 2n_0$ and 2
$\lesssim T\lesssim$ 5 MeV. The high-density region above the direct
Urca threshold does not have a significant impact on the damping of
density oscillations because the Urca processes are so fast that the
system is not driven far from equilibrium. This result again differs
from those of Ref.~\cite{Alford2020}. At $f=10$~kHz, the damping
timescale reaches down to the ms range also at high densities.  On the
long-term evolution timescale, the damping is efficient at all
densities. Correspondingly, the range of temperatures where the bulk
viscosity would play a role is larger than in the case of DDME2 model.

A comparison of Figs.~\ref{fig:tau_DDME2} and \ref{fig:tau_NL3} shows
that the damping timescale is a few times shorter for model NL3
although the bulk viscosity for NL3 is larger by an order of
magnitude.  This is because NL3 matter is stiffer and has larger
incompressibility, so density oscillations store more energy and this
outweighs the larger bulk viscosity (see Eq.~\eqref{eq:damping_time}).

The damping times shown in Figs.~\ref{fig:tau_DDME2} and
\ref{fig:tau_NL3} are for isothermal oscillations. We have also
performed calculations for adiabatic oscillations, and we find that at
low densities $n_B\leq 2n_0$ and sufficiently high temperatures
$T\geq 5$~MeV the adiabatic density oscillations have slightly shorter
damping timescales. The maximal difference between adiabatic and
isothermal nuclear incompressibilities is about 15\% for DDME2 and 7\%
for NL3 models at $T=10$~MeV, see also Ref.~\cite{Alford2019a}.


\end{widetext}

\section{Conclusions}
\label{sec:conclusions}

We studied the Urca-process-driven bulk viscosity of
neutrino-transparent, relativistic $npe\mu$ matter in the temperature
range $1\leq T\leq 10$~MeV and density range
$0.5 n_0\leq n_B\leq 5n_0$ which is relevant for BNS mergers. This
parallels (and complements) our recent work~\cite{Alford2021c} where
similar calculations were performed for relativistic neutrino-trapped
$npe\mu$ matter.  Using the analytic expressions for the relativistic
beta-equilibration rates derived in Ref.~\cite{Alford2021c} we compute
numerically the direct Urca neutron decay and lepton capture process
rates with two (DDME2 and NL3) EoS models within the relativistic
density functional theory for nuclear matter.

Imposing the $\beta$-equilibrium conditions $\mu_n=\mu_p+\mu_l$,
strictly valid at low temperatures, we find that in the case of DDME2
model, which does not allow for a low-temperature direct Urca process
(as the proton fraction stays always below the threshold), the neutron
decay rate is strongly suppressed as compared to the lepton capture
rate in the whole temperature-density range of interest and is
completely damped at high densities.  The qualitative picture is
similar in the case of NL3 model at densities below the direct Urca
threshold, whereas at higher densities above the threshold the neutron
decay and the lepton capture rates are almost equal. We also find that
the previous nonrelativistic approximation~\cite{Alford2019b}
underestimates the relativistic electron capture rates by factors from
1 to 10 depending on the density.

In contrast to the neutrino-trapped matter, where the beta-relaxation
rates $\gamma_e$ were always higher than the typical frequencies of
density oscillations, in the neutrino-transparent matter the
relaxation rate resonates with the typical frequencies
$1\leq f\leq 10$~kHz at a temperature that lies in the range
$4\leq T\leq 7$~MeV for DDME2 and $2\leq T\leq 5$~MeV for NL3, the
exact value depending on the density and oscillation frequency. As a
result, the bulk viscosity reaches a resonant maximum at that
temperature. As compared to the non-relativistic case, the location of
the maximum is shifted to lower temperatures, the shift being larger
at densities above the direct Urca threshold. We also find that, as
noted in Ref.~\cite{Alford2021c}, the nonrelativistic treatment of
nucleons strongly overestimates the maximal values of the bulk
viscosity because of an overestimate of susceptibilities in the
nonrelativistic approximation.

Another way in which this computation adds to earlier treatments is
the proper inclusion of muonic weak-equilibrium reactions in the bulk
viscosity. As in Ref.~\cite{Alford2021c}, we analyze the relative
rates of electronic and muonic Urca processes as well as the rates of
pure leptonic processes, which is the muon decay in this case.  The
muon decay rates are found to be smaller than the Urca process rates
almost in the whole temperature-density range, therefore, the bulk
viscosity of $npe\mu$ matter can be computed neglecting the muon decay
process. Thus, the bulk viscosity arises from two independent
equilibration channels (\ie, electronic and muonic Urca channels),
which results in a ``flattened" structure in the temperature
dependence of the bulk viscosity, which is in contrast to the bulk
viscosity of $npe$ matter with a single peak at low densities.
The ``flattened" structure is clearly seen in the left panels
  of Figs.~\ref{fig:zeta_slow_temp_DDME2}, and
  \ref{fig:zeta_slow_temp_NL3}, the relevant bulk viscosity being
  shown by solid lines corresponding to $n_B/n_0=1$.   The bulk
viscosity of $npe\mu$ matter is higher than that of $npe$ matter by
factors from 2.5 to 8 above the maximum temperature if the density is
below the direct Urca threshold. Above the threshold, we find
$\zeta\simeq 2\zeta_e$ below the maximum and $\zeta\lesssim \zeta_e$
above the maximum.

Using the results of the bulk viscosity we estimate the bulk viscous
damping times of density oscillations for frequencies $f=1$~kHz and
$f=10$~kHz.  The damping timescale has a minimum as a function of
temperature between $5\leq T\leq 7$~MeV (DDME2 model) and
$2\leq T\leq 5$~MeV (NL3 model) for various densities.

For a typical frequency $f=1$~kHz, the DDME2 model predicts that the
bulk viscous damping would be efficient only for long-living remnants
with $\tau_\zeta\geq 10$~ms at any density and in the temperature
range $2\leq T\leq 10$~MeV. The damping timescale reaches its minimum
at $n_B\leq n_0$ and $T\simeq 5$~MeV, where $\tau\simeq 10$~ms reaches
the short-term evolution timescale.  For higher frequencies, there is
already a window of densities and temperatures where the damping
timescales are shorter than the short-term evolution timescale of
mergers. For $f=10$~kHz the short-term damping is efficient at
densities $n_B\geq 3n_0$ and for temperatures between
$4\leq T\leq 10$~MeV.

In the case of NL3 model there is always a range of densities and
temperatures where the bulk viscous damping is efficient within the
short-term evolution timescale. At $f=1$~kHz the relevant parameter
range is $n_B\leq 2n_0$ and 2 $\leq T\leq$ 5 MeV. The high-density
region above the direct Urca threshold does not have a significant
impact on the damping of density oscillations (in contrast to findings
of Ref.~\cite{Alford2020}).  At $f=10$~kHz the damping timescale
reaches down to the ms range also at high densities.  On the long-term
evolution timescale, the damping is efficient at all
densities. Correspondingly, the range of temperatures where the bulk
viscosity would play a role is larger.

Our results for the bulk viscosity are most useful for
  estimating the damping of small-amplitude post-merger oscillations,
  see, \eg, Ref.~\cite{Alford2020}.  These results show the
  likely importance of bulk viscous damping arising from beta
  equilibration via weak interactions, and provide motivation for this
  physics to be included in simulations.  Merger simulation groups are
  already exploring different approaches to the inclusion of beta
  equilibration, for example using  the framework of the
  second-order Israel-Stewart relativistic
  hydrodynamics~\cite{Camelio2023a,Camelio2023b}; its relation to the
  one defined within the approach of Ref.~\cite{Sawyer1989} and used
  here is discussed in Ref.~\cite{Gavassino2021CQGra}.  Another
  possibility is to use an equation of state that includes the
  dependence on the particle fractions, and evolve those quantities,
  using the relevant reaction rates, in the simulation
  \cite{Most2022}.
  
 It should be noted that complete second-order multi-fluid
  formulations of relativistic hydrodynamics will contain additional
  transport coefficients describing the relaxation of dissipative
  fluxes, in particular, the bulk-viscous flux; for review and
  references see~Ref.~\cite{Harutyunyan2023Symm}.  

\section*{Acknowledgments}

M.~A. is partly supported by the U.S. Department of Energy, Office of
Science, Office of Nuclear Physics under Award No. DE-FG02-05ER41375.
A.~H. and A.~S. were supported by the Volkswagen Foundation (Hannover,
Germany) grant No.  96 839.  The research of A.~S. was funded by
Deutsche Forschungsgemeinschaft Grant No. SE 1836/5-2 and the Polish
NCN Grant No. 2020/37/B/ST9/01937 at Wroc\l{}aw University.

\appendix

\begin{widetext}

\section{Low-temperature limit of Urca process rates}
\label{app:rates_low}

Here we present the details of the  calculation 
of $\beta$-equilibration rates given by 
Eqs.~\eqref{eq:Gamma1_def} and \eqref{eq:Gamma2_def}. 
Writing the energy conservation in the form $\delta(k_0+p_0 \pm k'_0-p'_0)
=\delta(\ep_l+\ep_p-\ep_n\pm \ep_{\bar{\nu}_l/\nu_l}-\mu_{\Delta_l})$,
where $\ep_i$ are the energies of the particles computed from their 
(effective) chemical potentials (\eg, $\ep_p=\sqrt{p^2+m^{*2}_p}-\mu_p^*$), 
and substituting Eq.~\eqref{eq:matrix_el_full} into Eqs.~\eqref{eq:Gamma1_def} 
and \eqref{eq:Gamma2_def} we obtain [$G=G_F \cos\theta_c(1+g_A) $]
\bea\label{eq:Gamma1_dummy} 
\Gamma_{n\to pl\bar\nu} (\mu_{\Delta_l}) &=& 2 G^2
\int d^4q \int\!\! \frac{d^3p}{(2\pi)^3p_0} \int\!\!
\frac{d^3p'}{(2\pi)^3p'_0} \int\!\! \frac{d^3k}{(2\pi)^3k_0}
\int\!\! \frac{d^3k'}{(2 \pi)^3k'_0}(k\cdot p) (k'\cdot p') \nonumber\\
& \times & \bar{f}(k) \bar{f}(p) \bar{f}(k')  f(p') (2\pi)^4
\delta^{(4)}(k+p-q)\delta^{(4)}(k'-p'+q) =2 {G}^2 \int d^4q\, I_1(q)\, I_2(q),\\
\label{eq:Gamma2_dummy}
\Gamma_{pl\to n\nu} (\mu_{\Delta_l}) &=& 2 G^2
\int d^4q \int\!\! \frac{d^3p}{(2\pi)^3p_0} \int\!\!
\frac{d^3p'}{(2\pi)^3p'_0} \int\!\! \frac{d^3k}{(2\pi)^3k_0}
\int\!\! \frac{d^3k'}{(2 \pi)^3k'_0}(k\cdot p) (k'\cdot p') \nonumber\\
& \times & {f}(k) {f}(p) \bar{f}(k') \bar{f}(p') (2\pi)^4
\delta^{(4)}(k+p-q)\delta^{(4)}(-k'-p'+q) =2 {G}^2 \int d^4q\, 
\bar{I}_1(q)\, \bar{I}_3(q),
\eea
where 
\bea
\label{eq:I1}
I_1(q) &=& \int\!\! \frac{d^3p}{(2\pi)^3 p_0}\int\!\!
  \frac{d^3k}{(2\pi)^3 k_0} \bar{f}(k)\bar{f}(p) \,
  (k\cdot p)\, (2\pi)^4\, \delta^{(4)}(k+p-q),\\
\label{eq:I2}
I_2(q) &=& \int\!\! \frac{d^3p'}{(2\pi)^3 p'_0} \int\!\!
  \frac{d^3k'}{(2\pi)^3 k'_0} \bar{f}(k')f(p')\, 
  (k'\cdot p')\,\delta^{(4)}(k'-p'+q) ,\\
\label{eq:I3}
\bar{I}_3(q) &=& \int\!\! \frac{d^3p'}{(2\pi)^3 p'_0} \int\!\!
  \frac{d^3k'}{(2\pi)^3 k'_0} \bar{f}(k') \bar{f}(p')\, 
  (k'\cdot p')\, \delta^{(4)}(-k'-p'+q) ,
\eea
and $\bar{I}_1$ is obtained from $I_1$ by replacing 
$\bar{f}(k)\bar{f}(p)\to {f}(k){f}(p)$. Here
 $\delta^{(4)}(k+p-q)=\delta(\veck+\vecp-\vecq)
\delta(\ep_{k}+ \ep_{p}-\omega-\mu_{\Delta_l})$, and
 $\delta^{(4)}(\pm k'-p'+q)=\delta(\pm \veck'-\vecp'+\vecq)
\delta(\pm \ep_{k'}-\ep_{p'}+\omega)$. The calculation of 
integrals~\eqref{eq:I1}--\eqref{eq:I3} at finite temperatures 
was detailed in Ref.~\cite{Alford2021b}. Here we will derive 
only their low-temperature limit for the neutrino-transparent matter. 
In this limit, the antineutrino and neutrino distributions are zero in 
integrals $I_2$ and $\bar{I}_3$, respectively, the neutrino 
momentum in $\delta$-functions can be dropped and the magnitude of neutron momentum can be fixed to its value at the Fermi surface$p'=p_{Fn}$. We then find 
\bea
I_2(q) &=& (2\pi)^{-6}\int_0^\infty\!\! \frac{dp'}{p'_0} 
\int_0^\infty\!\!  \frac{k'^2dk'}{k'_0}\int d\Omega_{k'} 
 f(\ep_{p'})\, (p'_0 k'_0 -\vecq \cdot\veck')
\delta(\ep_{k'}-\ep_{p'}+\omega) \delta(q-p_{Fn})\nonumber\\
&=& \frac{1}{(2\pi)^{5}} \delta(q-p_{Fn})
\int_0^\infty\!\! k'^2dk'\, f(k'+\omega)\frac{1}{p'} 
\int_{-1}^1 dy\, (p'_0 -qy)\nonumber\\
&=& \frac{2}{(2\pi)^{5}} \delta(q-p_{Fn})\frac{\mu_n^*}{p_{Fn}}
\int_0^\infty\!\! dk'\, k'^2 f(k'+\omega),\\
\bar{I}_3(q) &=& (2\pi)^{-6}\int_0^\infty\!\! \frac{dp'}{ p'_0} 
\int_0^\infty\!\! \frac{k'^2k'}{ k'_0}\int d\Omega_{k'} 
\bar{f}(\ep_{p'})\, (p'_0 k'_0 -\vecq \cdot\veck')
\delta(-\ep_{k'}-\ep_{p'}+\omega) \delta(q-p_{Fn})\nonumber\\
&=& \frac{1}{(2\pi)^{5}} \delta(q-p_{Fn})
 \int_0^\infty\!\! k'^2dk' \bar{f}(\omega-\ep_{k'}) 
 \frac{1}{p'}\int_{-1}^1 dy\,(p'_0 - q y)\nonumber\\
&=& \frac{2}{(2\pi)^{5}} \delta(q-p_{Fn})\frac{\mu_n^*}{p_{Fn}}
 \int_0^\infty\!\! dk'\, k'^2 {f}(k'-\omega),
\eea
where $y$ is the cosine of the angle between $\bm q$ and $\bm k'$,
and we used $\ep_{k'}=k'_0=k'$, as $m_\nu=\mu_\nu=0$. The low-$T$ 
limits of the integrals $I_1$ and $\bar{I}_1$ are given by~\cite{Alford2021b}
\bea\label{eq:I1_lowT}
I_1(q) &\simeq & 
-\frac{{\omega} g(-{\omega})}{4\pi q} \,
\theta(p_{Fl}+{p}_{Fp}-q)\theta(q-\vert p_{Fl}- {p}_{Fp}\vert)\,
({p}_{Fp}^2+p_{Fl}^2+2\mu_l\mu_{p}^*-q^2),\\
\label{eq:I1_bar_lowT}
\bar{I}_1(q) &\simeq & \frac{{\omega} g({\omega})}{4\pi q} \,
\theta(p_{Fl}+{p}_{Fp}-q)\theta(q-\vert p_{Fl}- {p}_{Fp}\vert)\,
({p}_{Fp}^2+p_{Fl}^2+2\mu_l\mu_{p}^*-q^2).
\eea 
Then for the rates~\eqref{eq:Gamma1_dummy} 
and \eqref{eq:Gamma2_dummy} we obtain
\bea\label{eq:Gamma_lowT_trans}
\Gamma_{n\to pl\bar\nu /pl\to n\nu} &=& 2 {G}^2 4\pi 
\frac{\mu_n^*}{p_{Fn}} \int_{-\infty}^\infty\!\!\!
d\omega \int_0^\infty\!\!\! q^2 dq\, 
\frac{ \mp \omega g(\mp {\omega})}{4\pi q} 
\theta(p_{Fl}+{p}_{Fp}-q)\theta(q-\vert p_{Fl}-{p}_{Fp}\vert)
\nonumber\\
&&\times ({p}_{Fp}^2+p_{Fl}^2+2\mu_l\mu_{p}^*-q^2)
\frac{2}{(2\pi)^{5}} \delta(q-p_{Fn})
\int_0^\infty\!\! dk'\,k'^2 f(k'\pm \omega)\nonumber\\
&=& \frac{{G}^2T^5}{8\pi^{5}} \mu_n^* \theta(p_{Fl}+{p}_{Fp}-p_{Fn})
({p}_{Fp}^2+p_{Fl}^2+2\mu_l \mu_{p}^*-p_{Fn}^2)
\int_{-\infty}^\infty\!\!\! dy \, (\mp y) g(\mp{y}) 
\int_0^\infty\!\! dx\,x^2  f(x\pm y)\nonumber\\
&=& \frac{\alpha}{2} {G}^2T^5 \mu_n^* \theta(p_{Fl}+{p}_{Fp}-p_{Fn})
({p}_{Fp}^2+p_{Fl}^2+2\mu_l \mu_{p}^*-p_{Fn}^2),
\eea
where $\alpha=3\left[\pi^2 \zeta(3) + 15 \zeta(5)\right]
/16\pi^5\simeq 0.0168$. In the limit of nonrelativistic 
nucleons we keep only the term $2\mu_l\mu_{p}^*$ in the 
brackets and, approximating $\mu_N^*\approx m_N^*$, we obtain
\bea
\Gamma_{n\to pl\bar\nu }=\Gamma_{pl\to n\nu}=
\alpha m_{n}^* m_{p}^* \mu_l
{G}^2T^5 \theta(p_{Fl}+{p}_{Fp}-p_{Fn}),
\eea
which coincides with the results of Refs.~\cite{Haensel2000,
Yakovlev2001,Alford2018b,Alford2019a,Alford2019b} 
if the lepton mass is neglected, \ie, $\mu_l=p_{Fl}$.

If matter is out of chemical equilibrium, \ie, $\mu_{\Delta_l}\neq 0$, one should replace $\omega\to\omega+\mu_{\Delta_l}$ 
in Eq.~\eqref{eq:I1_lowT} and \eqref{eq:I1_bar_lowT} as implied 
by the energy $\delta$-function after Eq.~\eqref{eq:I3}. 
Then the derivatives of~\eqref{eq:Gamma_lowT_trans} with respect 
to $\mu_{\Delta_l}$ at $\mu_{\Delta_l}=0$ are given by 
\bea\label{eq:lambda1_lowT_trans}
\frac{\partial\Gamma_{n\to pl\bar\nu}}{\partial\mu_{\Delta_l}} 
\bigg\vert_{\mu_{\Delta_l}=0}
&=& \frac{{G}^2T^5}{8\pi^{5}} \mu_n^* 
\theta(p_{Fl}+{p}_{Fp}-p_{Fn})
({p}_{Fp}^2+p_{Fl}^2+2\mu_l\mu_{p}^*-p_{Fn}^2)
\nonumber\\
 &&\times \frac{\partial }{\partial\mu_{\Delta_l}}
 \int_{-\infty}^\infty\!\!\! dy \, (-\bar{y})g(-\bar{y}) 
\int_0^\infty\!\! dx\,x^2 f(x+ y)\nonumber\\
&=& \frac{{G}^2T^4}{8\pi^{5}} \mu_n^* \theta(p_{Fl}+{p}_{Fp}-p_{Fn})
({p}_{Fp}^2+p_{Fl}^2+2\mu_l\mu_{p}^*-p_{Fn}^2)\nonumber\\
 &&\times \int_{-\infty}^\infty\!\!\! dy \, 
 [1+g({y})][1-{y}g({y})] \int_0^\infty\!\! dx\,x^2 f(x+ y),\\
\label{eq:lambda2_lowT_trans}
-\frac{\partial\Gamma_{pl\to n\nu}}
{\partial\mu_{\Delta_l}}
\bigg\vert_{\mu_{\Delta_l}=0} & =&
 -\frac{{G}^2T^5}{8\pi^{5}} \mu_n^* 
\theta(p_{Fl}+{p}_{Fp}-p_{Fn})
({p}_{Fp}^2+p_{Fl}^2+2\mu_l \mu_{p}^*-p_{Fn}^2)
\nonumber\\
 &&\times \frac{\partial }{\partial\mu_{\Delta_l}}
 \int_{-\infty}^\infty\!\!\! dy \, \bar{y}g(\bar{y}) 
\int_0^\infty\!\! dx\,x^2 f(x-y)\nonumber\\
&=&  \frac{{G}^2T^4}{8\pi^{5}} \mu_n^* \theta(p_{Fl}+{p}_{Fp}-p_{Fn})
({p}_{Fp}^2+p_{Fl}^2+2\mu_l\mu_{p}^*-p_{Fn}^2)\nonumber\\
 &&\times \int_{-\infty}^\infty\!\!\! dy \, 
 g({y})[{y}(1+g({y}))-1] \int_0^\infty\!\! dx\,x^2 f(x-y),
\eea
where $\bar{y}=y+\mu_{\Delta_l}/T$. 
The two-dimensional integrals in these expressions are 
the same and are equal to $17\pi^4/120$, therefore
\bea\label{eq:lambda12_lowT_trans}
\lambda_l=\left(\frac{\partial\Gamma_{n\to pl\bar\nu}}
{\partial\mu_{\Delta_l}}-\frac{\partial\Gamma_{pl\to n\nu}}
{\partial\mu_{\Delta_l}}\right)\bigg\vert_{\mu_{\Delta_l}=0}
= \frac{17}{480\pi}{G}^2T^4 \mu_n^* \theta(p_{Fl}+{p}_{Fp}-p_{Fn})
({p}_{Fp}^2+p_{Fl}^2+2\mu_l \mu_{p}^*-p_{Fn}^2).
\eea
In the limit $\mu_N^*\simeq m_N^*\gg p_{FN}$, these results lead to
\bea\label{eq:lambda_lowT_trans}
\lambda_{l} =\frac{17}{240\pi} m_n^* m_{p}^*\mu_l
{G}^2T^4 \theta(p_{Fl}+{p}_{Fp}-p_{Fn}),
\eea
which is consistent with previous nonrelativistic calculations of Refs.~\cite{Haensel1992PhRvD,Haensel2000,Alford2019a,Alford2019b}.

\section{Computation of susceptibilities}
\label{app:A_coeff}

\subsection{Isothermal susceptibilities}
\label{app:A_is}

To compute the isothermal susceptibilities $A_{ij}^T= \frac{\partial \mu_i}
{\partial n_j}\big\vert_T$ we use the following formula for the particle densities
\bea\label{eq:dens}
n_i =\frac{1}{\pi^2}\int_0^\infty\! p^2dp\, [f_i(p)-f^*_i(p)],
\eea 
where $f_i(p)$ and ${f}^*_i(p)$ are the distribution functions 
for particles and antiparticles, respectively. To compute 
first the nucleon susceptibilities we differentiate the left and 
right sides of Eq.~\eqref{eq:dens} with respect to $n_j$ at 
{\it constant temperature} and use the relations
\bea\label{eq:fermi_i}
\frac{\partial f_i}{\partial n_j}
\bigg\vert_{T} &=& -f_i(1-f_i)\frac{1}{T}
\left(\frac{m^*}{E_{p}}\frac{\partial m^*}{\partial n_j} 
-\frac{\partial \mu^*_i}{\partial n_j}\right),\\
\frac{\partial {f}^*_i}{\partial n_j}
\bigg\vert_{T} &=& -{f}^*_i(1-{f}^*_i)\frac{1}{T}
\left(\frac{m^*}{E_{p}}\frac{\partial m^*}{\partial n_j} 
+\frac{\partial \mu^*_i}{\partial n_j}\right),
\eea
to obtain
\bea\label{eq:matrix_eq}
\delta_{ij}=-\left(\frac{\partial m^*}{\partial n_j}\right)
{I}_{i}^{1,0} +\left(\frac{\partial
\mu^*_i}{\partial n_j}\right) {I}_{i}^{0,0},
\eea
where $m^*\equiv m_n^*=m_p^*$ is a short-hand notation for 
effective nucleon mass, $E_{p}=\sqrt{m^{*2}+p^2}$, and
\bea\label{eq:I_def}
I^{q,s}_{i} = \frac{1}{\pi^2 T}\int_0^{\infty}\!  p^2 dp 
\left(\frac{m^*}{E_p}\right)^q 
 \Big[z_{pi}^s\, f_i(1-f_i) + (-1)^{q+s}\, 
 {z}_{pi}^{*s}\, {f}^*_i(1-\bar{f}_i)\Big],
\eea 
with $z_{pi} = (E_p-\mu^*_i)/T$, ${z}^*_{pi}=(E_p+\mu^*_i)/T$
(the integrals with $s\neq 0$ will be used in the next subsection). 
Recall that all derivatives above are computed at $T={\rm const}$. 
Using the relation $\mu^*_i = \mu_i-g_{\omega}\omega_0 -g_{\rho}
\rho_{03}I_{3i}-\Sigma_r$ and the equations for vector meson mean fields
\bea\label{eq:mean_fields}
g_\omega\omega_0 = \left(\frac{g_\omega}{m_\omega}\right)^2 
(n_n+n_p),\qquad
g_\rho \rho_{03} = \frac{1}{2}\left(\frac{g_\rho}{m_\rho}\right)^2
(n_p -n_n),
\eea
we obtain
\bea\label{eq:b_ij_def}
B_{ij}\equiv \frac{\partial \mu^*_i}{\partial n_j}\bigg\vert_{T} = 
A_{ij}^T -\left(\frac{g_\omega}{m_\omega}\right)^2\left[1+
\frac{2n_B}{g_\omega}\frac{\partial g_\omega}{\partial n_B}\right] -
 I_{3i}\left(\frac{g_\rho}{m_\rho}\right)^2 \left[I_{3j}+
 \frac{n_n -n_p}{n_0}a_\rho\right]-\frac{\partial\Sigma_r}{\partial n_j}.
\eea

Next, we use the following equation for the scalar mean field 
\bea\label{eq:sigma}
g_\sigma \sigma =m-m^*=-\frac{g_\sigma}{m_\sigma^2}
\frac{\partial U(\sigma)}{\partial \sigma}+\frac{1}{\pi^2}
\left(\frac{g_\sigma}{m_\sigma}\right)^2\sum_{i=n,p}\int_0^\infty 
p^2dp\, \frac{m^*}{E_p}\! \left[f_i(p)+{f}^*_i(p)\right],
\eea 
to obtain (up to terms $\partial g_\sigma/\partial n_B$ 
which are small in the regime of interest and can be neglected)
\bea
\frac{\partial m^*}{\partial n_j}= 
-\frac{1}{m_\sigma^2}
\frac{\partial^2 U(\sigma)}{\partial \sigma^2}
\frac{\partial m^*}{\partial n_j}+
\left(\frac{g_\sigma}{m_\sigma}\right)^2
\left(\frac{\partial m^*}{\partial n_j}\right)
\left({I}_{n}^{2,0} +{I}_{p}^{2,0}\right) -
\left(\frac{g_\sigma}{m_\sigma}\right)^2
\left(B_{nj}{I}_{n}^{1,0} +B_{pj}{I}_{p}^{1,0}\right)\nonumber\\
-\left(\frac{g_\sigma}{m_\sigma}\right)^2
\left(\frac{\partial m^*}{\partial n_j}\right)
\sum_{i=n,p}\frac{1}{\pi^2}\int_0^\infty\!\! dp\, 
\frac{p^4}{E_p^{3}}\! \left[f_i(p)+{f}^*_i(p)\right].
\eea 
Introducing the short-hand notations
\bea
\tilde{I}_{i} = {I}_{i}^{2,0} -
\frac{1}{\pi^2}\int_0^\infty\!\! dp\,
\frac{p^4}{E_p^{3}}\! \left[f_i(p)+{f}^*_i(p)\right],\qquad
I_\sigma =\left(\frac{m_\sigma}{g_\sigma}\right)^2
\left(1+\frac{1}{m_\sigma^2}\frac{\partial^2 U}
{\partial \sigma^2}\right),
\eea 
and
\bea\label{eq:gamma_def}
\gamma = \frac{1}{\tilde{I}_{n} +\tilde{I}_{p}-I_\sigma},
\eea
we obtain 
\bea\label{eq:mass_diff}
\frac{\partial m^*}{\partial n_j}
=\gamma(B_{nj}{I}_{n}^{1,0} + B_{pj}{I}_{p}^{1,0}).
\eea 
Substituting this into Eq.~\eqref{eq:matrix_eq} we obtain 
the following equations for coefficients $B_{ij}$
\bea\label{eq:matrix_eq1}
B_{ij} {I}_{i}^{0,0} -\gamma\left(B_{nj}{I}_{n}^{1,0}+ 
B_{pj}{I}_{p}^{1,0}\right){I}_{i}^{1,0} =\delta_{ij}.
\eea
In the case of $i\neq j$ we find from Eq.~\eqref{eq:matrix_eq1} 
\bea
B_{np}=\gamma B_{pp}\frac{I_{p}^{1,0} I_{n}^{1,0}}
{I_{n}^{0,0} -\gamma \left(I_{n}^{1,0}\right)^2},\qquad
B_{pn}=\gamma B_{nn}\frac{I_{n}^{1,0} I_{p}^{1,0}}
{I_{p}^{0,0} -\gamma \left(I_{p}^{1,0}\right)^{2}}.
\eea
Substituting these expressions into 
Eq.~\eqref{eq:matrix_eq1} for $i=j$ we obtain
\bea\label{eq:b_diag}
B_{nn}=
\frac{I_{p}^{0,0} -\gamma \left(I_{p}^{1,0}\right)^{2}}
{I_{n}^{0,0}I_{p}^{0,0}-\gamma I_{p}^{0,0}
\left(I_{n}^{1,0}\right)^{2}-\gamma I_{n}^{0,0}
\left(I_{p}^{1,0}\right)^{2}},\qquad
B_{pp}=
\frac{I_{n}^{0,0} -\gamma \left(I_{n}^{1,0}\right)^{2}}
{I_{n}^{0,0}I_{p}^{0,0}-\gamma I_{p}^{0,0}
\left(I_{n}^{1,0}\right)^{2}-\gamma I_{n}^{0,0}
\left(I_{p}^{1,0}\right)^{2}},
\eea
and
\bea
\label{eq:b_mix}
B_{np}=B_{pn}=
\frac{\gamma I_{p}^{1,0} I_{n}^{1,0}}{I_{n}^{0,0}I_{p}^{0,0}
-\gamma I_{p}^{0,0}\left(I_{n}^{1,0}\right)^{2}-
\gamma I_{n}^{0,0}\left(I_{p}^{1,0}\right)^{2}}.
\eea
Finally, substituting Eqs.~\eqref{eq:b_diag} and \eqref{eq:b_mix} in 
Eq.~\eqref{eq:b_ij_def} and recalling the definitions $A_n=A_{nn}-A_{pn}$, 
$A_p=A_{pp}-A_{np}$ we obtain for isothermal susceptibilites
\bea\label{eq:A_n_final}
A_{n}^T=\frac{I_{p}^{0,0} -\gamma I_{p}^{1,0}
\left(I_{n}^{1,0}+ I_{p}^{1,0}\right)}
{I_{n}^{0,0}I_{p}^{0,0}-\gamma I_{p}^{0,0}
\left(I_{n}^{1,0}\right)^{2}-\gamma I_{n}^{0,0}
\left(I_{p}^{1,0}\right)^{2}}
+\left(\frac{g_\rho}{m_\rho}\right)^2\left(\frac{1}{2}
-\frac{n_n -n_p}{n_0}a_\rho \right),\\
\label{eq:A_p_final}
A_{p}^T=\frac{I_{n}^{0,0} -\gamma I_{n}^{1,0}
\left(I_{n}^{1,0}+ I_{p}^{1,0}\right)}
{I_{n}^{0,0}I_{p}^{0,0}-\gamma I_{p}^{0,0}
\left(I_{n}^{1,0}\right)^{2}-\gamma I_{n}^{0,0}
\left(I_{p}^{1,0}\right)^{2}}
+\left(\frac{g_\rho}{m_\rho}\right)^2\left(\frac{1}{2}
+\frac{n_n -n_p}{n_0}a_\rho \right).
\eea
For lepton susceptibilities we have simply
$A_{l}^T = 1/{I}_{l}^{0,0}$, $l=\{e,\mu\}$.

\subsection{Adiabatic susceptibilities}
\label{app:A_ad}

The adiabatic susceptibilities can be obtained by using the following chain rule for partial derivatives
\bea\label{eq:mu_deriv_chain}
A_{ij}^s\equiv \frac{\partial \mu_i}{\partial n_j}\bigg\vert_{s} =
\frac{\partial \mu_i}{\partial n_j}\bigg\vert_{T}
+\frac{\partial \mu_i}{\partial T}\,
\frac{\partial T}{\partial n_j}\bigg\vert_{s}
=A_{ij}^T -\frac{\partial \mu_i}{\partial T}
\left(\frac{\partial s}{\partial T}\right)^{-1}
\frac{\partial s}{\partial n_j}\bigg\vert_{T},
\eea
where $s$ is the entropy per baryon
\bea \label{eq:entropy}
s = - \sum_{i}\frac{1}{\pi^2 n_B}\int_0^{\infty}\! p^2 dp 
\Big[ f_i\ln f_i + (1-f_i)\ln (1-f_i)+ 
{f}^*_i\ln {f}^*_i + (1-{f}^*_i)\ln (1-{f}^*_i)\Big],
\eea
where the summation goes over all particle species, \ie, 
nucleons, and leptons. In the second step in Eq.~\eqref{eq:mu_deriv_chain}
we used the relation 
\bea
\frac{\partial s}{\partial n_j}\bigg\vert_{T}
=-\frac{\partial s}{\partial T}\,
\frac{\partial T}{\partial n_j}\bigg\vert_{s},
\eea
which can be obtained if one applies an analogous to 
Eq.~\eqref{eq:mu_deriv_chain} chain rule to $s$. Note 
that all particle densities are assumed to be kept constant
 in the partial derivatives with respect to $T$ in 
 Eq.~\eqref{eq:mu_deriv_chain}.

From Eq.~\eqref{eq:entropy} we obtain
\bea\label{eq:entropy_nj}
n_B\frac{\partial s}{\partial n_j} \bigg\vert_{T} &=&
  \sum_i\frac{1}{\pi^2}\int_0^{\infty}\! p^2 dp 
  \left[\frac{\partial f_i}{\partial n_j}
 \ln \frac{1-f_i}{f_i}+\frac{\partial {f}^*_i}
 {\partial n_j}\ln \frac{1-{f}^*_i}{{f}^*_i} \right]
 - s\frac{\partial n_B}{\partial n_j}\bigg\vert_{T}.
\eea
Substituting here Eqs.~\eqref{eq:fermi_i} and 
\eqref{eq:mass_diff} for nucleons we obtain
\bea\label{eq:entropy_nj1}
n_B\frac{\partial s}{\partial n_j} \bigg\vert_{T} &=&
 -s + \sum_N\frac{1}{\pi^2 T}\int_0^{\infty}\!
 p^2 dp \,\biggl\{ B_{ij}\left[z_{pi}\, f_i(1-f_i) 
 -{z}^*_{pi}\, {f}^*_i(1-{f}^*_i) \right] \nonumber\\
&& -\gamma\left(B_{nj}{I}_{n}^{1,0}+ B_{pj}{I}_{p}^{1,0} \right) 
 \frac{m^*}{E_p}\! \left[z_{pi}\, f_i(1-f_i) +{z}^*_{pi}\,
 {f}^*_i(1-{f}^*_i) \right]\biggr\} \nonumber\\
&=& -s + \left(B_{nj}{I}^{0,1}_{n}+B_{pj}{I}^{0,1}_{p}\right)
 -\gamma \left(B_{nj}{I}_{n}^{1,0} + B_{pj}{I}_{p}^{1,0} \right)
\left({I}^{1,1}_{n}+{I}^{1,1}_{p}\right),
\eea
where we used the identities $\ln (1-f_i)/f_i= z_{pi}$, 
$\ln (1-{f}^*_i)/{f}^*_i={z}^*_{pi}$, and recalled
the definitions~\eqref{eq:I_def}.

Substitutuing now $B_{ij}$ from Eqs.~\eqref{eq:b_diag} 
and \eqref{eq:b_mix} in Eq.~\eqref{eq:entropy_nj1} we find
\bea\label{eq:entropy_n}
n_B\frac{\partial s}{\partial n_n} \bigg\vert_{T} &=& -s+
 \left(B_{nn}{I}^{0,1}_{n}+B_{pn}{I}^{0,1}_{p}\right)  
-\gamma \left(B_{nn}{I}_{n}^{1,0}+ B_{pn}{I}_{p}^{1,0} \right)
\left({I}^{1,1}_{n}+{I}^{1,1}_{p}\right) \nonumber\\
&=& -s+ \frac{I_{p}^{0,0} {I}^{0,1}_{n}-\gamma 
I_{p}^{1,0}\left(I_{p}^{1,0} {I}^{0,1}_{n}-
 I_{n}^{1,0} {I}^{0,1}_{p}\right)
-\gamma I_{p}^{0,0} {I}_{n}^{1,0}\left({I}^{1,1}_{n}
+{I}^{1,1}_{p}\right) }{I_{n}^{0,0}I_{p}^{0,0}-
\gamma I_{p}^{0,0} \left(I_{n}^{1,0}\right)^2-
\gamma I_{n}^{0,0} \left(I_{p}^{1,0}\right)^2} ,\\
\label{eq:entropy_p}
n_B\frac{\partial s}{\partial n_p} \bigg\vert_{T}
&=& -s+ \frac{I_{n}^{0,0} {I}^{0,1}_{p}-\gamma 
I_{n}^{1,0}\left(I_{n}^{1,0} {I}^{0,1}_{p}-
 I_{p}^{1,0} {I}^{0,1}_{n}\right)
-\gamma I_{n}^{0,0} {I}_{p}^{1,0}\left({I}^{1,1}_{n}
+{I}^{1,1}_{p}\right) }{I_{n}^{0,0}I_{p}^{0,0}-
\gamma I_{p}^{0,0} \left(I_{n}^{1,0}\right)^2-
\gamma I_{n}^{0,0} \left(I_{p}^{1,0}\right)^2} .
\eea

For leptons we have 
\bea\label{eq:fermi_l}
\frac{\partial f_l}{\partial n_l}
\bigg\vert_{T} =
\frac{f_l(1-f_l)}{T {I}_{l}^{0,0}},\qquad
\frac{\partial {f}^*_l}{\partial n_l}
\bigg\vert_{T} =
-\frac{{f}^*_l(1-{f}^*_l)}{T {I}_{l}^{0,0}},
\eea
and
\bea\label{eq:entropy_l}
n_B\frac{\partial s}{\partial n_l}\bigg\vert_{T}
 &=& \frac{1}{\pi^2}\int_0^{\infty}\!
 p^2 dp \left[ \frac{\partial f_l}{\partial n_l} 
 \ln \frac{1-f_l}{f_l}+\frac{\partial {f}^*_l} 
 {\partial n_l}\ln \frac{1-{f}^*_l}{{f}^*_l} \right]
 \bigg\vert_{T} \nonumber\\
 &=&
  \frac{1}{\pi^2 T {I}_{l}^{0,0}}\int_0^{\infty}\! p^2 dp 
  \left[z_{pl}\, f_l(1-f_l) - {z}^*_{pl}\,{f}^*_l(1-{f}^*_l)
  \right] = \frac{I^{0,1}_{l}}{{I}_{l}^{0,0}}.
\eea
Next, we compute the temperature derivatives. Differentiating 
the left and right sides of Eq.~\eqref{eq:dens} with respect 
to $T$ at constant $n_j$ and exploiting the expressions
\bea\label{eq:fermi_T}
\frac{\partial f_i}{\partial T} =f_i(1-f_i)\frac{1}{T}
\left(z_{pi}+\frac{\partial \mu^*_i}{\partial T}\right),\qquad
\frac{\partial {f}^*_i}{\partial T} ={f}^*_i(1-{f}^*_i) 
\frac{1}{T}\left({z}^*_{pi}-\frac{\partial \mu^*_i}{\partial T}\right),
\eea
we obtain
\bea\label{eq:mu_T}
\frac{\partial\mu_i}{\partial T}\simeq
\frac{\partial\mu^*_i}{\partial T}
=-\frac{{I}_{i}^{0,1}} {{I}_{i}^{0,0}},
\eea
where we took into account that the nucleon masses and mesonic 
mean fields are almost independent of the temperature. Then from 
Eqs.~\eqref{eq:entropy}, \eqref{eq:fermi_T} and \eqref{eq:mu_T} we find
\bea\label{eq:entropy_T}
n_B\frac{\partial s}{\partial T}  &=&
  \sum_i\frac{1}{\pi^2}\int_0^{\infty}\!
 p^2 dp \left[\frac{\partial f_i}{\partial T}
 \ln \frac{1-f_i}{f_i}+\frac{\partial \bar{f}_i}
 {\partial T}\ln \frac{1-\bar{f}_i}{\bar{f}_i} \right]\nonumber\\
 &=&
 \sum_i\frac{1}{\pi^2 T}\frac{\partial \mu^*_i}{\partial T}
 \int_0^{\infty}\! p^2 dp \left[z_{pi}\, f_i(1-f_i)
 -\bar{z}_{pi}\,\bar{f}_i(1-\bar{f}_i)\right]\nonumber\\
 &+& \sum_i\frac{1}{\pi^2 T}\int_0^{\infty}\!
 p^2 dp \left[z_{pi}^2\, f_i(1-f_i)+ \bar{z}_{pi}^2\,
 \bar{f}_i(1-\bar{f}_i)\right]\nonumber\\
&=&  \sum_k \left[I^{0,2}_k-\frac{\left({I}_{k}^{0,1} 
\right)^2} {{I}_{k}^{0,0}}  \right]\equiv S.
\eea
Then from Eqs.~\eqref{eq:mu_deriv_chain}, \eqref{eq:mu_T} 
and \eqref{eq:entropy_T} we find
\bea\label{eq:A_ij_ad}
A_{ij}^s =A_{ij}^T + \frac{n_B}{S}
\frac{{I}_{i}^{0,1}} {{I}_{i}^{0,0}}
\frac{\partial s}{\partial n_j}\bigg\vert_T,
\eea
which leads to 
\bea\label{eq:A_n_ad}
A_{n}^s &=& A_{n}^T + \frac{n_B}{S}
\left(\frac{{I}_{n}^{0,1}} {{I}_{n}^{0,0}}
-\frac{{I}_{p}^{0,1}} {{I}_{p}^{0,0}}\right)
\frac{\partial s}{\partial n_n}\bigg\vert_T,\\
\label{eq:A_p_ad}
A_{p}^s &=& A_{p}^T -\frac{n_B}{S}
\left(\frac{{I}_{n}^{0,1}} {{I}_{n}^{0,0}}
-\frac{{I}_{p}^{0,1}} {{I}_{p}^{0,0}}\right)
\frac{\partial s}{\partial n_p}\bigg\vert_T,\\
\label{eq:A_l_ad}
A_{l}^s &=& A_{l}^T + \frac{1}{S}
\left(\frac{I_{l}^{0,1}}{I_{l}^{0,0}}\right)^2.
\eea

Note that due to the second term in Eq.~\eqref{eq:mu_deriv_chain} 
there are additional cross-terms between different particle species,
\eg, between baryons and leptons for adiabatic 
susceptibilities. However, these terms are found to be smaller than 
the diagonal terms in the whole regime of interest and can be neglected.

\end{widetext}

\providecommand{\href}[2]{#2}\begingroup\raggedright\endgroup


\begin{thebibliography}{10}

\bibitem{Oertel2017}
M.~{Oertel}, M.~{Hempel}, T.~{Kl{\"a}hn} and S.~{Typel}, \emph{{Equations of
  state for supernovae and compact stars}},
  \href{http://dx.doi.org/10.1103/RevModPhys.89.015007}{\emph{Reviews of Modern
  Physics} {\bf 89} (2017) 015007},
  [\href{http://arxiv.org/abs/1610.03361}{{\tt 1610.03361}}].

\bibitem{Lovato2022}
A.~{Lovato}, T.~{Dore}, R.~D. {Pisarski}, B.~{Schenke}, K.~{Chatziioannou},
  J.~S. {Read} et~al., \emph{{Long Range Plan: Dense matter theory for
  heavy-ion collisions and neutron stars}},
  \href{http://dx.doi.org/10.48550/arXiv.2211.02224}{\emph{arXiv e-prints}
  (2022) arXiv:2211.02224}, [\href{http://arxiv.org/abs/2211.02224}{{\tt
  2211.02224}}].

\bibitem{Sedrakian2023}
A.~{Sedrakian}, J.~J. {Li} and F.~{Weber}, \emph{{Heavy baryons in compact
  stars}}, \href{http://dx.doi.org/10.1016/j.ppnp.2023.104041}{\emph{Progress
  in Particle and Nuclear Physics} {\bf 131} (2023) 104041},
  [\href{http://arxiv.org/abs/2212.01086}{{\tt 2212.01086}}].

\bibitem{Dexheimer2022}
V.~{Dexheimer}, M.~{Mancini}, M.~{Oertel}, C.~{Provid{\^e}ncia}, L.~{Tolos} and
  S.~{Typel}, \emph{{Quick Guides for Use of the CompOSE Data Base}},
  \href{http://dx.doi.org/10.3390/particles5030028}{\emph{Particles} {\bf 5}
  (2022) 346--360}.

\bibitem{Perego:2019adq}
A.~Perego, S.~Bernuzzi and D.~Radice, \emph{{Thermodynamics conditions of
  matter in neutron star mergers}},
  \href{http://dx.doi.org/10.1140/epja/i2019-12810-7}{\emph{Eur. Phys. J. A}
  {\bf 55} (2019) 124}, [\href{http://arxiv.org/abs/1903.07898}{{\tt
  1903.07898}}].

\bibitem{Hanauske:2019qgs}
M.~Hanauske, J.~Steinheimer, A.~Motornenko, V.~Vovchenko, L.~Bovard, E.~R. Most
  et~al., \emph{{Neutron Star Mergers: Probing the EoS of Hot, Dense Matter by
  Gravitational Waves}},
  \href{http://dx.doi.org/10.3390/particles2010004}{\emph{Particles} {\bf 2}
  (2019) 44--56}.

\bibitem{Kastaun:2016elu}
W.~Kastaun, R.~Ciolfi, A.~Endrizzi and B.~Giacomazzo, \emph{{Structure of
  Stable Binary Neutron Star Merger Remnants: Role of Initial Spin}},
  \href{http://dx.doi.org/10.1103/PhysRevD.96.043019}{\emph{Phys. Rev.} {\bf
  D96} (2017) 043019}, [\href{http://arxiv.org/abs/1612.03671}{{\tt
  1612.03671}}].

\bibitem{Bernuzzi:2015opx}
S.~Bernuzzi, D.~Radice, C.~D. Ott, L.~F. Roberts, P.~Moesta and F.~Galeazzi,
  \emph{{How loud are neutron star mergers?}},
  \href{http://dx.doi.org/10.1103/PhysRevD.94.024023}{\emph{Phys. Rev.} {\bf
  D94} (2016) 024023}, [\href{http://arxiv.org/abs/1512.06397}{{\tt
  1512.06397}}].

\bibitem{Foucart:2015gaa}
F.~Foucart, R.~Haas, M.~D. Duez, E.~O'Connor, C.~D. Ott, L.~Roberts et~al.,
  \emph{{Low mass binary neutron star mergers: gravitational waves and neutrino
  emission}}, \href{http://dx.doi.org/10.1103/PhysRevD.93.044019}{\emph{Phys.
  Rev.} {\bf D93} (2016) 044019}, [\href{http://arxiv.org/abs/1510.06398}{{\tt
  1510.06398}}].

\bibitem{Kiuchi:2012mk}
K.~Kiuchi, Y.~Sekiguchi, K.~Kyutoku and M.~Shibata, \emph{{Gravitational waves,
  neutrino emissions, and effects of hyperons in binary neutron star mergers}},
  \href{http://dx.doi.org/10.1088/0264-9381/29/12/124003}{\emph{Class. Quant.
  Grav.} {\bf 29} (2012) 124003}, [\href{http://arxiv.org/abs/1206.0509}{{\tt
  1206.0509}}].

\bibitem{Sekiguchi:2011zd}
Y.~Sekiguchi, K.~Kiuchi, K.~Kyutoku and M.~Shibata, \emph{{Gravitational waves
  and neutrino emission from the merger of binary neutron stars}},
  \href{http://dx.doi.org/10.1103/PhysRevLett.107.051102}{\emph{Phys. Rev.
  Lett.} {\bf 107} (2011) 051102}, [\href{http://arxiv.org/abs/1105.2125}{{\tt
  1105.2125}}].

\bibitem{Ruiz2016}
M.~{Ruiz}, R.~N. {Lang}, V.~{Paschalidis} and S.~L. {Shapiro}, \emph{{Binary
  Neutron Star Mergers: A Jet Engine for Short Gamma-Ray Bursts}},
  \href{http://dx.doi.org/10.3847/2041-8205/824/1/L6}{\emph{Astrophys. J.
  Lett.} {\bf 824} (2016) L6}, [\href{http://arxiv.org/abs/1604.02455}{{\tt
  1604.02455}}].

\bibitem{East:2016}
W.~E. {East}, V.~{Paschalidis}, F.~{Pretorius} and S.~L. {Shapiro},
  \emph{{Relativistic simulations of eccentric binary neutron star mergers:
  One-arm spir\ al instability and effects of neutron star spin}},
  \href{http://dx.doi.org/10.1103/PhysRevD.93.024011}{\emph{\prd} {\bf 93}
  (2016) 024011}, [\href{http://arxiv.org/abs/1511.01093}{{\tt 1511.01093}}].

\bibitem{Most2019}
E.~R. {Most}, L.~J. {Papenfort} and L.~{Rezzolla}, \emph{{Beyond second-order
  convergence in simulations of magnetized binary neutron stars with realistic
  microphysics}}, \href{http://dx.doi.org/10.1093/mnras/stz2809}{\emph{\mnras}
  {\bf 490} (2019) 3588--3600}, [\href{http://arxiv.org/abs/1907.10328}{{\tt
  1907.10328}}].

\bibitem{Bauswein2019}
A.~{Bauswein}, N.-U.~F. {Bastian}, D.~B. {Blaschke}, K.~{Chatziioannou}, J.~A.
  {Clark}, T.~{Fischer} et~al., \emph{{Identifying a First-Order Phase
  Transition in Neutron-Star Mergers through Gravitational Waves}},
  \href{http://dx.doi.org/10.1103/PhysRevLett.122.061102}{\emph{\prl} {\bf 122}
  (2019) 061102}, [\href{http://arxiv.org/abs/1809.01116}{{\tt 1809.01116}}].

\bibitem{Endrizzi2018}
A.~{Endrizzi}, D.~{Logoteta}, B.~{Giacomazzo}, I.~{Bombaci}, W.~{Kastaun} and
  R.~{Ciolfi}, \emph{{Effects of chiral effective field theory equation of
  state on binary neutron star mergers}},
  \href{http://dx.doi.org/10.1103/PhysRevD.98.043015}{\emph{\prd} {\bf 98}
  (2018) 043015}, [\href{http://arxiv.org/abs/1806.09832}{{\tt 1806.09832}}].

\bibitem{Ciolfi2019}
R.~{Ciolfi}, W.~{Kastaun}, J.~V. {Kalinani} and B.~{Giacomazzo}, \emph{{First
  100 ms of a long-lived magnetized neutron star formed in a binary neutron
  star merger}},
  \href{http://dx.doi.org/10.1103/PhysRevD.100.023005}{\emph{\prd} {\bf 100}
  (2019) 023005}, [\href{http://arxiv.org/abs/1904.10222}{{\tt 1904.10222}}].

\bibitem{Tsokaros2019}
A.~{Tsokaros}, M.~{Ruiz}, V.~{Paschalidis}, S.~L. {Shapiro} and
  K.~{Ury{\={u}}}, \emph{{Effect of spin on the inspiral of binary neutron
  stars}}, \href{http://dx.doi.org/10.1103/PhysRevD.100.024061}{\emph{\prd}
  {\bf 100} (2019) 024061}, [\href{http://arxiv.org/abs/1906.00011}{{\tt
  1906.00011}}].

\bibitem{Baiotti:2016qnr}
L.~Baiotti and L.~Rezzolla, \emph{{Binary neutron star mergers: a review of
  Einstein’s richest laboratory}},
  \href{http://dx.doi.org/10.1088/1361-6633/aa67bb}{\emph{Rept. Prog. Phys.}
  {\bf 80} (2017) 096901}, [\href{http://arxiv.org/abs/1607.03540}{{\tt
  1607.03540}}].

\bibitem{Baiotti2019}
L.~{Baiotti}, \emph{{Gravitational waves from neutron star mergers and their
  relation to the nuclear equation of state}},
  \href{http://dx.doi.org/10.1016/j.ppnp.2019.103714}{\emph{Progress in
  Particle and Nuclear Physics} {\bf 109} (2019) 103714},
  [\href{http://arxiv.org/abs/1907.08534}{{\tt 1907.08534}}].

\bibitem{Faber2012:lrr}
J.~A. {Faber} and F.~A. {Rasio}, \emph{{Binary Neutron Star Mergers}},
  \href{http://dx.doi.org/10.12942/lrr-2012-8}{\emph{Living Reviews in
  Relativity} {\bf 15} (2012) 8}, [\href{http://arxiv.org/abs/1204.3858}{{\tt
  1204.3858}}].

\bibitem{Alford2018a}
M.~G. {Alford}, L.~{Bovard}, M.~{Hanauske}, L.~{Rezzolla} and K.~{Schwenzer},
  \emph{{Viscous Dissipation and Heat Conduction in Binary Neutron-Star
  Mergers}},
  \href{http://dx.doi.org/10.1103/PhysRevLett.120.041101}{\emph{\prl} {\bf 120}
  (2018) 041101}, [\href{http://arxiv.org/abs/1707.09475}{{\tt 1707.09475}}].

\bibitem{Celora2022}
T.~{Celora}, I.~{Hawke}, P.~C. {Hammond}, N.~{Andersson} and G.~L. {Comer},
  \emph{{Formulating bulk viscosity for neutron star simulations}},
  \href{http://dx.doi.org/10.1103/PhysRevD.105.103016}{\emph{\prd} {\bf 105}
  (2022) 103016}, [\href{http://arxiv.org/abs/2202.01576}{{\tt 2202.01576}}].

\bibitem{Most2022}
E.~R. {Most}, A.~{Haber}, S.~P. {Harris}, Z.~{Zhang}, M.~G. {Alford} and
  J.~{Noronha}, \emph{{Emergence of microphysical viscosity in binary neutron
  star post-merger dynamics}},
  \href{http://dx.doi.org/10.48550/arXiv.2207.00442}{\emph{arXiv e-prints}
  (2022) arXiv:2207.00442}, [\href{http://arxiv.org/abs/2207.00442}{{\tt
  2207.00442}}].

\bibitem{Camelio2023a}
G.~{Camelio}, L.~{Gavassino}, M.~{Antonelli}, S.~{Bernuzzi} and B.~{Haskell},
  \emph{{Simulating bulk viscosity in neutron stars. I. Formalism}},
  \href{http://dx.doi.org/10.1103/PhysRevD.107.103031}{\emph{\prd} {\bf 107}
  (2023) 103031}, [\href{http://arxiv.org/abs/2204.11809}{{\tt 2204.11809}}].

\bibitem{Camelio2023b}
G.~{Camelio}, L.~{Gavassino}, M.~{Antonelli}, S.~{Bernuzzi} and B.~{Haskell},
  \emph{{Simulating bulk viscosity in neutron stars. II. Evolution in spherical
  symmetry}}, \href{http://dx.doi.org/10.1103/PhysRevD.107.103032}{\emph{\prd}
  {\bf 107} (2023) 103032}, [\href{http://arxiv.org/abs/2204.11810}{{\tt
  2204.11810}}].

\bibitem{Hammond2023}
P.~{Hammond}, I.~{Hawke} and N.~{Andersson}, \emph{{Impact of nuclear reactions
  on gravitational waves from neutron star mergers}},
  \href{http://dx.doi.org/10.1103/PhysRevD.107.043023}{\emph{\prd} {\bf 107}
  (2023) 043023}.

\bibitem{Sawyer1989}
R.~F. {Sawyer}, \emph{{Bulk viscosity of hot neutron-star matter and the
  maximum rotation \ rates of neutron stars}},
  \href{http://dx.doi.org/10.1103/PhysRevD.39.3804}{\emph{\prd} {\bf 39} (1989)
  3804--3806}.

\bibitem{Alford2019a}
M.~G. {Alford} and S.~P. {Harris}, \emph{{Damping of density oscillations in
  neutrino-transparent nuclear matter}},
  \href{http://dx.doi.org/10.1103/PhysRevC.100.035803}{\emph{\prc} {\bf 100}
  (2019) 035803}, [\href{http://arxiv.org/abs/1907.03795}{{\tt 1907.03795}}].

\bibitem{Alford2019b}
M.~{Alford}, A.~{Harutyunyan} and A.~{Sedrakian}, \emph{{Bulk viscosity of
  baryonic matter with trapped neutrinos}},
  \href{http://dx.doi.org/10.1103/PhysRevD.100.103021}{\emph{\prd} {\bf 100}
  (2019) 103021}, [\href{http://arxiv.org/abs/1907.04192}{{\tt 1907.04192}}].

\bibitem{Alford2021a}
M.~G. {Alford} and A.~{Haber}, \emph{{Strangeness-changing rates and hyperonic
  bulk viscosity in neutron star mergers}},
  \href{http://dx.doi.org/10.1103/PhysRevC.103.045810}{\emph{\prc} {\bf 103}
  (2021) 045810}, [\href{http://arxiv.org/abs/2009.05181}{{\tt 2009.05181}}].

\bibitem{Alford2021c}
M.~{Alford}, A.~{Harutyunyan} and A.~{Sedrakian}, \emph{{Bulk viscosity from
  Urca processes: n p e {\ensuremath{\mu}} matter in the neutrino-trapped
  regime}}, \href{http://dx.doi.org/10.1103/PhysRevD.104.103027}{\emph{\prd}
  {\bf 104} (2021) 103027}, [\href{http://arxiv.org/abs/2108.07523}{{\tt
  2108.07523}}].

\bibitem{Alford2022}
M.~{Alford}, A.~{Harutyunyan} and A.~{Sedrakian}, \emph{{Bulk Viscosity of
  Relativistic npe{\ensuremath{\mu}} Matter in Neutron-Star Mergers}},
  \href{http://dx.doi.org/10.3390/particles5030029}{\emph{Particles} {\bf 5}
  (2022) 361--376}, [\href{http://arxiv.org/abs/2209.04717}{{\tt 2209.04717}}].

\bibitem{Alford:2023gxq}
M.~G. Alford, A.~Haber and Z.~Zhang, \emph{{Isospin Equilibration in Neutron
  Star Mergers}},  \href{http://arxiv.org/abs/2306.06180}{{\tt 2306.06180}}.

\bibitem{Alford2020}
M.~Alford, A.~Harutyunyan and A.~Sedrakian, \emph{{Bulk Viscous Damping of
  Density Oscillations in Neutron Star Mergers}},
  \href{http://dx.doi.org/10.3390/particles3020034}{\emph{Particles} {\bf 3}
  (2020) 500--517}, [\href{http://arxiv.org/abs/2006.07975}{{\tt 2006.07975}}].

\bibitem{Shapiro1983}
S.~L. {Shapiro} and S.~A. {Teukolsky}, \emph{{Black holes, white dwarfs and
  neutron stars. The physics of compact objects}}.
\newblock  WILEY‐VCH Verlag GmbH, 1983.

\bibitem{Greiner2000gauge}
W.~Greiner and B.~M{\"u}ller, \emph{Gauge Theory of Weak Interactions}.
\newblock Physics and Astronomy Online Library. Springer, 2000, pp. 30, 260.

\bibitem{Alford2018b}
M.~G. {Alford} and S.~P. {Harris}, \emph{{{$\beta$} equilibrium in neutron-star
  mergers}}, \href{http://dx.doi.org/10.1103/PhysRevC.98.065806}{\emph{\prc}
  {\bf 98} (2018) 065806}, [\href{http://arxiv.org/abs/1803.00662}{{\tt
  1803.00662}}].

\bibitem{Guo:2020tgx}
G.~Guo, G.~Mart\'\i{}nez-Pinedo, A.~Lohs and T.~Fischer, \emph{{Charged-Current
  Muonic Reactions in Core-Collapse Supernovae}},
  \href{http://dx.doi.org/10.1103/PhysRevD.102.023037}{\emph{Phys. Rev. D} {\bf
  102} (2020) 023037}, [\href{http://arxiv.org/abs/2006.12051}{{\tt
  2006.12051}}].

\bibitem{Alford:2010jf}
M.~G. Alford and G.~Good, \emph{{Leptonic contribution to the bulk viscosity of
  nuclear matter}},
  \href{http://dx.doi.org/10.1103/PhysRevC.82.055805}{\emph{Phys. Rev.} {\bf
  C82} (2010) 055805}, [\href{http://arxiv.org/abs/1003.1093}{{\tt
  1003.1093}}].

\bibitem{Lalazissis2005}
G.~A. {Lalazissis}, T.~{Nik{\v s}i{\'c}}, D.~{Vretenar} and P.~{Ring},
  \emph{{New relativistic mean-field interaction with density-dependent
  meson-nucleon couplings}},
  \href{http://dx.doi.org/10.1103/PhysRevC.71.024312}{\emph{\prc} {\bf 71}
  (2005) 024312}.

\bibitem{Lalazissis1997}
G.~A. {Lalazissis}, J.~{K{\"o}nig} and P.~{Ring}, \emph{{New parametrization
  for the Lagrangian density of relativistic mean field theory}},
  \href{http://dx.doi.org/10.1103/PhysRevC.55.540}{\emph{\prc} {\bf 55} (1997)
  540--543}, [\href{http://arxiv.org/abs/nucl-th/9607039}{{\tt
  nucl-th/9607039}}].

\bibitem{Glendenning_book}
N.~K. {Glendenning}, \emph{{Compact stars: nuclear physics, particle physics,
  and general relativity}}.
\newblock Springer, New York, N.Y., 2000.

\bibitem{Typel1999}
S.~{Typel} and H.~H. {Wolter}, \emph{{Relativistic mean field calculations with
  density-dependent meson-nucleon coupling}},
  \href{http://dx.doi.org/10.1016/S0375-9474(99)00310-3}{\emph{\nphysa} {\bf
  656} (1999) 331--364}.

\bibitem{Alford2021b}
M.~G. {Alford}, A.~{Haber}, S.~P. {Harris} and Z.~{Zhang}, \emph{{Beta
  Equilibrium Under Neutron Star Merger Conditions}},
  \href{http://dx.doi.org/10.3390/universe7110399}{\emph{Universe} {\bf 7}
  (2021) 399}, [\href{http://arxiv.org/abs/2108.03324}{{\tt 2108.03324}}].

\bibitem{Haensel2000}
P.~{Haensel}, K.~P. {Levenfish} and D.~G. {Yakovlev}, \emph{{Bulk viscosity in
  superfluid neutron star cores. I. Direct Urca processes in npemu matter}},
  {\emph{\aap} {\bf 357} (2000) 1157--1169},
  [\href{http://arxiv.org/abs/astro-ph/0004183}{{\tt astro-ph/0004183}}].

\bibitem{Gavassino2021CQGra}
L.~{Gavassino}, M.~{Antonelli} and B.~{Haskell}, \emph{{Bulk viscosity in
  relativistic fluids: from thermodynamics to hydrodynamics}},
  \href{http://dx.doi.org/10.1088/1361-6382/abe588}{\emph{Classical and Quantum
  Gravity} {\bf 38} (2021) 075001},
  [\href{http://arxiv.org/abs/2003.04609}{{\tt 2003.04609}}].

\bibitem{Harutyunyan2023Symm}
A.~{Harutyunyan} and A.~{Sedrakian}, \emph{{Phenomenological Relativistic
  Second-Order Hydrodynamics for Multiflavor Fluids}},
  \href{http://dx.doi.org/10.3390/sym15020494}{\emph{Symmetry} {\bf 15} (2023)
  494}, [\href{http://arxiv.org/abs/2302.09596}{{\tt 2302.09596}}].

\bibitem{Yakovlev2001}
D.~G. {Yakovlev}, A.~D. {Kaminker}, O.~Y. {Gnedin} and P.~{Haensel},
  \emph{{Neutrino emission from neutron stars}},
  \href{http://dx.doi.org/10.1016/S0370-1573(00)00131-9}{\emph{Physics Reports}
  {\bf 354} (2001) 1--155}, [\href{http://arxiv.org/abs/astro-ph/0012122}{{\tt
  astro-ph/0012122}}].

\bibitem{Haensel1992PhRvD}
P.~{Haensel} and R.~{Schaeffer}, \emph{{Bulk viscosity of hot-neutron-star
  matter from direct URCA processes}},
  \href{http://dx.doi.org/10.1103/PhysRevD.45.4708}{\emph{\prd} {\bf 45} (1992)
  4708--4712}.

\end{thebibliography}

\end{document}